%% file: main.tex
\documentclass[runningheads]{llncs}

\usepackage{eccv}

\usepackage{eccvabbrv}

\newcommand{\myparatight}[1]{\smallskip\noindent{\bf {#1}:}~}

\usepackage{amsmath}
\usepackage{amssymb}

\usepackage{float}
 \usepackage{algorithm} 
\usepackage[noend]{algorithmic}
\renewcommand{\algorithmicrequire}{\textbf{Input:}}
\renewcommand{\algorithmicensure}{\textbf{Output:}}

\DeclareMathOperator*{\argmax}{arg\,max}
\DeclareMathOperator*{\argmin}{arg\,min}

\usepackage{graphicx}
\usepackage{booktabs}
\usepackage[accsupp]{axessibility}

\usepackage[breaklinks,colorlinks,citecolor=eccvblue]{hyperref}

\begin{document}

\title{Certifiably Robust Image Watermark} 


\author{Zhengyuan Jiang\inst{1} \and
Moyang Guo\inst{1} \and
Yuepeng Hu\inst{1} \and \\
Jinyuan Jia\inst{2} \and
Neil Zhenqiang Gong\inst{1}
}

\authorrunning{Z. Jiang et al.}

\institute{$^1$Duke University, $^2$Pennsylvania State University}

\maketitle
\input{0_abstract}
\input{1_introduction}
\input{2_background}
\input{3_problem}
\input{4_method}
\input{5_evaluation}
\input{7_conclusion}

\input{9_acknowledgement}

\bibliographystyle{splncs04}
\bibliography{refs}

\input{8_appendix}

\end{document}

%% file: 0_abstract.tex
\begin{abstract}
Generative AI raises many societal concerns such as boosting disinformation and propaganda campaigns. Watermarking AI-generated content is a key technology to address these concerns and has been widely deployed in industry. However, watermarking is vulnerable to \emph{removal attacks} and \emph{forgery attacks}. In this work, we propose the \emph{first} image watermarks with certified robustness guarantees against removal and forgery attacks. Our method leverages \emph{randomized smoothing}, a popular technique to build certifiably robust classifiers and regression models. Our major technical contributions include extending randomized smoothing to watermarking by considering its unique characteristics, deriving the certified robustness guarantees, and designing algorithms to estimate them. Moreover, we extensively evaluate our image watermarks in terms of both certified and empirical robustness. Our code is available at \url{https://github.com/zhengyuan-jiang/Watermark-Library}.
\keywords{Watermark, Certified Robustness, Watermark Removal/Forgery}
\end{abstract}

%% file: 1_introduction.tex
\section{Introduction}
Generative AI (\emph{GenAI})--such as Stable Diffusion~\cite{stablediffusion}, Midjourney~\cite{midjourney}, and DALL-E~\cite{dalle}--brings revolutionary capabilities in producing images. This technological advancement offers immense possibilities for  content creation. However, it also introduces critical ethical concerns. For instance, the ease of generating realistic content raises alarms about its potential misuse in spreading disinformation and propaganda. These ethical dilemmas highlight the urgent need to manage GenAI responsibly, balancing its creative benefits against the risks of misuse. 

To deal with these risks, watermarking has been leveraged to detect AI-generated images~\cite{jiang2024watermark}. The Executive Order on trustworthy AI issued by the White House lists watermarking AI-generated content as a key technology~\cite{Executive-Order}. Indeed, many AI companies--such as OpenAI, Google, and Stability AI--have deployed watermarking in their GenAI services to mark AI-generated images~\cite{openai-c2pa,google-synthid,stable-diffusion-watermark}. Specifically, a bitstring watermark (called \emph{ground-truth watermark}) is embedded into  AI-generated images at generation using a \emph{watermarking encoder}; and an image is detected as AI-generated if the corresponding \emph{watermarking decoder} can decode a similar watermark from it. Formally, if the \emph{bitwise accuracy (BA)} of the watermark decoded from an image is no smaller than a \emph{detection threshold} $\tau$, the image is predicted as AI-generated. BA of a watermark is the fraction of its bits that match with those of the ground-truth watermark. 

However, existing watermarking methods are not robust to \emph{removal attacks} and \emph{forgery attacks}~\cite{jiang2023evading,wang2021watermark}, which aim to remove the watermark from a watermarked image and forge the watermark in a non-watermarked image, respectively. Specifically, a removal attack (or forgery attack) adds a perturbation to a watermarked image (or non-watermarked image) to remove (or forge) the watermark, i.e., BA of the watermark decoded from the perturbed image is smaller (or no smaller) than $\tau$. Existing defenses against removal/forgery attacks mainly rely on \emph{adversarial training}~\cite{zhu2018hidden}, which considers removal/forgery attacks when training the  encoder and decoder. However, adversarial training is only robust to the removal/forgery attacks that are considered during training, and can still be defeated by strong, adaptive  attacks. For instance, Jiang et al.~\cite{jiang2023evading} showed that a strong removal attack can still remove the watermark from a watermarked image without sacrificing its visual quality even if adversarial training is used. 

We propose the \emph{first} image watermark that is \emph{certifiably robust} against removal and forgery attacks. Certifiable robustness means that our watermark is robust to \emph{any} removal (or forgery) attacks that add $\ell_2$-norm bounded perturbations to watermarked (or non-watermarked) images. Our method extends \emph{randomized smoothing}~\cite{cohen2019certified}, a popular technique to build certifiably robust classifiers and regression models, to watermark by considering its unique characteristics, e.g., watermark is a bitstring. 

Given any watermarking decoder, we build a \emph{smoothed decoder} via adding random Gaussian noise to an image. Specifically, we propose three ways to build a smoothed decoder, i.e., \emph{multi-class smoothing}, \emph{multi-label smoothing}, and \emph{regression smoothing}. In multi-class smoothing, we treat decoding each bit of the  watermark from an image as a binary classification problem; in multi-label smoothing, we treat decoding a watermark from an image as a multi-label classification problem, where the $i$th bit of the decoded watermark is 1 means that the watermark has label $i$; and in regression smoothing, we treat decoding a watermark from an image as a regression problem, where  BA of the decoded watermark is treated as the regression outcome.     

We derive certified robustness guarantees of our smoothed decoder for the three ways of smoothing. Specifically, given any image, we derive a lower bound and an upper bound for the BA of the watermark decoded by our smoothed decoder, no matter what perturbation a removal or forgery attack adds to the image once the $\ell_2$-norm of the perturbation is bounded. Our lower bound guarantees that no removal attacks with $\ell_2$-norm bounded perturbations can remove the watermark from a watermarked image; and our upper bound guarantees that no forgery attacks  with $\ell_2$-norm bounded perturbations can forge the watermark in a non-watermarked image. We also propose randomized algorithms to estimate the lower bound and upper bound for any given image. Our randomized algorithm adds multiple Gaussian noise to the image and estimates the lower/upper bounds with probabilistic guarantees. 

We conduct empirical evaluation on three AI-generated image datasets and non-AI-generated images. We adopt  HiDDeN~\cite{zhu2018hidden} as a base watermarking method and use our method to smooth it. We evaluate both certified and empirical robustness. Certified robustness measures the detection performance under \emph{any} removal and forgery attacks, which is only applicable to our smoothed decoder; while empirical robustness measures the detection performance under state-of-the-art attacks, which is applicable to both base decoder and our smoothed decoder. We find that regression smoothing outperforms multi-class and multi-label smoothing. This is because regression smoothing better takes the correlations between bits of the watermark into consideration. Moreover, other than achieving certified robustness, we find that  smoothing also improves empirical robustness, i.e., smoothed HiDDeN outperforms HiDDeN.

%% file: 2_background.tex
\section{Related Works}
\subsection{Watermarking}
We consider a watermarking method~\cite{pereira2000robust,zhu2018hidden,tancik2020stegastamp,luo2020distortion,zhang2020udh,yoo2022deep,fernandez2023stable,zhao2023protecting,wen2024tree,fang2023denol}  defined by a triple ($w_t$, $E$, $D$). The ground-truth watermark $w_t$ is a bitstring with $m$ bits; the encoder  $E$ embeds $w_t$ into an image to produce a watermarked image; and the decoder $D$ decodes a watermark from an image. Our method can transform \emph{any} such watermarking method to be certifiably robust. Note that $D$ can decode a watermark from any (watermarked or non-watermarked) image, and the decoded watermark is supposed to be similar to $w_t$ when the image is watermarked. The encoder and $w_t$ can also be integrated into the parameters of a GenAI model such that its generated images are inherently watermarked with $w_t$, e.g., Stable Signature is such a watermarking method~\cite{fernandez2023stable}. Our method is also applicable in such scenarios because decoding watermarks from images only involves the decoder $D$ which our method smooths. In state-of-the-art watermarking methods~\cite{zhu2018hidden}, $E$ and $D$ are jointly trained using an image dataset such that 1) a watermarked image produced by $E$ is visually similar to the image before watermarking, and 2) $D$ can accurately decode the watermark in a watermarked image produced by $E$.

\subsection{Watermark Removal and Forgery Attacks}
Removal attacks~\cite{jiang2023evading,lukas2023leveraging,saberi2023robustness,an2024benchmarking,zhao2023invisible,nie2022diffusion,hu2024transfer,hu2024stable} aim to remove the watermark in a watermarked image via adding a small perturbation to it; while forgery attacks~\cite{saberi2023robustness} aim to forge the watermark in a non-watermarked image via adding a small perturbation to it. A removal attack can often be adapted as a forgery attack. In particular, we can adapt the objective of a removal attack when finding the perturbation such that the watermark is falsely detected in the perturbed image. Different attacks use different methods to  find the perturbations. For instance, commonly used image editing methods--such as JPEG compression and Gaussian noise--can be used to find the perturbations. An attacker can also use more advanced, optimization-based  methods. For instance, Jiang et al.~\cite{jiang2023evading} proposed an optimization-based method to find the perturbations in the white-box setting where the attacker has access to the decoder; and they also proposed a query-based method to find the perturbations in the black-box setting where the attacker can repeatedly query the watermark detector API, which returns a binary prediction (i.e., watermarked or non-watermarked) for an image.

\subsection{Randomized Smoothing}
Randomized smoothing~\cite{cohen2019certified,jia2019certified,jia2022multiguard,chiang2020detection,jiang2023ipcert} is a state-of-the-art technique to build certifiably robust classifiers and regression models. Roughly speaking, given a classifier (called \emph{base classifier}) or regression model (called \emph{base regression model}), randomized smoothing builds a smoothed classifier or regression model by adding isotropic Gaussian noise to the input (i.e., an image in this work) of the base classifier or regression model. The smoothed classifier (or smoothed regression model) is guaranteed to predict the same label (or a similar response) for an image when the $\ell_2$-norm of the perturbation added to it is bounded.

\myparatight{Multi-class smoothing} In this smoothing~\cite{cohen2019certified,jia2019certified,jiang2023ipcert}, the base classifier $f$ is a multi-class classifier. Given an image $x$, the smoothed classifier $g$ predicts the label that is the most likely to be predicted by $f$ when adding isotropic Gaussian noise to $x$. Formally,  the predicted label is $g(x) = \argmax_{c \in \mathcal{Y}} \text{Pr}(f(x+\epsilon)=c)$, where $\epsilon \sim \mathcal{N}(0, \sigma^{2} I)$ is an isotropic Gaussian noise and $\mathcal{Y}$ is the set of labels.  $g$ predicts the same label for $x$ once the $\ell_2$-norm of the perturbation added to it is bounded by $r(x)$. Formally, we have: 
\begin{align}
\label{equ:cohen}
    g(x+\delta) &= g(x)=l, \quad \forall \|\delta\|_2 < r(x), \\
    r(x) &= \sigma \Phi^{-1}(\underline{p_l}),
\end{align}
where $\underline{p_l}$ is a lower bound of $\text{Pr}(f(x+\epsilon)=l)$, i.e., $\underline{p_l}\leq \text{Pr}(f(x+\epsilon)=l)$, and $\Phi^{-1}$  is the inverse cumulative distribution function of the standard Gaussian.

\myparatight{Multi-label smoothing} In this smoothing~\cite{jia2022multiguard}, the base classifier $f$ is a multi-label classifier, which predicts a set of $k'$ labels for an image $x$. The smoothed classifier $g$ predicts the $k$ labels that are most likely to be predicted by $f$ when adding isotropic Gaussian noise to $x$. Formally, the predicted labels are $g(x) = \text{argk-max}_{c \in \mathcal{Y}} \text{Pr} (c \in f(x+\epsilon))$, where $\epsilon \sim \mathcal{N}(0, \sigma^{2} I)$ and argk-max means the $k$ labels that are most likely to be predicted by $f(x+\epsilon)$. When the perturbation $\delta$ added to $x$ is $\ell_2$-norm bounded by $R$, the intersection size between the  predicted labels $g(x+\delta)$ and the set of ground-truth labels $L$ of $x$  is at least $e(x)$, i.e., we have the following~\cite{jia2022multiguard}: 
\begin{align}
    \left|L \cap g(x+\delta)\right| \geq e(x), \quad \forall \|\delta\|_2 < R,
    \label{equ:multiguard}
\end{align}
where $e(x)$ depends on a lower bound of the probability $\text{Pr} (c \in f(x+\epsilon))$ for each $c\in L$  and an upper bound of the probability $\text{Pr} (c \in f(x+\epsilon))$ for each $c\in \mathcal{Y}/L$. The complete form of $e(x)$ is rather complex and omitted for simplicity.

\myparatight{Regression smoothing} In this smoothing~\cite{chiang2020detection,bansal2022certified},  the base regression model $f$ predicts a continuous value for a given image $x$.  The smoothed regression model $g$ predicts the median of all possible values that can be predicted by $f$ when adding  isotropic Gaussian noise to $x$. Formally, the predicted value is $g(x)=\argmin_y \text{Pr}(f(x+\epsilon)\leq y)\geq 0.5$, where $\epsilon \sim \mathcal{N}(0, \sigma^{2} I)$.  When the $l_2$-norm of the perturbation added to $x$ is bounded by $R$, $g(x+\delta)$ is bounded as follows:
\begin{align}
    \underline{g}(x) &\leq g(x+\delta) \leq \overline{g}(x), \quad \forall \|\delta\|_2 < R, 
\label{equ:median smoothing}
\end{align}
where $\underline{g}(x) = \sup \{y \in \mathbb{R} \mid \text{Pr}(f(x+\epsilon) \leq y) \leq \Phi(-\frac{R}{\sigma}) \}$,  $\overline{g}(x) = \inf \{y \in \mathbb{R} \mid \text{Pr}(f(x+\epsilon) \leq y) \geq {\Phi(\frac{R}{\sigma})}\}$, and $\Phi$ is the cumulative distribution function of the standard Gaussian.

%% file: 3_problem.tex
\section{Problem Formulation}
\label{threat model}

\vspace{-2mm}
\myparatight{Notations}
We use $x$, $x_w$, and $x_n$ to represent an image, a watermarked image, and a non-watermarked image, respectively. $x$ can be either a watermarked or non-watermarked image. The ground-truth watermark $w_t$ has  $m$ bits and $w_t[i]$ is the $i$th bit of $w_t$, where $i=1,2,\cdots,m$. $E(x_n,w_t)$ means embedding $w_t$ into $x_n$ to produce $x_w$; while $D(x)$ is the watermark decoded from $x$. $BA(w, w_t)$ is the bitwise accuracy of watermark $w$, which is the fraction of its bits that match with those of $w_t$. Formally, $BA(w, w_t) = \frac{1}{m}\sum_{i=1}^m \mathbb{I}(w[i]=w_t[i])$, where $\mathbb{I}$ is an indicator function whose output is 1 if the condition is satisfied and 0 otherwise. 
An image $x$ is detected as watermarked if $BA(D(x),w_t)\geq \tau$. 

\myparatight{Threat model} In a removal attack, an attacker aims to add a small perturbation $\delta$ to a watermarked image $x_w$ to remove the watermark, i.e., $BA(D(x_w+\delta),w_t)< \tau$; while in a forgery attack, an attacker aims to add a small perturbation $\delta$ to a non-watermarked image $x_n$ to forge the watermark, i.e., $BA(D(x_n+\delta),w_t)\geq \tau$. We assume the attacker can use \emph{any} removal or forgery attack to find the perturbation $\delta$. Moreover, the attacker knows everything about the watermarking method, e.g., its ground-truth watermark, encoder parameters,  decoder parameters, and the smoothing process.

\myparatight{Certifiably robust watermark} A watermarking method $(w_t, E, D)$ is certifiably robust if BA of the watermark decoded from any image $x$ has a lower bound and upper bound when the $\ell_2$-norm of the perturbation added to it is bounded by $R$. Formally, we have the following definition:
\begin{definition}[Certifiably Robust Watermark]
\label{definitionrobust}
Given a watermarking method $(w_t, E, D)$ and any image $x$. Suppose a perturbation $\delta$, whose $\ell_2$-norm is bounded by $R$, is added to $x$. We say the watermarking method is  certifiably robust if the following is satisfied:
\begin{align}
\label{robustwatermarkdefinition}
    &  \underline{BA}(x) \leq BA(D(x+\delta), w_t) \leq \overline{BA}(x), \quad \forall \|\delta\|_2 < R,
\end{align}
\end{definition}
where $\underline{BA}(x)$ is a lower bound and $\overline{BA}(x)$ is an upper bound of BA under perturbation. For a watermarked image $x_w$, a certifiably robust watermark defends against any removal attacks with at most $R$ $\ell_2$-norm perturbations, once the lower bound $\underline{BA}(x_w)$ is no smaller than $\tau$; and  for a non-watermarked image $x_n$, a certifiably robust watermark defends against any forgery attacks with at most $R$ $\ell_2$-norm perturbations, once the upper bound $\overline{BA}(x_n)$ is smaller than $\tau$. Note that Equation~\ref{robustwatermarkdefinition} only involves $D$ and $w_t$ of the watermarking method, but not $E$ explicitly. However, when $E$ and $D$ are jointly trained well, the lower bound  $\underline{BA}(x_w)$ for a watermarked image $x_w=E(x_n, w_t)$ is larger, while the upper bound $\overline{BA}(x_n)$ for a non-watermarked image $x_n$ is smaller, making the watermarking method more certifiably robust. 

%% file: 4_method.tex
\section{\label{method} Our Smoothing Framework}
\subsection{Overview}
\begin{figure*}[!t]
    \centering
    \includegraphics[width=1\linewidth]{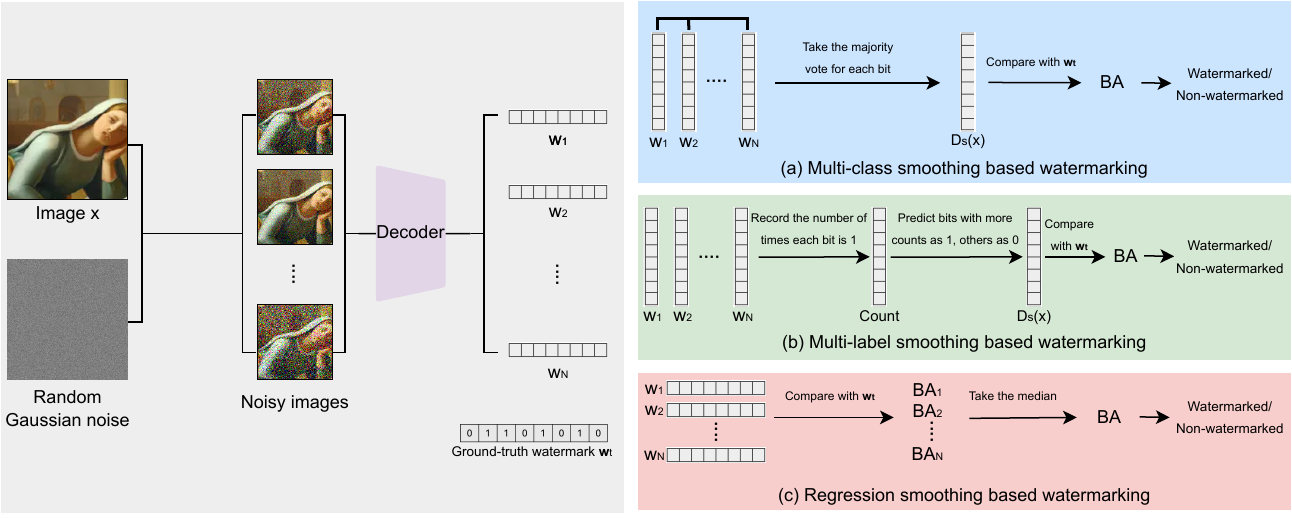}
    \vspace{-2mm}
    \caption{Illustration of our smoothing framework with three variants.}
    \label{fig:pipeline}
    \vspace{-4mm}
\end{figure*}

Given any watermarking method $(w_t,E,D)$, we build a certifiably robust watermarking method $(w_t,E,D_s)$ by smoothing $D$ as $D_s$. We smooth $D$ but not $E$ because only $D$ is involved during watermark detection. Specifically, given an image $x$, we add $N$ isotropic Gaussian noise $\epsilon_1, \epsilon_2, \cdots, \epsilon_N$ to it to construct $N$ noisy images  $x+\epsilon_1, x+\epsilon_2, \cdots, x+\epsilon_N$. Then, we use $D$ to decode a watermark for each noisy image. We propose three smoothing methods to aggregate the $N$ decoded watermarks to calculate bitwise accuracy. In multi-class smoothing, we treat decoding each bit as a binary classification problem and aggregate bits of the watermark separately. In multi-label smoothing, we treat decoding a watermark from a (noisy) image as a multi-label classification problem, where the $i$th bit is 1 means that the watermark has label $i$. In regression smoothing, we treat the bitwise accuracy of a decoded watermark for a (noisy) image as a regression response and directly obtain a smoothed bitwise accuracy. Figure~\ref{fig:pipeline} illustrates our three  methods to smooth $D$ to obtain a bitwise accuracy for an image $x$. 

\subsection{\label{building smoothed decoder}Building a Smoothed Decoder $D_s$}
\vspace{-2mm}
\myparatight{Multi-class smoothing based watermarking} In our first smoothing method, we treat decoding each bit of a watermark from an image $x$ as a binary classification problem and leverage multi-label smoothing to build a smoothed decoder $D_s$ based on $D$. Specifically, we define a binary base classifier $f_i$ for each $i$th bit, i.e., $f_i(x)=D(x)[i]$, where $i=1,2,\cdots,m$ and $D(x)[i]$ is the $i$th bit of the decoded watermark $D(x)$. We add Gaussian noise $\epsilon$ to $x$. Our $i$th smoothed classifier $g_i$ predicts a binary label for $x$ as follows:  $g_i(x)=\argmax_{c \in \{0,1\}} \text{Pr}(f_i(x+\epsilon)=c)$, where $\epsilon \sim \mathcal{N}(0, \sigma^2 I)$. We treat $g_i(x)$ as the $i$th bit of the watermark $D_{s}(x)$ decoded by the smoothed decoder $D_s$. Formally, we have: 
\begin{align} 
  D_{s}(x)[i] = g_i(x)=\argmax_{c \in \{0,1\}} \text{Pr}(D(x+\epsilon)[i]=c), i=1,2,\cdots,m. 
\end{align}

\myparatight{Multi-label smoothing based watermarking} In this smoothing method, we treat decoding a watermark from an image $x$ as a multi-label classification problem and utilize multi-label smoothing to build a smoothed decoder $D_s$ based on $D$. Specifically, we define the set of labels $\mathcal{Y}=\{1,2,\cdots,m\}$, where a label corresponds to the index of a bit. Given $D$, we define a base multi-label classifier $f$, which predicts $k'$ labels for an image $x$ whose corresponding bits of the decoded watermark $D(x)$ are the most likely to be 1. Note that we can also define $f$ to predict the $k'$ labels whose corresponding bits of $D(x)$ are most likely to be 0 (i.e., having a label is mapped to a bit 0), which we found to achieve similar certified robustness. Formally, we denote by $Z_i(x)$ the logit of the $i$th bit outputted by $D$ for an image $x$, i.e., $Z$ is the second-to-last layer of $D$. $f$ predicts $k'$ labels for $x$ as follows: $f(x) = \text{argk}'\text{-max}_{i \in \mathcal{Y}} Z_i(x)$, where  $\text{argk}'\text{-max}$ is the set of $k'$ labels that have the largest values of $Z_i(x)$. We add Gaussian noise $\epsilon$ to $x$. Our smoothed multi-label classifier $g$ predicts $k$ labels for $x$ as follows: $g(x) = \text{argk-max}_{i \in \mathcal{Y}} \text{Pr} (i \in f(x+\epsilon))$, where $\epsilon \sim \mathcal{N}(0, \sigma^2 I)$. Based on $g$, we formally define the watermark $D_s(x)$ decoded by $D_s$ for $x$ as follows:
\begin{align}
    D_s(x)[i] =  
    \begin{cases} 
    1, & \text{if } i \in g(x), \\
    0, & \text{if } i \notin g(x).
    \end{cases}
\end{align}

\myparatight{Regression smoothing based watermarking} In both multi-class and multi-label smoothing, we explicitly obtain the watermark $D_s(x)$ decoded by $D_s$ for an image $x$. When detecting watermarked images,  we further calculate BA of the decoded watermark $D_s(x)$ and predict $x$ to be watermarked if $BA(D_s(x), w_t)\geq \tau$. Since detection eventually relies on BA, in our third smoothing method, we directly compute $BA(D_s(x), w_t)$ without explicitly obtaining the watermark $D_s(x)$. Specifically, we treat computing BA for an image as a regression problem and leverage regression smoothing. As our experiments will show, directly computing $BA(D_s(x), w_t)$ via regression smoothing outperforms multi-class and multi-label smoothing, because it can better take the correlations between bits into consideration. Given $D$, we define a base regression model $f$ as follows: $f(x)=BA(D(x),w_t)$. We add Gaussian noise $\epsilon$ to $x$ and our smoothed regression model $g$ is as follows: $g(x)=\argmin_y \text{Pr}(f(x+\epsilon)\leq y)\geq 0.5$. Given $g$, we define $BA(D_s(x), w_t)$ as follows:
\begin{align}
    BA(D_s(x), w_t) = g(x) = \argmin_y \text{Pr}(BA(D(x+\epsilon),w_t)\leq y)\geq 0.5,
\end{align}
where $\epsilon \sim \mathcal{N}(0, \sigma^2 I)$.

\subsection{\label{deriving certified robustness}Deriving Certified Robustness}
We show that our smoothed watermarking method $(w_t,E,D_s)$ is certifiably robust (i.e., satisfies Definition~\ref{definitionrobust}) for each of the three smoothing methods. In particular, given any image $x$, we can derive a lower bound $\underline{BA}(x)$ and an upper bound $\overline{BA}(x)$ of $BA(D_s(x+\delta), w_t)$, which is BA of the watermark decoded by $D_s$ from a perturbed image $x+\delta$. Proofs of our theorems are shown in Appendix.

\begin{theorem}[\label{theorem:multi-class}Certified Robustness of Multi-class Smoothing based Watermarking]
 Our  watermarking method $(w_t,E,D_s)$ obtained by multi-class smoothing is certifiably robust for any image $x$. Specifically, when the perturbation $\delta$ added to $x$ is bounded by $R$, we can derive the following lower bound $\underline{BA}(x)$ and upper bound $\overline{BA}(x)$ for $BA(D_s(x+\delta),w_t)$: 
\begin{align}
    \underline{BA}(x) &= \frac{1}{m} \sum_{i=1}^m \mathbb{I}(D_s(x)[i]=w_t[i]) \cdot \mathbb{I}(r_i(x) \geq R), \\
    \overline{BA}(x) &= 1 - \frac{1}{m} \sum_{i=1}^m \mathbb{I}(D_s(x)[i]=\neg w_t[i]) \cdot \mathbb{I}(r_i(x) \geq R), \quad \forall \|\delta\|_2<R,
\end{align}
where $\mathbb{I}$ is the indicator function,  $r_i(x) = \sigma \Phi^{-1}(\underline{p_{l_i}})$, $\Phi^{-1}$ is the inverse cumulative distribution function of the standard Gaussian, $\underline{p_{l_i}}$ is a lower bound of $Pr(D(x+\epsilon)[i]=D_s(x)[i])$, i.e., $\underline{p_{l_i}} \leq Pr(D(x+\epsilon)[i]=D_s(x)[i])$, $\epsilon \sim \mathcal{N}(0, \sigma^2 I)$, and $\neg w_t$ means flipping each bit of the watermark $w_t$.
\end{theorem}

\begin{theorem}[\label{theorem:multi-label}Certified Robustness of Multi-label Smoothing based Watermarking]
Our watermarking method $(w_t,E,D_s)$ obtained by multi-label smoothing is certifiably robust for any image $x$. Specifically, when the perturbation $\delta$ added to $x$ is bounded by $R$, we can derive the following lower bound $\underline{BA}(x)$ and upper bound $\overline{BA}(x)$ for $BA(D_s(x+\delta),w_t)$: 
\begin{align}
    \underline{BA}(x) &= 1 - \frac{\|w_t\|_1 + k - 2\underline{e}(x)}{m}, \\
    \overline{BA}(x) &= 1 - \frac{\|w_t\|_1 - k + 2\overline{e}(x)}{m}, \quad \forall \|\delta\|_2<R,
\end{align}
where $\|w_t\|_1$ is $\ell_1$-norm of $w_t$, i.e., the number of ones in $w_t$, $\underline{e}(x) = \sup \{e \in \mathbb{N} \mid \left|L \cap g(x+\delta)\right| \geq e, \forall \|\delta\|_2<R \}$, $L$ is the set of indices of ones in $w_t$, i.e., $L=\{i\in \mathcal{Y}|w_t[i]=1\}$, and $\overline{e}(x) = \sup \{e \in \mathbb{N} \mid \left|\mathcal{Y}/L \cap g(x+\delta)\right| \geq e, \forall \|\delta\|_2<R  \}$. $\mathcal{Y}=\{1,2,\cdots,m\}$, $g$ is the smoothed multi-label classifier defined in Section~\ref{building smoothed decoder} for multi-label smoothing based watermarking and $g$ returns $k$ labels. 
\end{theorem}

\begin{theorem}[\label{theorem:regression}Certified Robustness of Regression Smoothing based Watermarking] 
Our watermarking method $(w_t,E,D_s)$ obtained by regression smoothing is certifiably robust for any image $x$. Specifically, when the perturbation $\delta$ added to $x$ is bounded by $R$, we can derive the following lower bound $\underline{BA}(x)$ and upper bound $\overline{BA}(x)$ for $BA(D_s(x+\delta),w_t)$: 
\begin{align}
    \underline{BA}(x) &= \sup \{y \in \mathbb{R} \mid \text{Pr}\left(BA(D(x+\epsilon), w_t) \leq y\right) \leq \Phi(-\frac{R}{\sigma})\}, \\
    \overline{BA}(x) &= \inf \{y \in \mathbb{R} \mid \text{Pr}\left(BA(D(x+\epsilon), w_t) \leq y\right) \geq \Phi(\frac{R}{\sigma})\}, \quad \forall \|\delta\|_2<R,
\end{align}
where $\epsilon \sim \mathcal{N}(0, \sigma^2 I)$, and $\Phi$ is the cumulative distribution function of the standard Gaussian.
\end{theorem}

\subsection{\label{estimating bound}Estimating $BA(D_s(x),w_t)$, $\underline{BA}(x)$, and $\overline{BA}(x)$}
In practice, given an image $x$ and a smoothed decoder $D_s$, we need to estimate $BA(D_s(x),w_t)$ in order to predict whether $x$ is watermarked or not when no perturbations are added to $x$. Moreover, we further compute $\underline{BA}(x)$ and $\overline{BA}(x)$ to measure certified robustness under attacks. 
Towards these goals, we sample $N$ isotropic Gaussian noise $\epsilon_1,\epsilon_2,\cdots,\epsilon_N$ uniformly at random from $\mathcal{N}(0, \sigma^2 I)$, and add them to $x$ to obtain $N$ noisy images $x+\epsilon_1,x+\epsilon_2,\cdots,x+\epsilon_N$. Then, we use the base decoder $D$ to decode a watermark from each noisy image. Finally,  we aggregate the $N$ decoded watermarks to estimate $BA(D_s(x),w_t)$, $\underline{BA}(x)$, and $\overline{BA}(x)$. Our estimations with $N$ random Gaussian noise are correct  with a confidence level $1-\alpha$, where $\alpha$ can be set to be arbitrarily small.

\myparatight{Multi-class smoothing based watermarking} We estimate the $i$th bit of $D_s(x)$ via majority vote of the $i$th bits in the $N$ decoded watermarks, i.e., $D_s(x)[i]=\argmax_{c \in \{0,1\}} \sum_{j=1}^N \mathbb{I}(D(x+\epsilon_j)[i]=c)$. Given $D_s(x)$, we can calculate $BA(D_s(x),w_t)$. We denote by $N_{i}=\sum_{j=1}^N \mathbb{I}(D(x+\epsilon_j)[i]=D_s(x)[i])$  the number of decoded watermarks whose $i$th bits are $D_s(x)[i]$. $N_{i}$ follows a binomial distribution $N_{i} \sim B(N, p_{l_i})$, where $p_{l_i}=Pr(D(x+\epsilon)[i]=D_s(x)[i])$. Therefore, based on the Clopper-Pearson method~\cite{clopper1934use}, we can estimate a lower bound $\underline{p_{l_i}}$ of $p_{l_i}$ for all $i=1,2,\cdots,m$ \emph{simultaneously} as follows:
\begin{align}
& \underline{p_{l_i}}=\operatorname{Beta}(\frac{\alpha}{m}; N_{i}, N-N_{i}+1),
\end{align}
where $1-\frac{\alpha}{m}$ is the confidence level for estimating one $\underline{p_{l_i}}$, and $\operatorname{Beta}(\alpha; u, v)$ is the $\alpha$th quantile of the Beta distribution with shape parameters $u$ and $v$. According to the \emph{Bonferroni correction}~\cite{bonferroni1936teoria}, the overall confidence level for estimating the $m$ lower bounds  is at least $1-\alpha$. Given the estimated $\underline{p_{l_1}}, \underline{p_{l_2}}, \cdots, \underline{p_{l_m}}$, we can calculate $\underline{BA}(x)$ and $\overline{BA}(x)$ in Theorem~\ref{theorem:multi-class}.

\myparatight{Multi-label smoothing based watermarking} We denote by $N_i = \sum_{j=1}^N \mathbb{I}(i \in f(x+\epsilon_j))$ the number of noisy images for which   the base multi-label classifier $f$ predicts label $i$. $f$ is defined in Section~\ref{building smoothed decoder} for multi-label smoothing based watermarking. We estimate the $i$th bit of $D_s(x)$ as follows:
\begin{align}
    D_s(x)[i] =  
    \begin{cases} 
    1, & \text{if } i \in \text{argk-max}_{i' \in \mathcal{Y}} N_{i'}, \\
    0, & \text{Otherwise}, 
    \end{cases}
\end{align}
where $\mathcal{Y}=\{1,2,\cdots,m\}$ and argk-max is the set of $k$ labels that have the largest values of $N_{i'}$. $N_i$ follows a binomial distribution $N_i \sim B(N, p_i)$, where $p_i=\text{Pr}(i \in f(x+\epsilon))$, $\epsilon \sim \mathcal{N}(0,\sigma^2 I)$. According to Jia et al.~\cite{jia2022multiguard}, we can estimate the lower/upper bound of $p_i$ for all $i=1,2,\cdots,m$ \emph{simultaneously} as follows:
\begin{align}
& \underline{p_s}=\operatorname{Beta}(\frac{\alpha}{m} ; N_s, N-N_s+1), \quad s \in \{i \in \mathcal{Y}|w_t[i]=1\}, \\
& \overline{p_t}=\operatorname{Beta}(1-\frac{\alpha}{m} ; N_t, N-N_t+1), \quad t \in \{i \in \mathcal{Y}|w_t[i]=0\}, 
\end{align}
where  $1-\frac{\alpha}{m}$ is the confidence level for estimating one lower/upper bound, and the overall confidence level of estimating the $m$ lower/upper bounds is at least $1-\alpha$ based on the \emph{Bonferroni correction}~\cite{bonferroni1936teoria}. Then, according to Jia et al.~\cite{jia2022multiguard} (we also show details in Appendix~\ref{appendix:calculate e}), we can calculate $\underline{e}$ and $\overline{e}$ in Theorem~\ref{theorem:multi-label} based on the lower bounds $\underline{p_s}$ and upper bounds $\overline{p_t}$. Given $\underline{e}$ and $\overline{e}$, we can further calculate $\underline{BA}(x)$ and $\overline{BA}(x)$.

\myparatight{Regression smoothing based watermarking} We denote by $BA_{j}=BA(D(x+\epsilon_j),w_t)$ the BA of the watermark decoded by $D$ from the $j$th noisy image. We sort $BA_{1},BA_{2},\cdots,BA_{N}$ in a descending order, and without loss of generality, we assume $BA_{1} \geq BA_{2} \geq \cdots \geq BA_{N}$. We estimate $BA(D_s(x),w_t)$ as the median of  the $N$ bitwise accuracy. Moreover,  we estimate $\underline{BA}(x)$ as $BA_{l^*}$ and $\overline{BA}(x)$ as $BA_{h^*}$, where  $l^*$ and $h^*$ are defined as follows:
\begin{align}
    l^* &= \argmax_{j \in \{1,2,\cdots,N\}} 1 - \sum_{i=1}^{j}\left(
    \begin{array}{l}
        N \\
        i
    \end{array}
    \right)(\Phi(-\frac{R}{\sigma}))^i(1-\Phi(-\frac{R}{\sigma}))^{N-i} \geq 1-\alpha, \\
    h^* &= \argmin_{j \in \{1,2,\cdots,N\}} \sum_{i=1}^{j}\left(
    \begin{array}{l}
        N \\
        i
    \end{array}
    \right)(\Phi(\frac{R}{\sigma}))^i(1-\Phi(\frac{R}{\sigma}))^{N-i} \geq 1-\alpha.
\end{align}
According to regression smoothing~\cite{chiang2020detection}, the confidence level of such  estimation of $\underline{BA}(x)$ and $\overline{BA}(x)$ is $1-\alpha$.

\subsection{Improving Certified Robustness via Adversarial Training} 
Given a watermarking method $(w_t, E, D)$, we build a smoothed watermarking method $(w_t, E, D_s)$. Our smoothed decoder $D_s$ relies on using the base decoder $D$ to decode watermarks from images $x+\delta$ with Gaussian noise. When $D$ can more accurately decode $w_t$ from noisy watermarked images,  $D_s$ is more accurate and robust. However, in \emph{standard training}, $E$ and $D$ are jointly trained such that $D$ is supposed to accurately decode $w_t$ from watermarked images, but not noisy ones. To address this challenge, we can use \emph{adversarial training}~\cite{zhu2018hidden} to jointly train $E$ and $D$. Specifically, we add a noisy layer between $E$ and $D$. During training, for each watermarked image produced by $E$, the noisy layer adds Gaussian noise to it and then passes it to $D$. With adversarial training,  $D$ can more accurately decode $w_t$ from watermarked images with Gaussian noise, making our smoothed decoder $D_s$ more robust, as shown in our experiments. 

%% file: 5_evaluation.tex
\section{Evaluation}
\subsection{Experimental Setup}
\label{sec:ExperimentalSetup}
\myparatight{Image datasets Stable Diffusion, Midjourney, and DALL-E} We use three image datasets, each of which consists of (10K training AI-generated images, 1K testing AI-generated images, 1K non-AI-generated images). The training/testing AI-generated images in the three datasets are produced by  Stable Diffusion~\cite{wang2022diffusiondb}, Midjourney~\cite{midjourney-database}, and DALL-E~\cite{dalle2-database}, respectively. The non-AI-generated images in the three datasets are sampled from the combined dataset of  COCO~\cite{lin2014microsoft}, ImageNet~\cite{deng2009imagenet}, and Conceptual Caption~\cite{sharma2018conceptual}. In each dataset, the training AI-generated images are used to train watermarking encoders and decoders, while the testing AI-generated images and non-AI-generated images are used for testing. Note that we embed the ground-truth watermark $w_t$ into each testing AI-generated image using the corresponding encoder, while the non-AI-generated images are non-watermarked. For consistency, we standardize the size of images across all datasets to 128 $\times$ 128 pixels.

\myparatight{Training base watermarking encoders and decoders} We use HiDDeN~\cite{zhu2018hidden}, whose code is publicly available, to train the base watermarking encoders and decoders for each  dataset. We consider HiDDeN because it achieves state-of-the-art performance and is the basis of more recent watermarks like Stable Signature~\cite{fernandez2023stable}. We use \emph{standard training} and \emph{adversarial training} to train different base watermarking encoders and decoders, and compare them. For adversarial training, the noisy layer samples a Gaussian noise from $\mathcal{N}(0, \sigma^2 I)$  for each watermarked image. We follow the training setting in the public code of  HiDDeN~\cite{zhu2018hidden}. The only difference is that we increase the weight of \emph{encoder loss} from 0.7 to 2 in adversarial training to better preserve the visual quality of watermarked images.

\myparatight{Evaluation metrics} An image is detected as watermarked/AI-generated if  BA of the watermark decoded from it is no smaller than $\tau$.  For \emph{empirical robustness}, we evaluate performance of a watermarking method under state-of-the-art removal and forgery attacks. Specifically, we use a removal attack to perturb the watermarked/AI-generated testing images, and we compute \emph{false negative rate (FNR)}, which is the fraction of perturbed watermarked images that are falsely detected as non-watermarked/non-AI-generated. Moreover, we use a forgery attack to perturb the non-watermarked/non-AI-generated images in a dataset, and we compute \emph{false positive rate (FPR)}, which is the fraction of perturbed non-watermarked images that are falsely detected as watermarked. We use FNR and FPR to measure the empirical robustness of both base and our smoothed watermarking methods. 

For \emph{certified robustness}, we evaluate performance of a watermarking method under \emph{any} removal and forgery attacks. FNR and FPR cannot be applied since the worst-case removal and forgery attacks are unknown. To address the challenge, we propose \emph{certified false negative rate (CFNR)} and \emph{certified false positive rate (CFPR)}, which respectively are upper bounds of FNR and FPR under any removal and forgery attacks. Note that  CFNR and CFPR are only applicable to our smoothed watermarking method that achieves certified robustness. Formally, given a $\ell_2$-norm perturbation size $R$ introduced by any removal or forgery attacks, CFNR and CFPR are defined as follows:
\begin{align}
    CFNR &= \frac{1}{|X_w|}\sum_{x_w\in X_w} \mathbb{I}(\underline{BA}(x_w)<\tau), CFPR &= \frac{1}{|X_n|}\sum_{x_n\in X_n} \mathbb{I}(\overline{BA}(x_n)\geq \tau), \nonumber
\end{align}
where $X_w$ is the set of testing watermarked/AI-generated images in a dataset, $X_n$ is the set of non-watermarked/non-AI-generated images in a dataset, $\mathbb{I}$ is the indicator function,  $\underline{BA}(x_w)$ is a lower bound of $BA(D_s(x_w+\delta), w_t)$,  $\overline{BA}(x_n)$ is an upper bound of $BA(D_s(x_n+\delta), w_t)$, and $\|\delta\|_2<R$. Intuitively, CFNR (or CFPR) is the fraction of watermarked (or non-watermarked) images whose lower bounds  $\underline{BA}(x_w)$ (or upper bounds $\overline{BA}(x_n)$) of bitwise accuracy under any removal attacks (or forgery attacks) are smaller (or no smaller) than $\tau$, i.e., such watermarked (or non-watermarked) images are likely to be falsely detected as non-watermarked (or watermarked) under attacks. Note that CFNR and CFPR depend on $R$.

\myparatight{Removal and forgery attacks} We consider 4 removal and forgery attacks to evaluate empirical robustness. These attacks are  \emph{JPEG compression}, compressing a watermarked or non-watermarked image using JPEG; \emph{black-box attack}~\cite{jiang2023evading}, finding a perturbation for a watermarked or non-watermarked image via repeatedly querying the detection API; \emph{white-box attack}~\cite{jiang2023evading}, finding a perturbation for a watermarked or non-watermarked image based on a decoder; and \emph{adaptive white-box attack}, which extends the white-box attack to find a perturbation for a watermarked or non-watermarked image via taking smoothing into consideration. The details of these 4 attacks are shown in Appendix~\ref{appendix:post-processing}. Note that each attack can be used as a removal or forgery attack.

\myparatight{Parameter settings} Unless otherwise mentioned,  $m=30$; $w_t$ is picked uniformly at random; $N=10,000$; $\tau=0.83$ (corresponding to FPR$<10^{-4}$ for the base watermarking method under no attacks~\cite{jiang2023evading});  the standard deviation of Gaussian noise in adversarial training is $\sigma^{\prime}=0.1$; the standard deviation of Gaussian noise in smoothing is $\sigma=0.1$; confidence level $1-\alpha=0.999$;  $k^{\prime}$ and $k$ for multi-label smoothing based watermarking are the number of ones in $w_t$; and we show results on regression smoothing based watermarking and adversarial training. In evaluation of empirical robustness, we set $N=100$ due to limited computational resources. 

\begin{figure}[!t]
\centering
\subfloat[CFNR\label{figure:diff smoothing1}]{\includegraphics[width=0.24\textwidth]{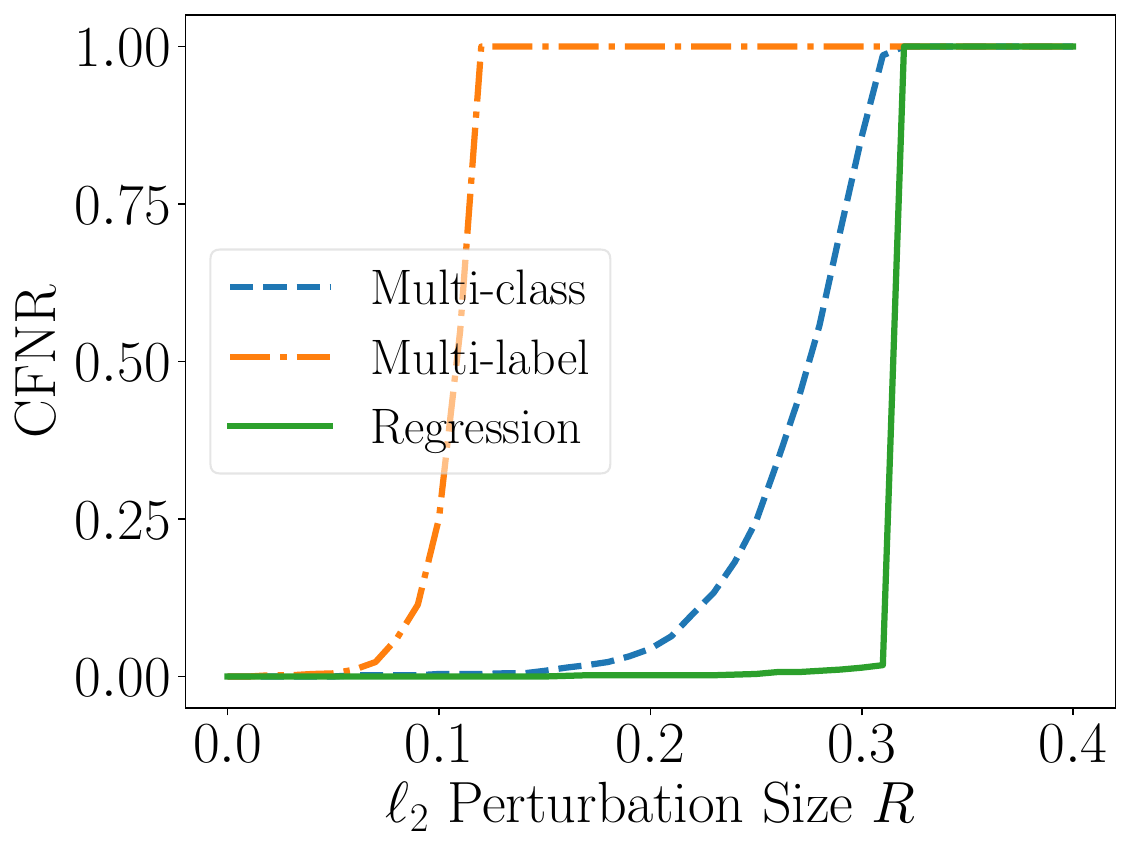}}
\subfloat[CFPR\label{figure:diff smoothing2}]{\includegraphics[width=0.24\textwidth]{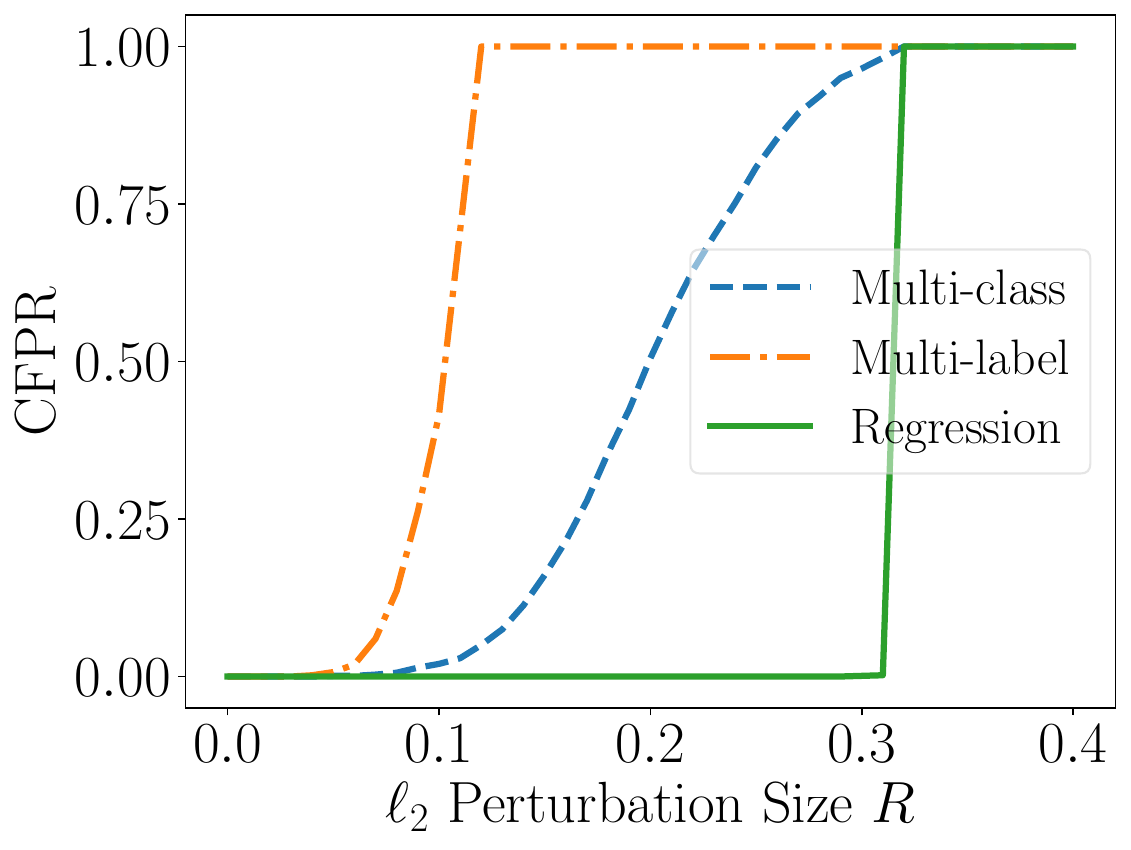}}
\subfloat[CFNR]{\includegraphics[width=0.24\textwidth]{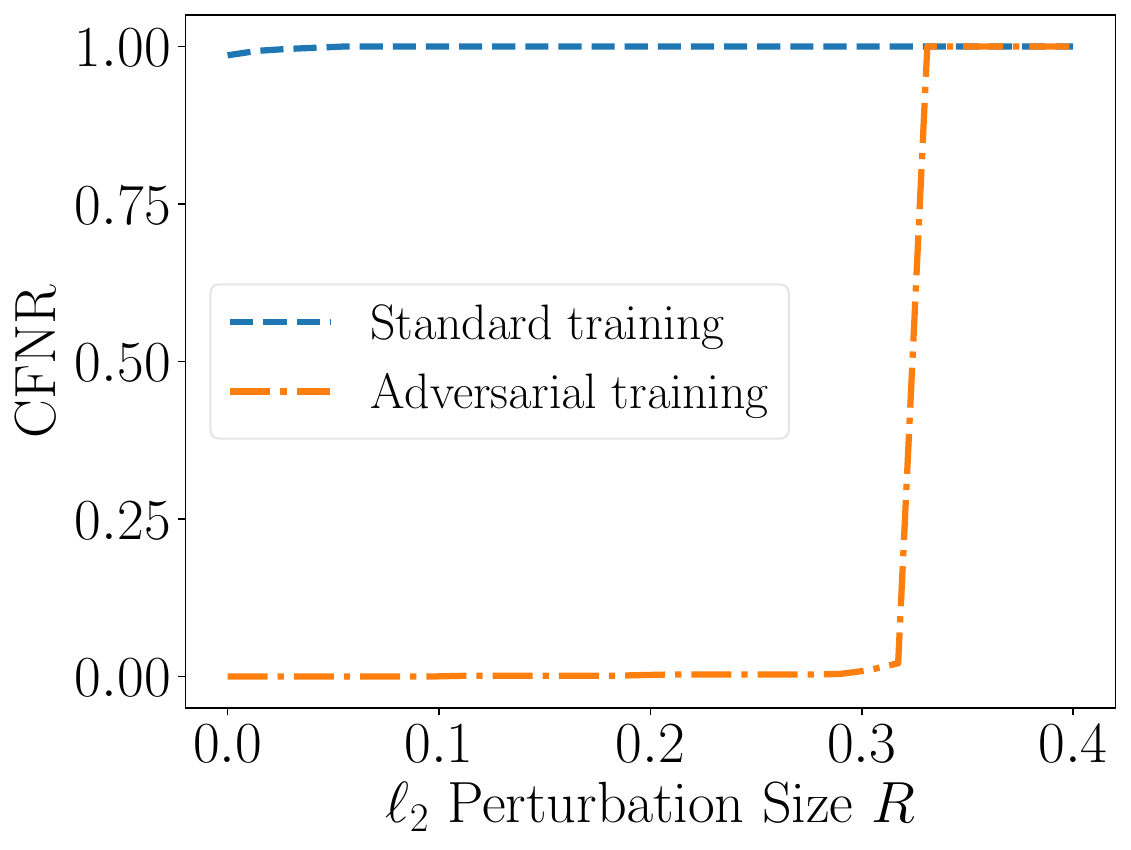}\label{figure:diff training:CFNR}}
\subfloat[CFPR]{\includegraphics[width=0.24\textwidth]{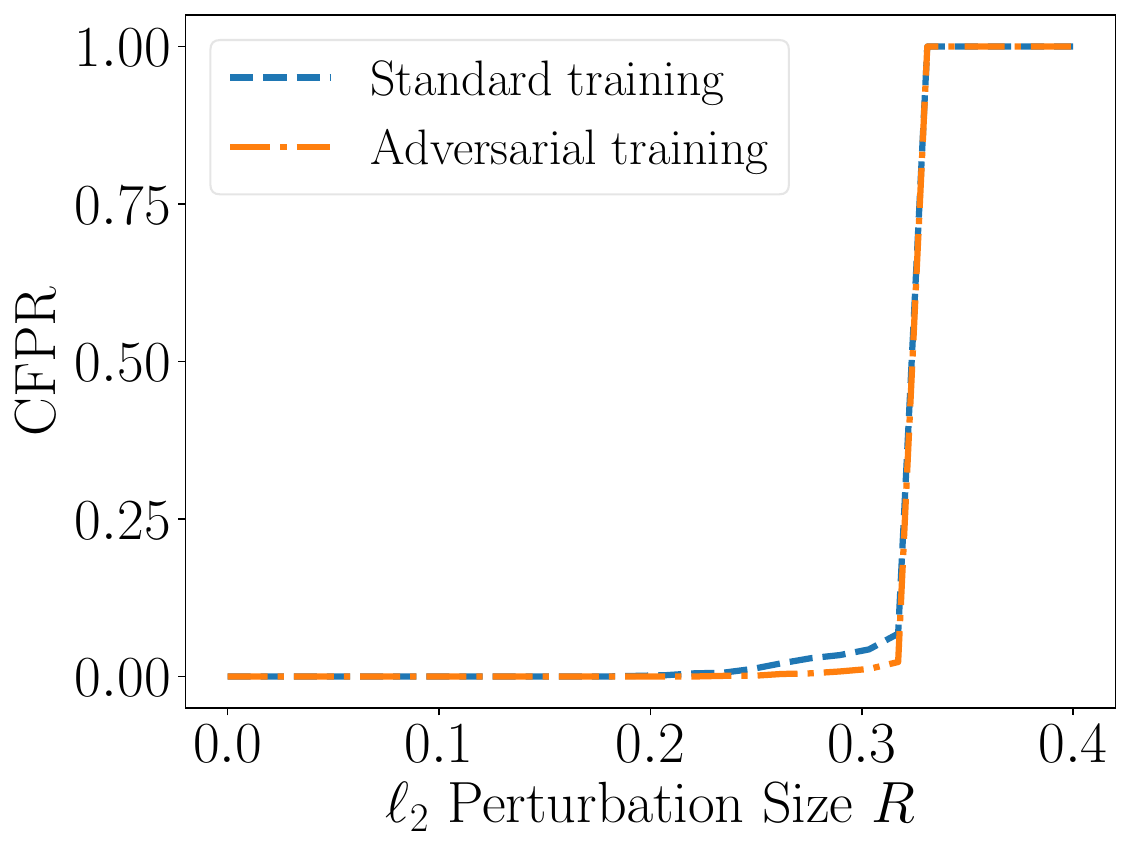}\label{figure:diff training:CFPR}}
\caption{(a) CFNR and (b) CFPR of our three smoothing based watermarking methods. (c) CFNR and (d) CFPR of our regression smoothing based watermarking when the base watermarking method is trained via standard or adversarial training.}
\vspace{-1mm}
\end{figure}

\subsection{Certified Robustness}
\vspace{-2mm}
\myparatight{Comparing our three smoothing based watermarking methods} Figure~\ref{figure:diff smoothing1} and~\ref{figure:diff smoothing2} compare our three smoothing based watermarking methods with respect to CFNR  and CFPR of Stable Diffusion dataset as the perturbation size $R$ increases. The results for the other two datasets are shown in Figure~\ref{figure:diff smoothing--other datasets} in Appendix. We observe that regression  smoothing based watermarking outperforms multi-class and multi-label smoothing based watermarking, i.e., regression  smoothing based watermarking achieves smaller CFNR and CFPR. The reason is that regression smoothing based watermarking accounts for the correlation between bits because bitwise accuracy is aggregated across all bits. Thus, in the remaining experiments, we focus on  regression smoothing based watermarking.

\myparatight{Standard vs. adversarial training} Figure~\ref{figure:diff training:CFNR} and~\ref{figure:diff training:CFPR}  compare standard and adversarial training with respect to  CFNR  and CFPR of Stable Diffusion dataset. The results for the other two datasets are shown in Figure~\ref{figure:diff training--midjourney} in Appendix. We observe that when the base watermarking method is trained via adversarial training, our smoothed watermarking achieves better certified robustness. In particular, adversarial training achieves much smaller CFNR and slightly smaller CFPR. Note that in order to fairly compare  standard and adversarial training, we tune their training settings as discussed in Section~\ref{sec:ExperimentalSetup} to achieve similar visual quality of watermarked images.  Specifically, the average \emph{SSIM} between images and their watermarked versions is 0.943 and 0.941 for standard training and adversarial training, respectively. Figure~\ref{figure:visulization} in Appendix shows some examples of watermarked images for the two training strategies.

\begin{figure}[!t]
\centering
\subfloat[CFNR\label{figure:tau1}]{\includegraphics[width=0.24\textwidth]{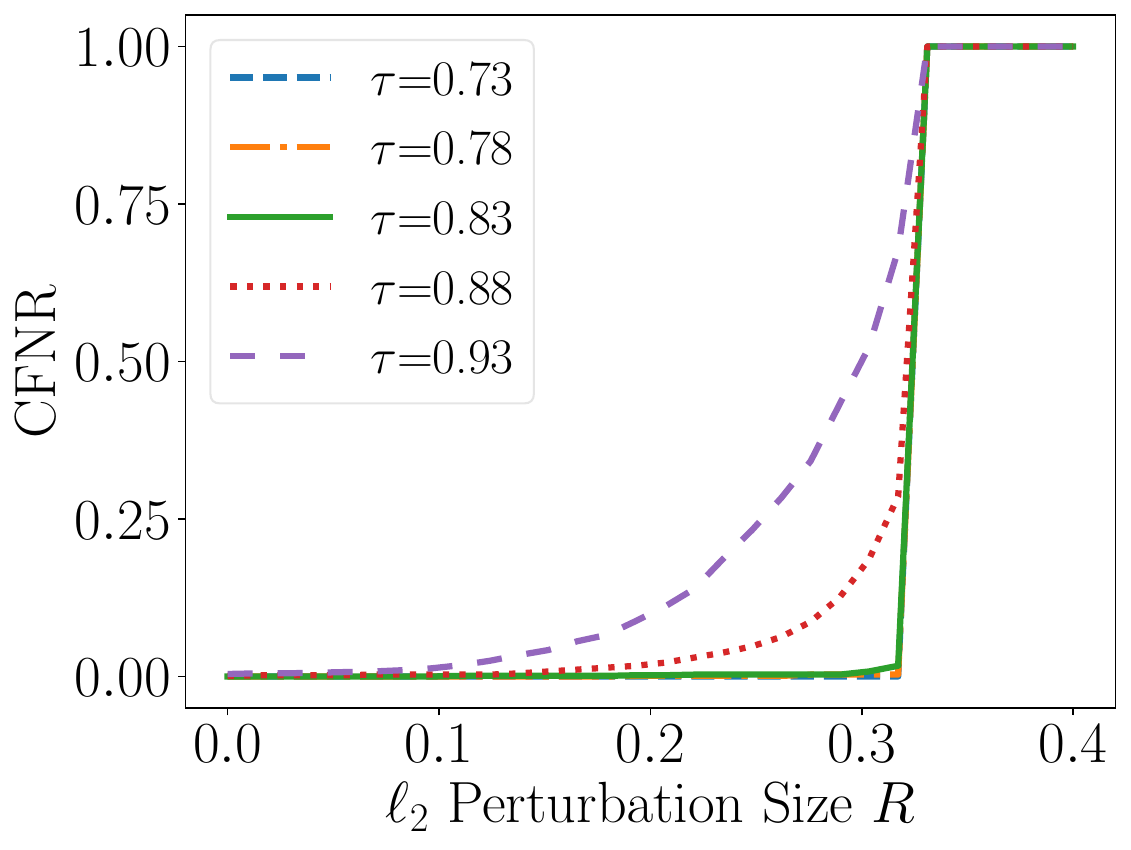}}
\subfloat[CFPR\label{figure:tau2}]{\includegraphics[width=0.24\textwidth]{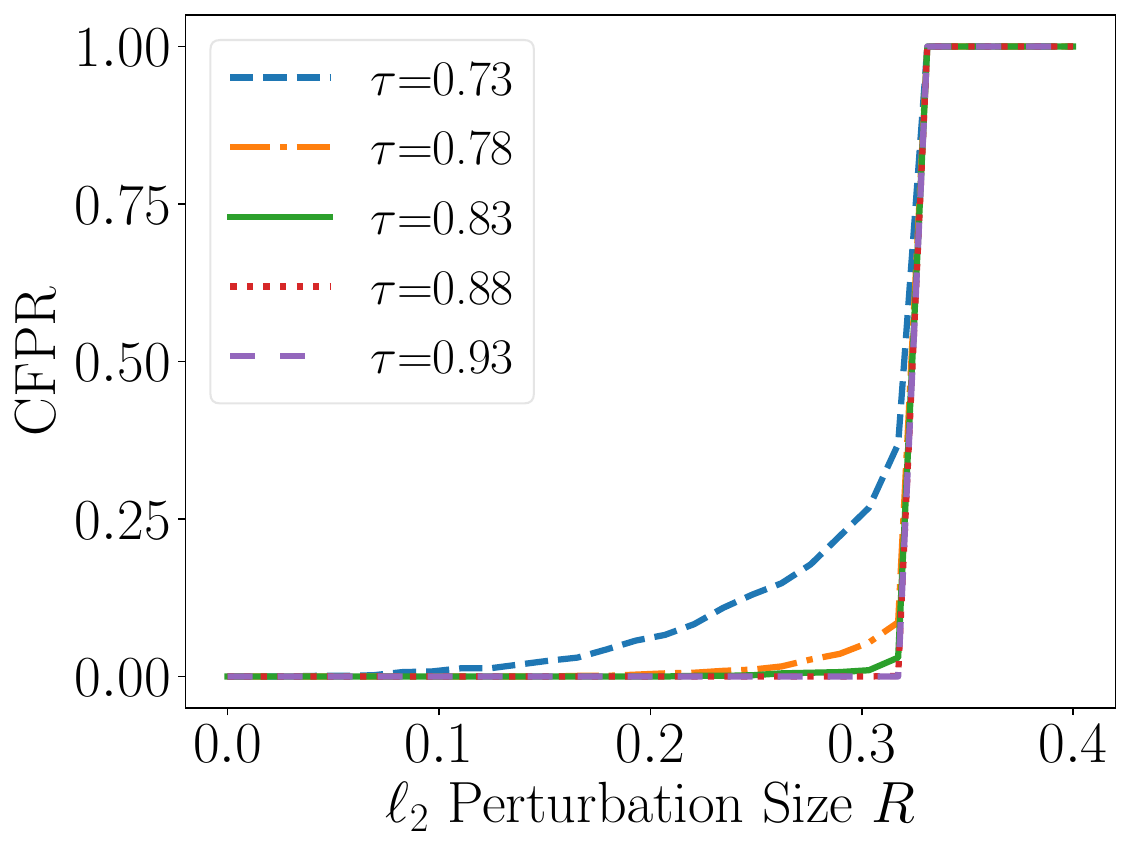}}
\subfloat[CFNR\label{figure:sigma1}]{\includegraphics[width=0.24\textwidth]{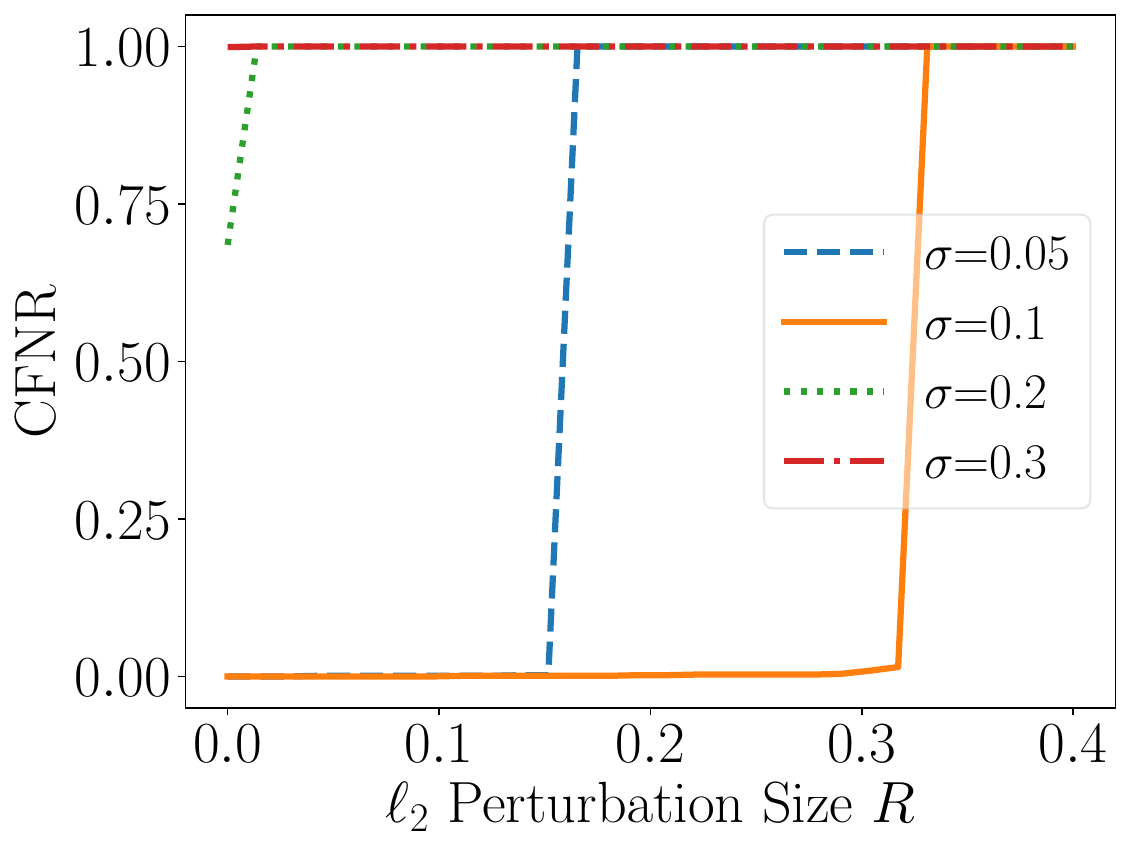}}
\subfloat[CFPR\label{figure:sigma2}]{\includegraphics[width=0.234\textwidth]{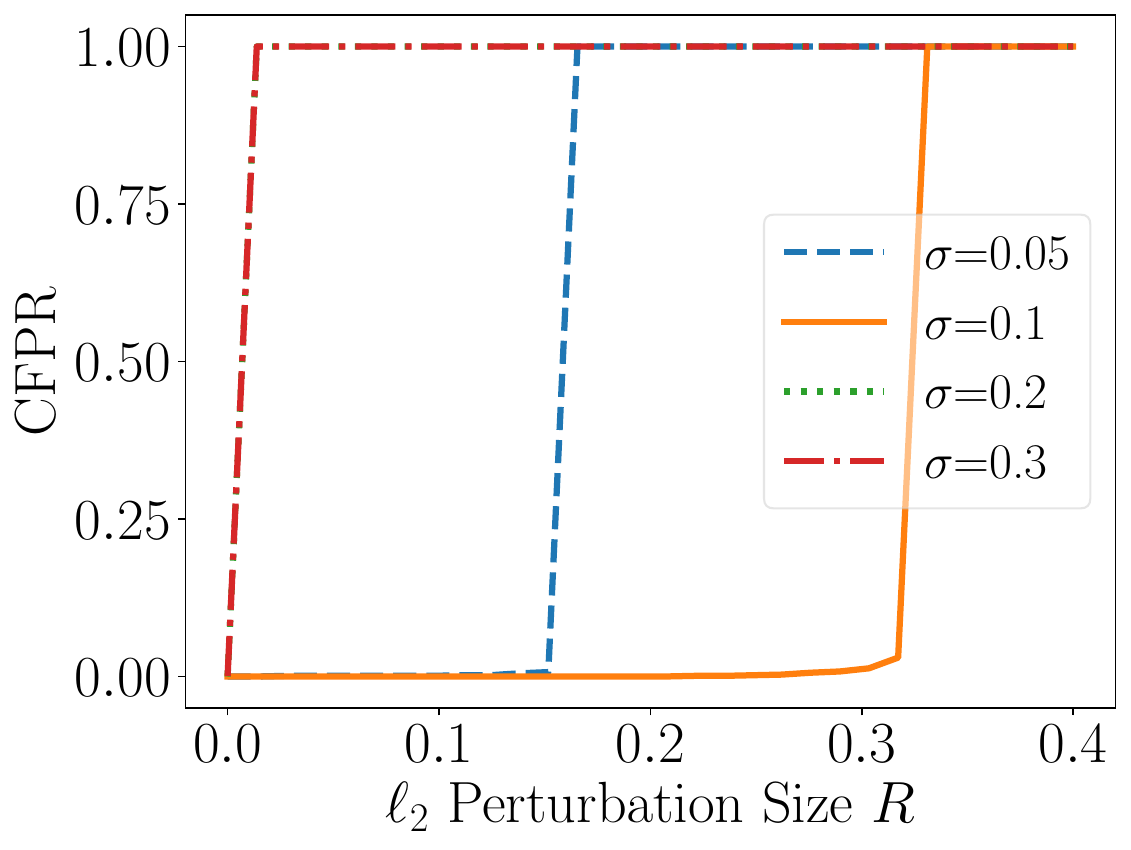}}
\caption{(a-b) Impact of detection threshold $\tau$. (c-d) Impact of smoothing Gaussian noise standard derivation $\sigma$.}
\vspace{-3mm}
\end{figure}

\myparatight{Impact of detection threshold $\tau$} Figure~\ref{figure:tau1} and~\ref{figure:tau2} compare different detection threshold $\tau$ with respect to CFNR  and CFPR of Stable Diffusion dataset. Figure~\ref{figure:tau--other datasets} in Appendix shows results on the other two datasets. We vary the default $\tau$=0.83 with a step size 0.05. We observe $\tau$ controls a trade-off between CFNR and CFPR: a smaller $\tau$ achieves a smaller CFNR but also a larger CFPR. 

\myparatight{Impact of smoothing Gaussian noise $\sigma$} Figure~\ref{figure:sigma1} and~\ref{figure:sigma2} compare different $\sigma$ with respect to CFNR and CFPR of Stable Diffusion dataset. Figure~\ref{figure:sigma other datasets} in Appendix shows results on the other two datasets. We observe that certified robustness is sub-optimal when $\sigma$ is too small or too large. This is because, when $\sigma$ is too small, the percentile $\Phi(-\frac{R}{\sigma})$ is small, leading to a small lower bound $\underline{BA}(x)$.  When $\sigma$ is too large, the introduced noise makes watermark decoding incorrect, leading to a smaller bitwise accuracy.

\subsection{Empirical Robustness}
\begin{figure}[!t]
\centering
\subfloat[JPEG compression]{\includegraphics[width=0.247\textwidth]{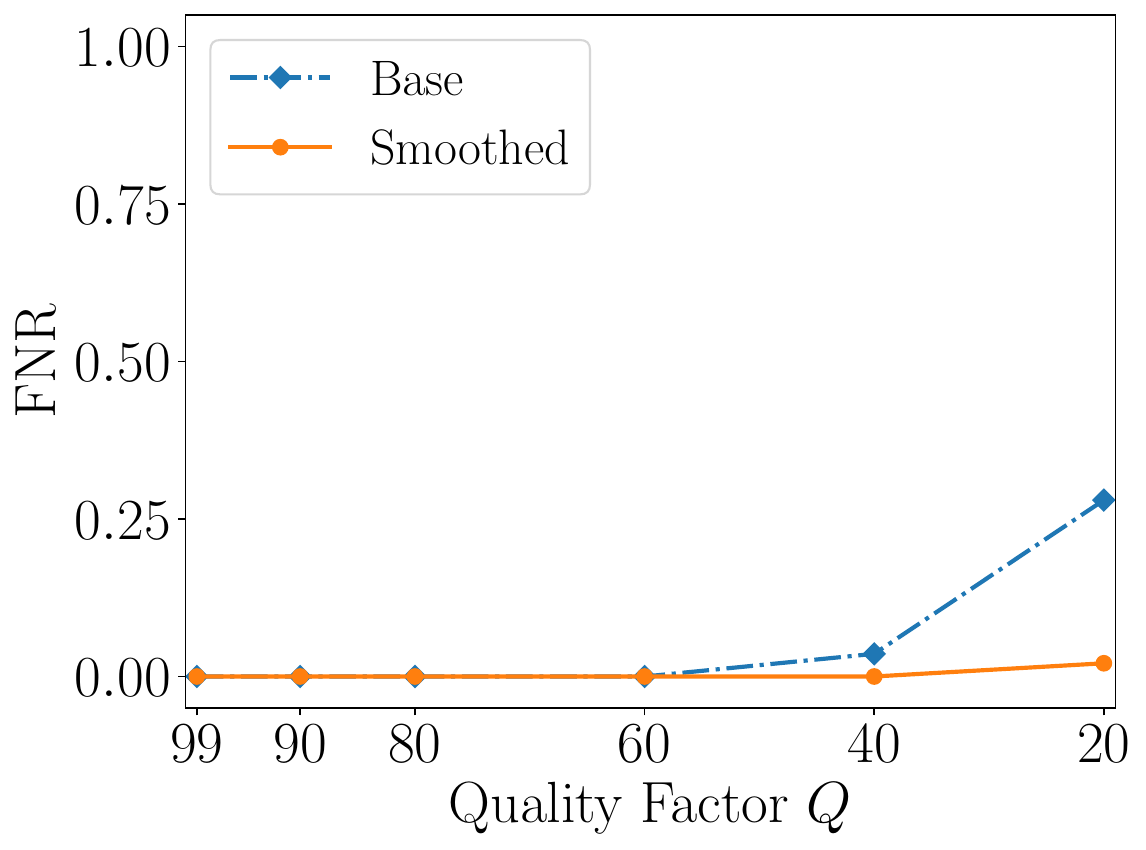}}
\subfloat[Black-box]{\includegraphics[width=0.235\textwidth]{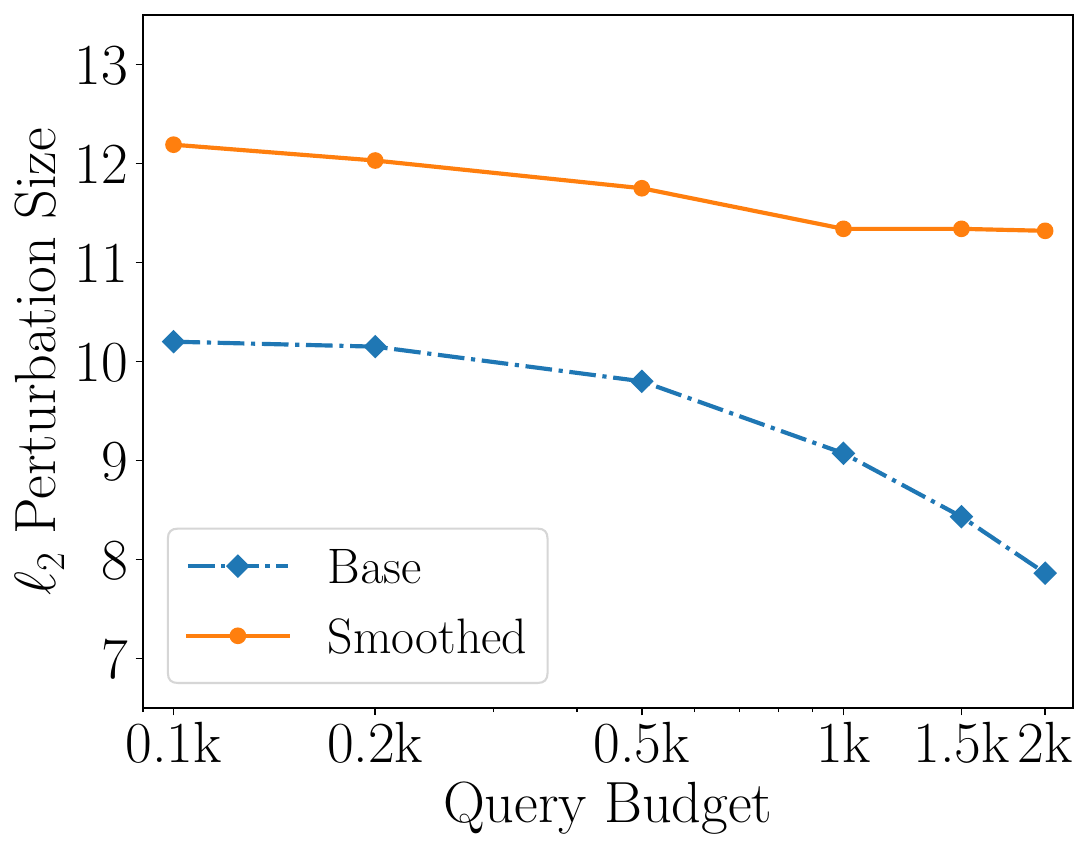}}
\subfloat[White-box]{\includegraphics[width=0.24\textwidth]{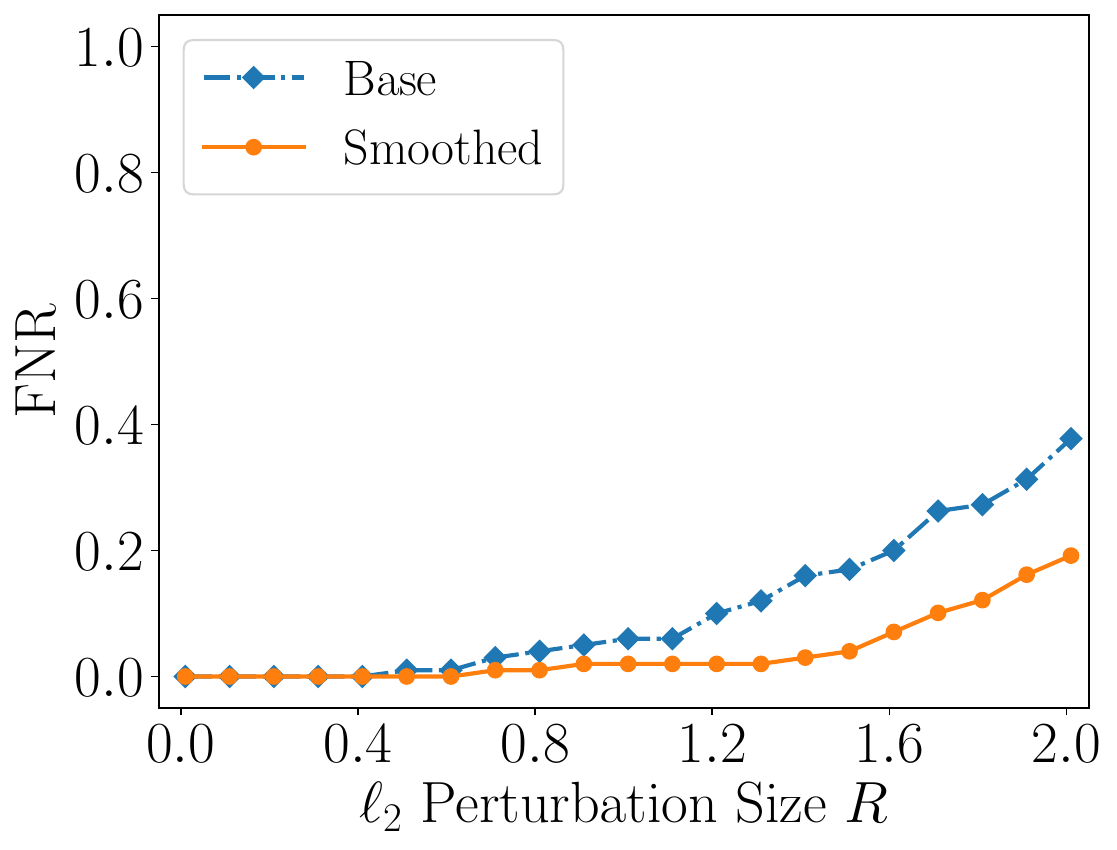}}
\subfloat[Adaptive white-box]{\includegraphics[width=0.24\textwidth]{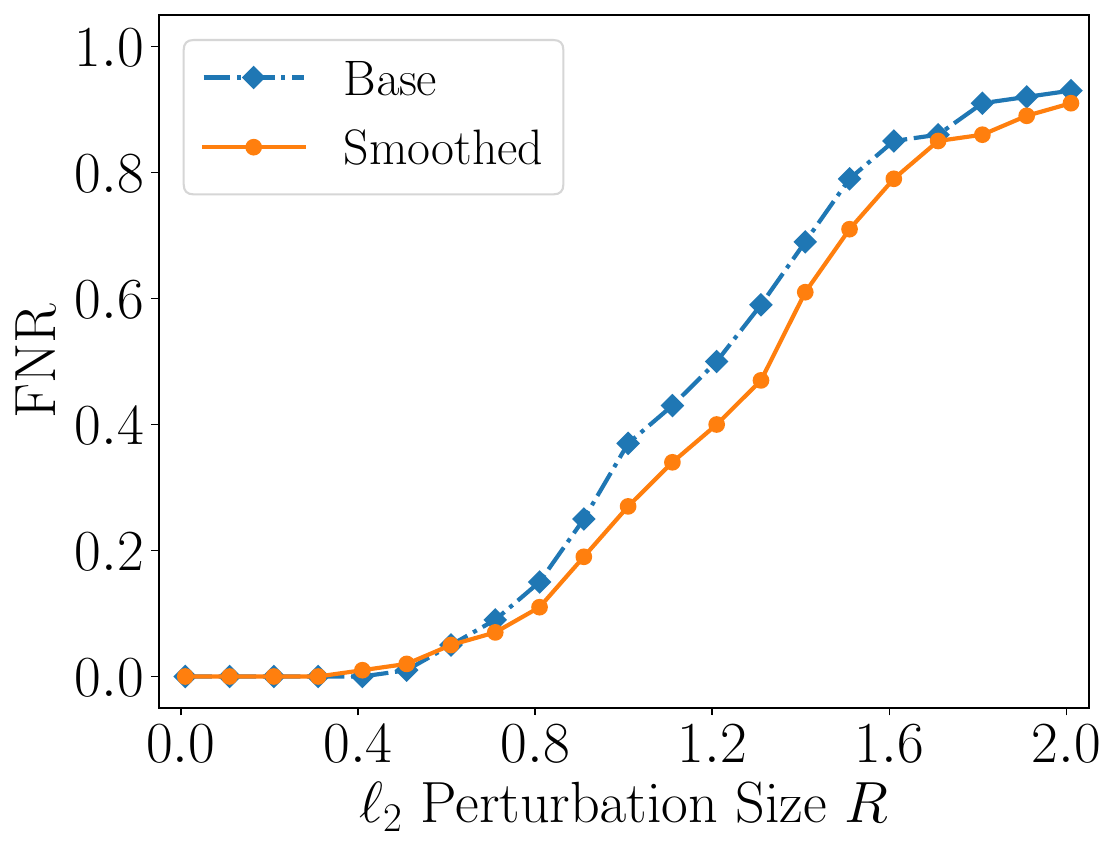}}
\caption{Results of base vs. smoothed watermarking under the 4 removal attacks.}
\label{empiricalFNR}
\vspace{-3mm}
\end{figure}

Figure~\ref{empiricalFNR} shows the results of base vs. our smoothed watermarking under the 4 removal attacks for Stable Diffusion dataset. Figure~\ref{empiricalFNR-othertwo} in Appendix shows the  results for the other two datasets under the 4 removal attacks, while  Figure~\ref{empiricalFPR} in Appendix shows the results for the three datasets under the 4 forgery attacks. 

We observe that our smoothed watermarking method achieves better empirical robustness than the base one under the 4 removal/forgery attacks. For instance, as the quality factor $Q$ of JPEG compression as a removal attack decreases from 99 to 20, our smoothed method always achieves FNR close to 0 while  FNR of the base method increases to 0.25.  Given the same number of queries (i.e., Query Budget) to the detector API,  perturbations found by the black-box removal/forgery attack are larger for our smoothed method, which implies that our smoothed method is more robust. Note that the black-box removal (or forgery) attack achieves FNR (or FPR) of 1 while minimizing the perturbations.   When the perturbation size $R$ is larger than some threshold, our smoothed method achieves smaller FNR (or FPR) than the base method under the white-box and adaptive white-box removal (or forgery) attacks; while both the base method and our smoothed method achieve  FNR and FPR close to 0 when the perturbation size $R$ is small. 

%% file: 7_conclusion.tex
\section{Conclusion and Future Work}
We show randomized smoothing can be extended to build image watermarks that are certifiably robust against removal and forgery attacks. Specifically, we can leverage multi-class, multi-label, and regression smoothing to build certifiably robust image watermarks. We find that regression smoothing based watermarking achieves the best robustness because it can better take the correlations between bits of a watermark into consideration. Other than achieving certified robustness, smoothed watermarking also achieves better empirical robustness than the base watermarking against removal and forgery attacks. An interesting future work is to extend our method to audio, text, and video watermarks, as well as watermarks that do not use bitstrings explicitly.

%% file: 9_acknowledgement.tex
\section{Acknowledgements}
We thank the anonymous reviewers for their constructive comments. This work was supported by NSF grant No. 2125977, 2112562, 1937786, and 1937787, as well as ARO grant No. W911NF2110182.

%% file: 8_appendix.tex
\newpage
\appendix

\begin{figure}[!t]
\centering
\subfloat[Midjourney]{\includegraphics[width=0.23\textwidth]{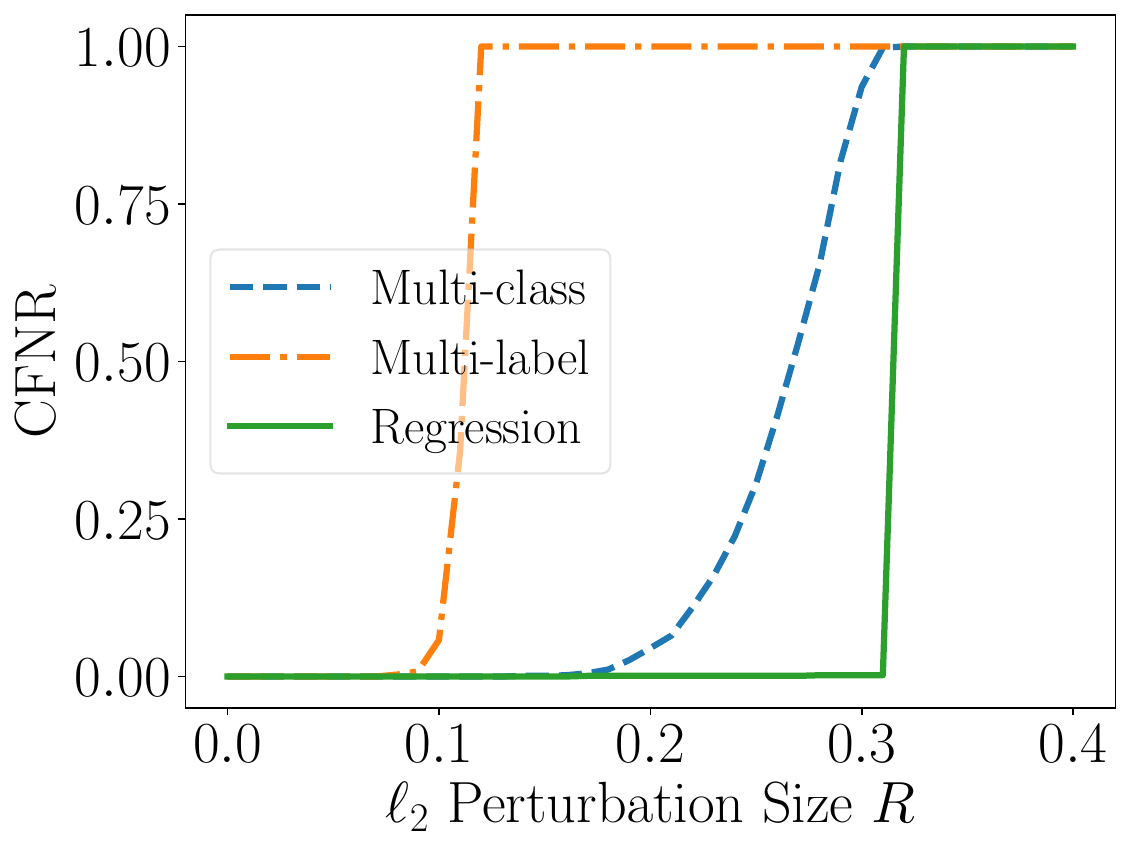}}
\subfloat[Midjourney]{\includegraphics[width=0.23\textwidth]{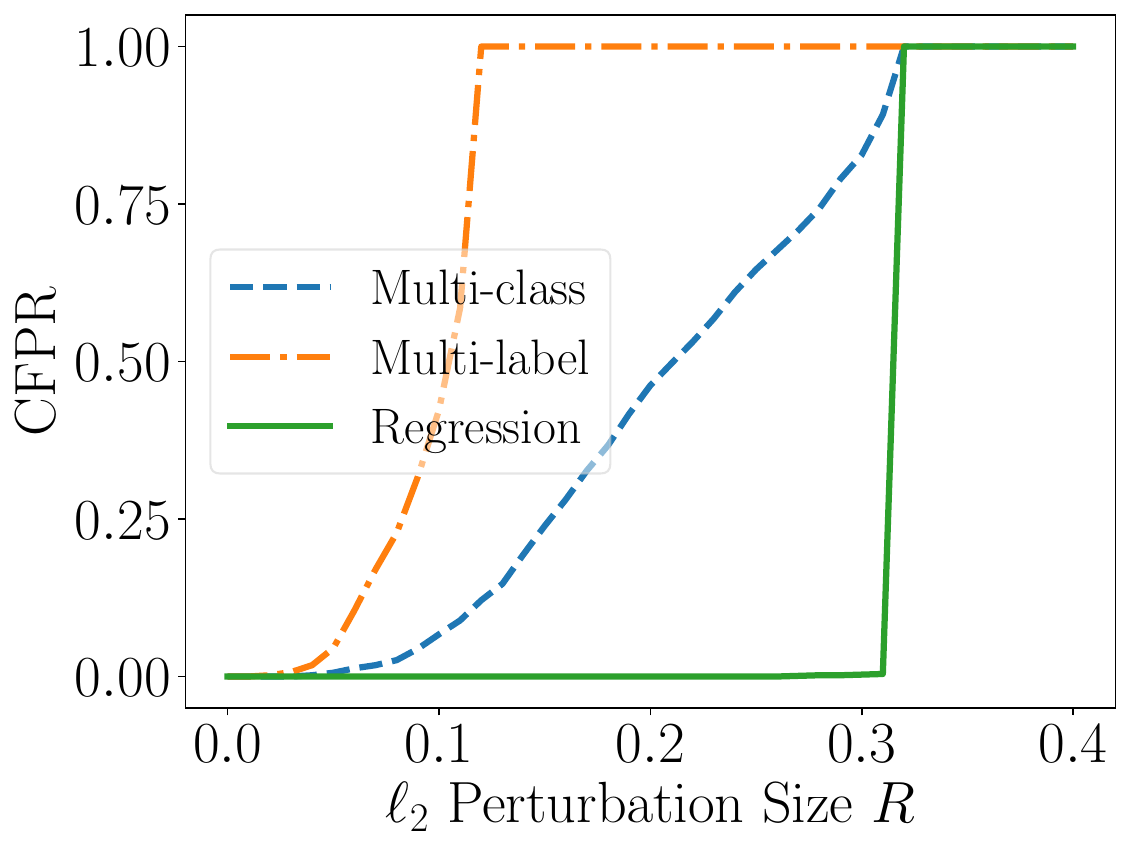}}
\subfloat[DALL-E]{\includegraphics[width=0.23\textwidth]{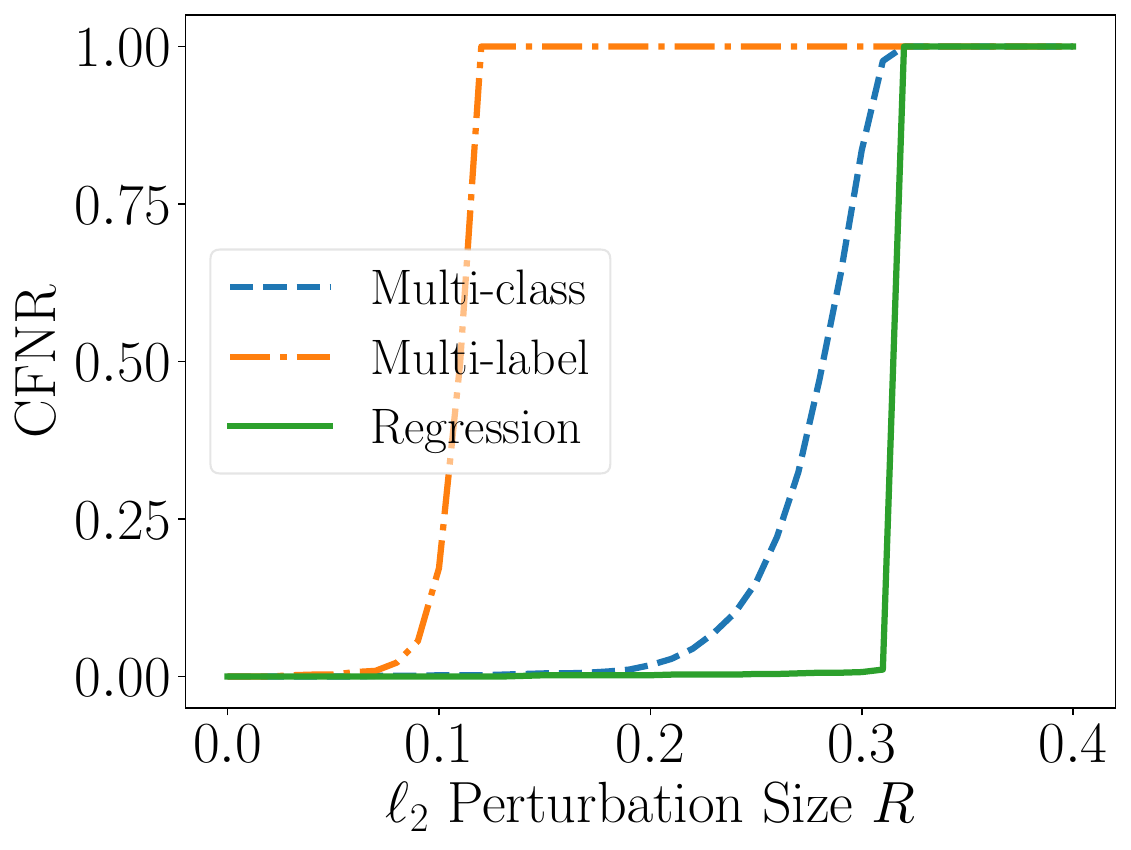}}
\subfloat[DALL-E]{\includegraphics[width=0.23\textwidth]{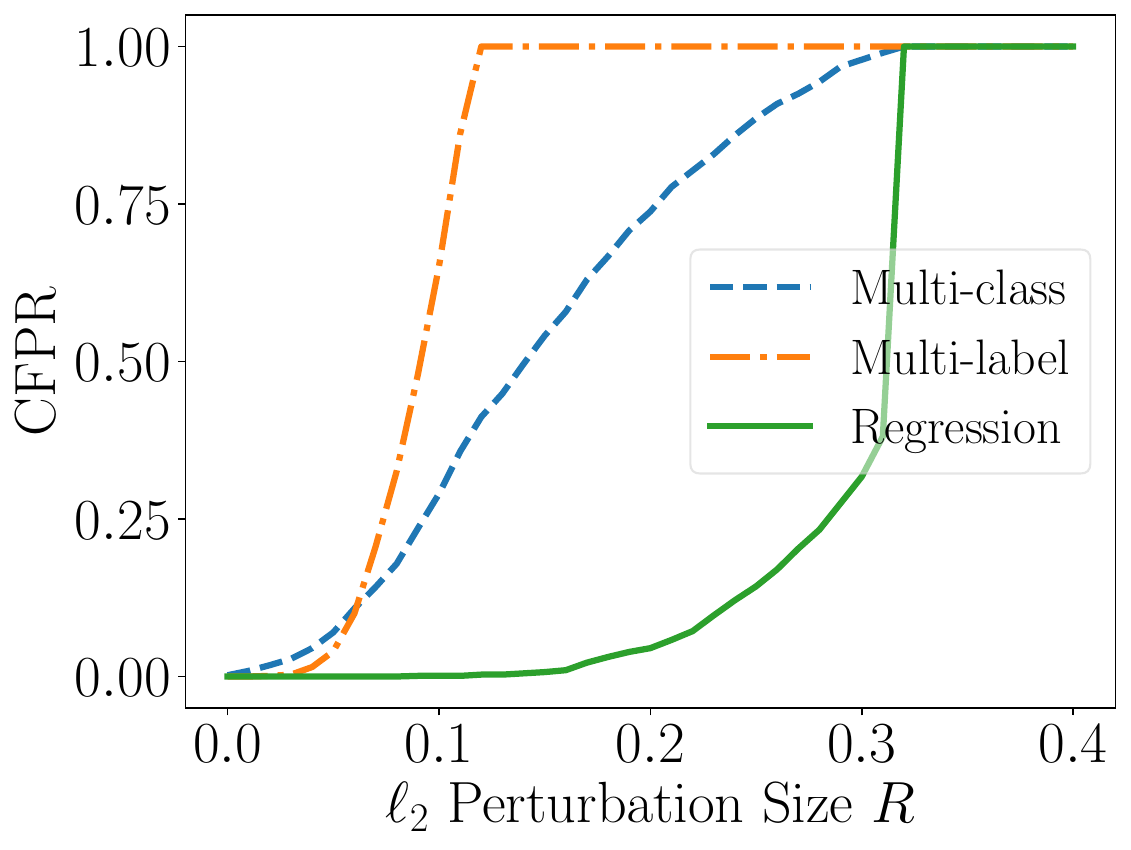}}
\caption{\label{figure:diff smoothing--other datasets} Comparing our three smoothing based watermarking methods on Midjourney and DALL-E datasets.}
\end{figure}

\begin{figure}[!t]
\centering
{\includegraphics[width=0.9\textwidth]{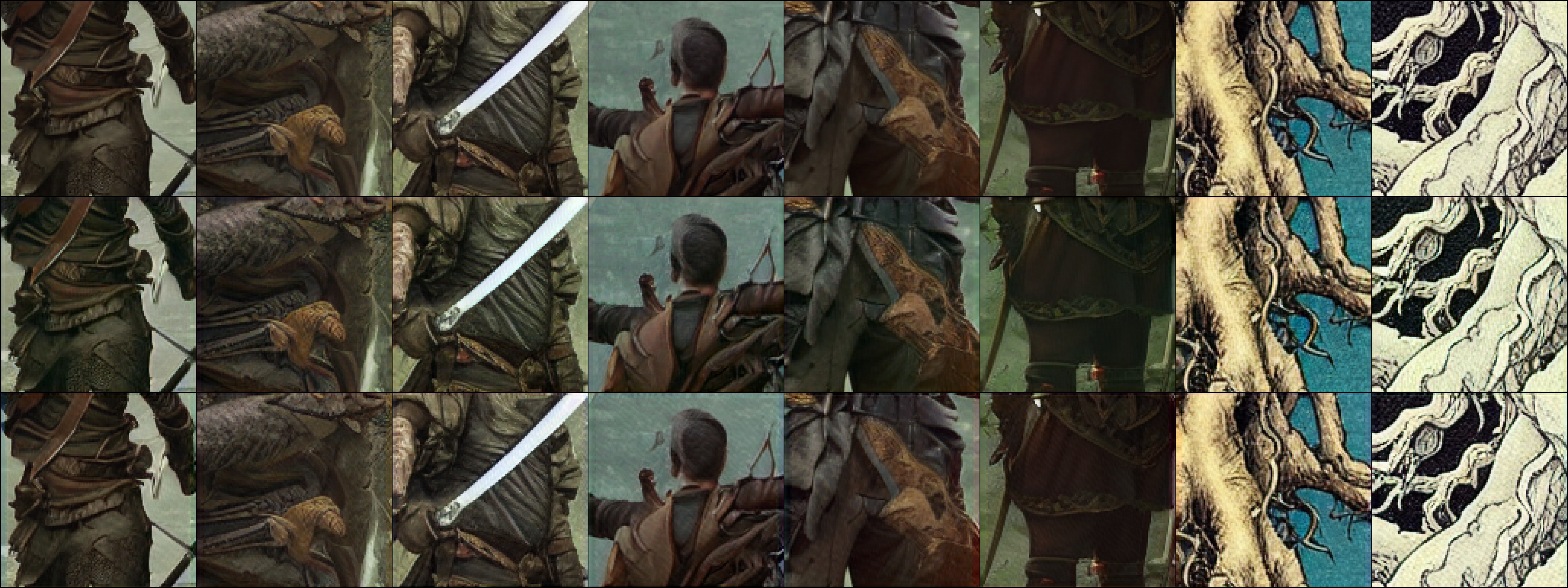}}
\caption{\label{figure:visulization} Standard and adversarial training achieve similar visual quality of watermarked images. The first row shows original AI-generated images, the second row shows their watermarked versions when standard training is used, and the third row shows their watermarked versions when adversarial training is used.}
\end{figure}

\begin{figure}[!t]
\centering
\subfloat[Midjourney]{\includegraphics[width=0.24\textwidth]{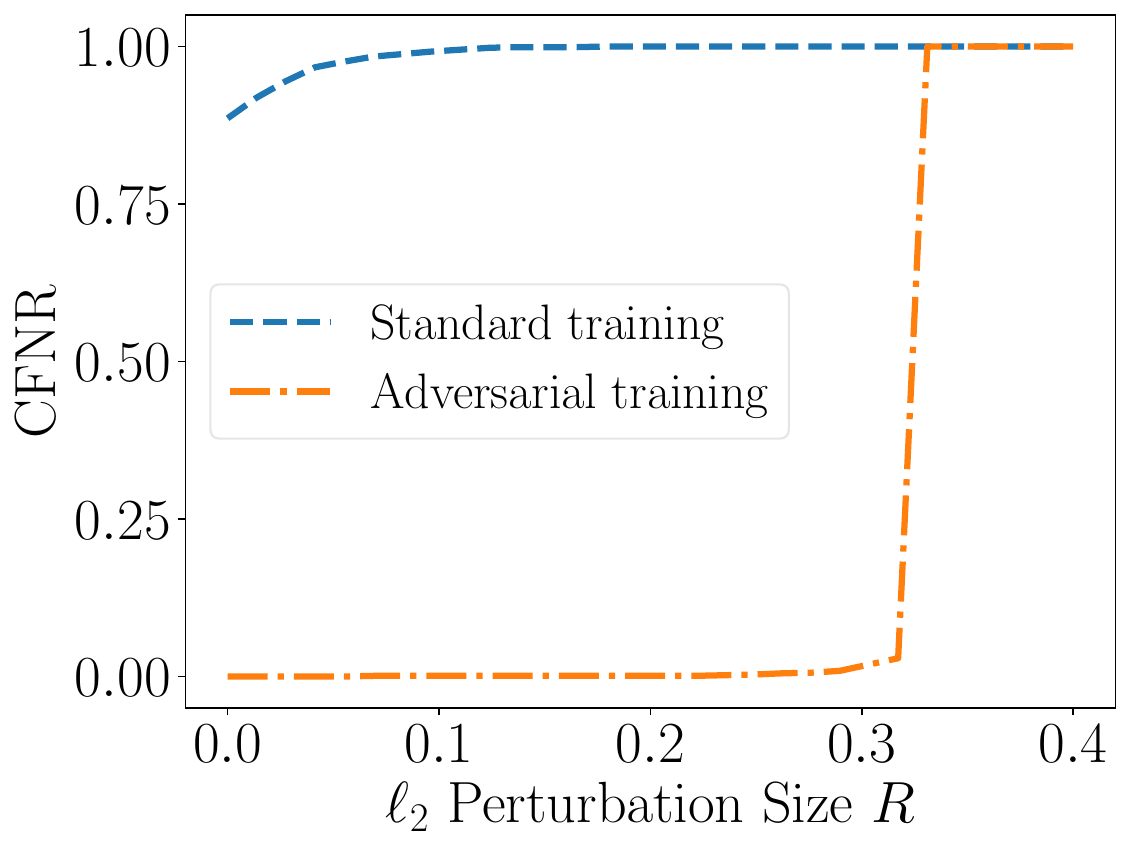}}
\subfloat[Midjourney]{\includegraphics[width=0.24\textwidth]{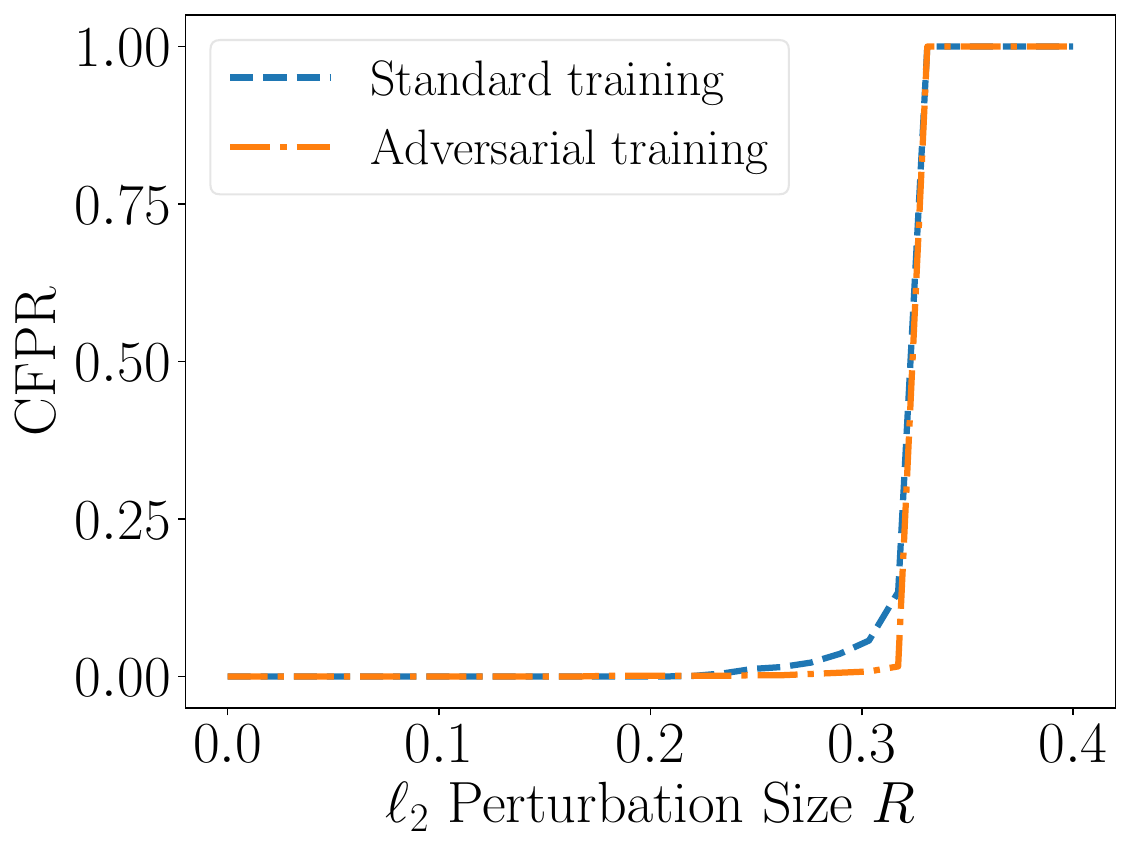}}
\subfloat[DALL-E]{\includegraphics[width=0.24\textwidth]{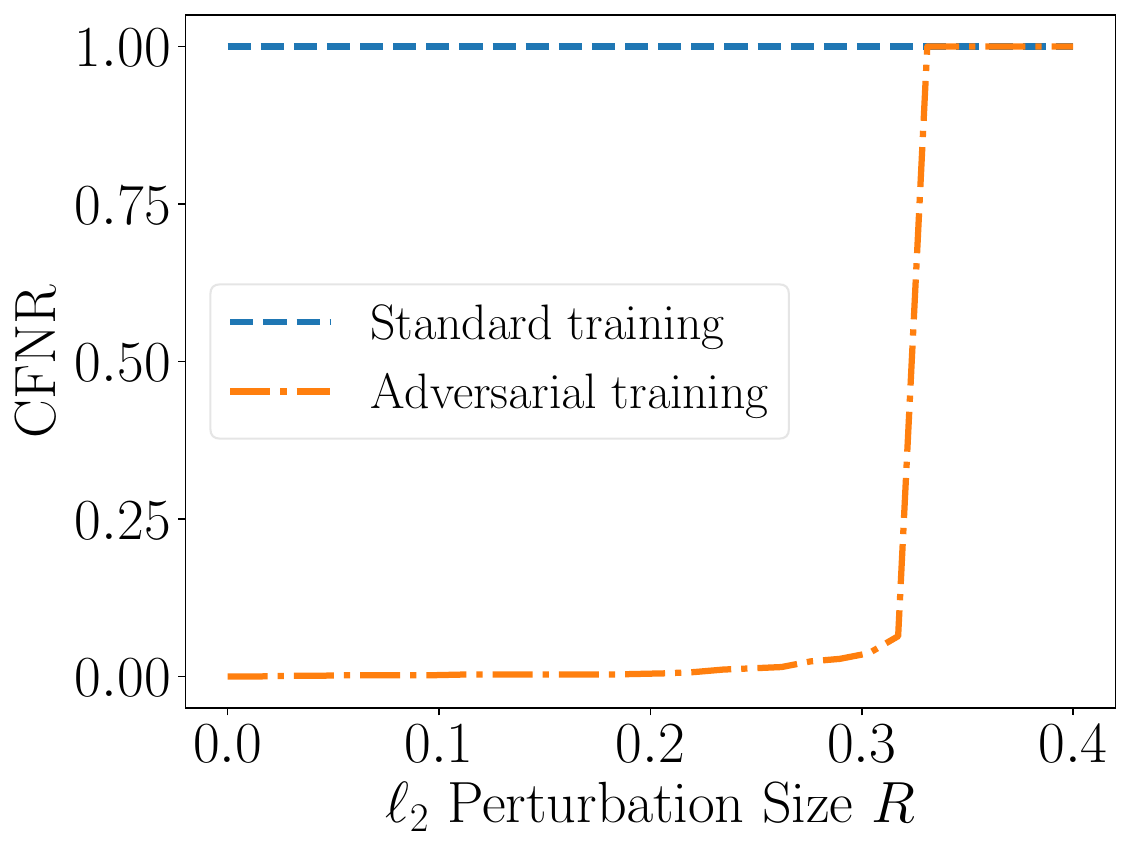}}
\subfloat[DALL-E]{\includegraphics[width=0.24\textwidth]{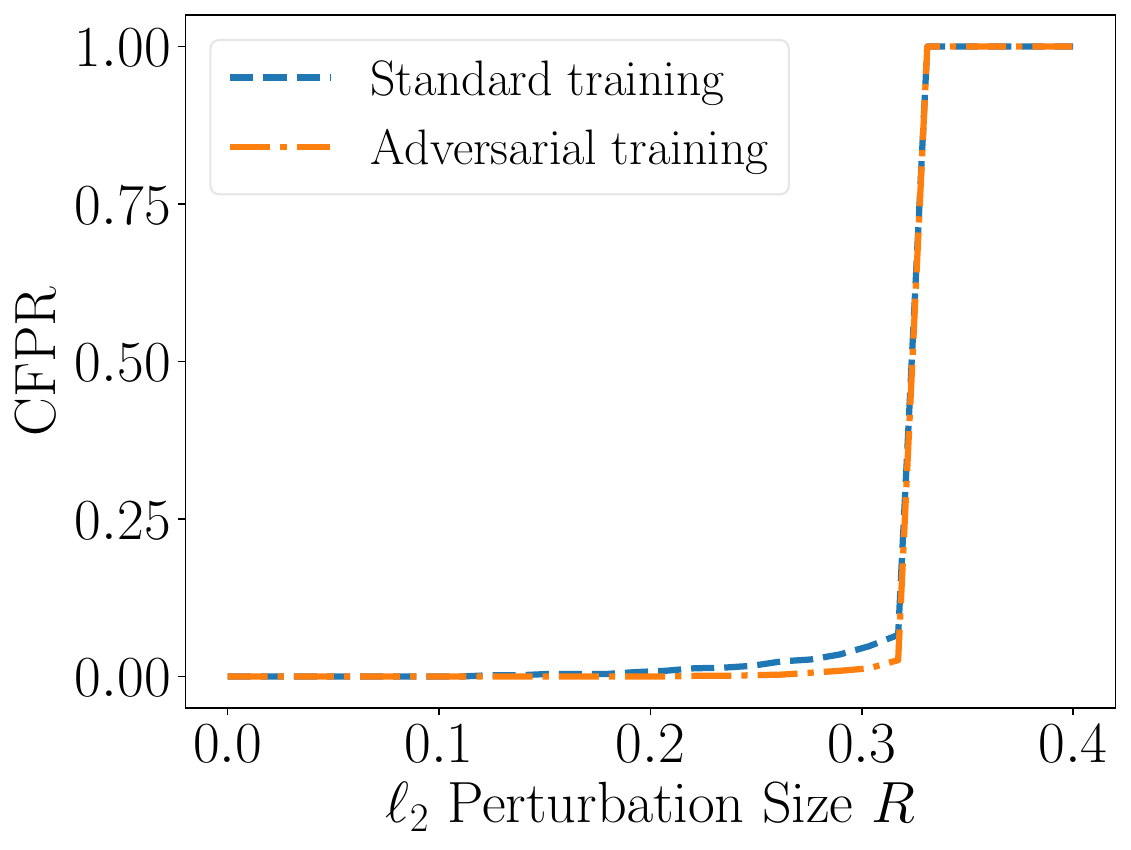}}
\caption{\label{figure:diff training--midjourney} Results of standard vs. adversarial training for  Midjourney and DALL-E datasets.}
\end{figure}

\begin{figure}[!t]
\centering
\subfloat[Midjourney]{\includegraphics[width=0.23\textwidth]{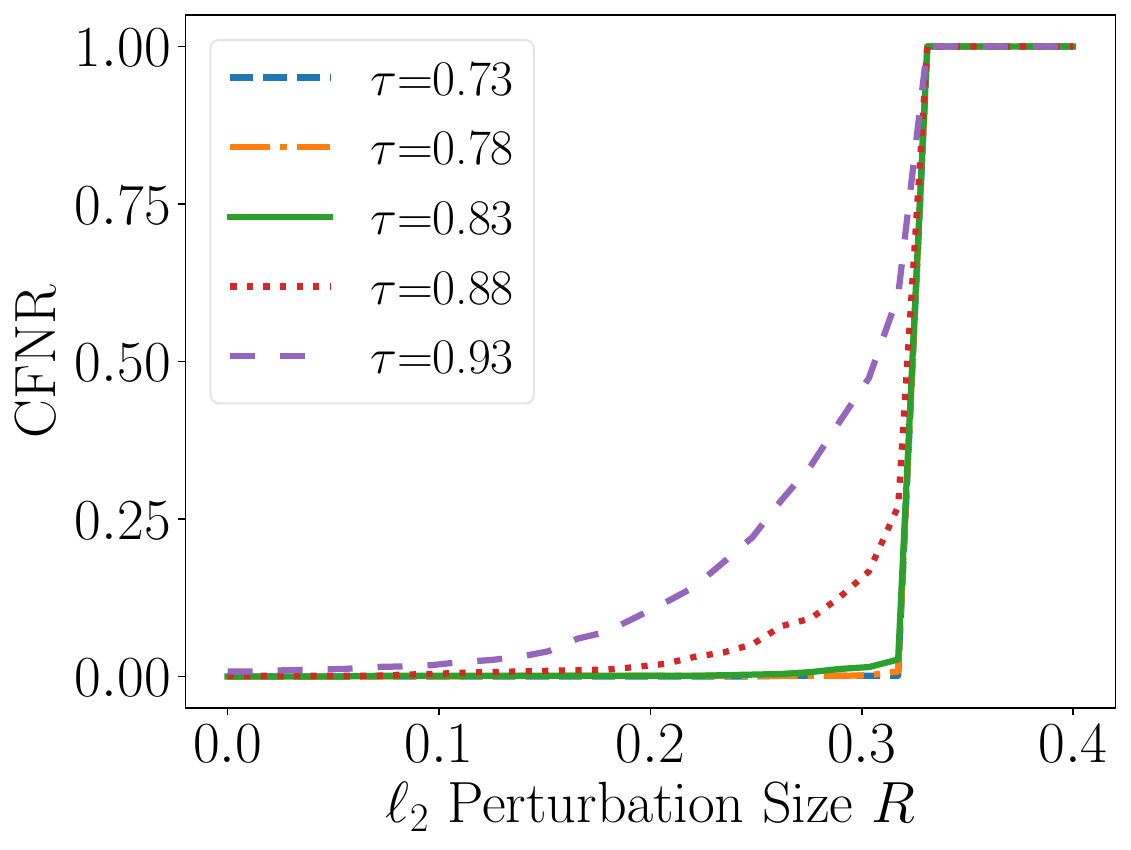}}
\subfloat[Midjourney]{\includegraphics[width=0.23\textwidth]{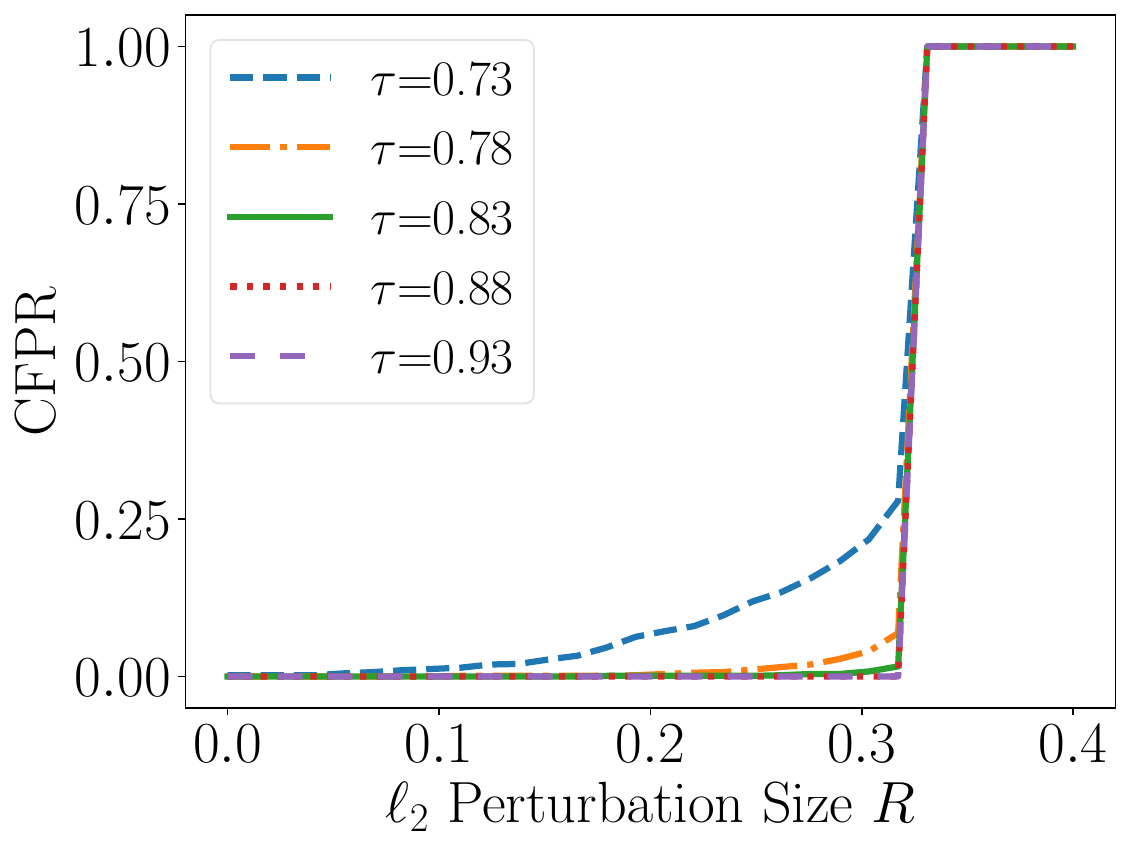}}
\subfloat[DALL-E]{\includegraphics[width=0.23\textwidth]{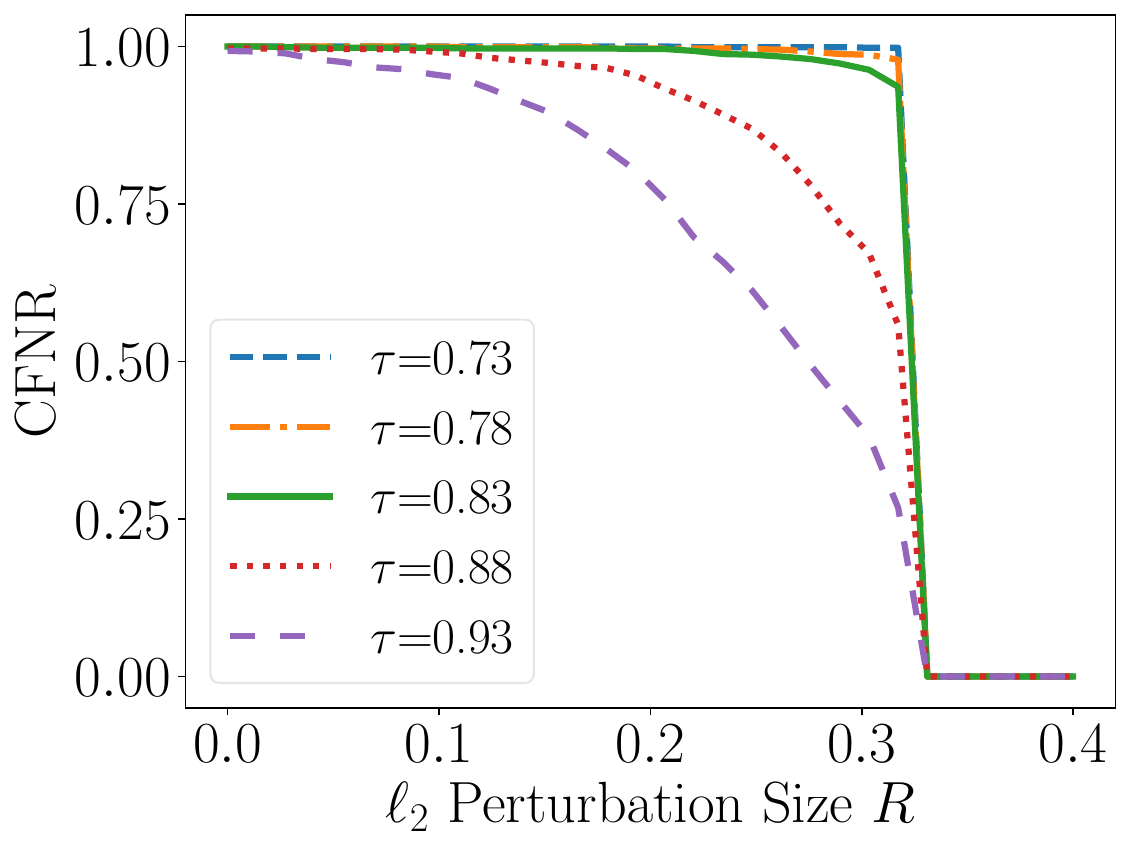}}
\subfloat[DALL-E]{\includegraphics[width=0.23\textwidth]{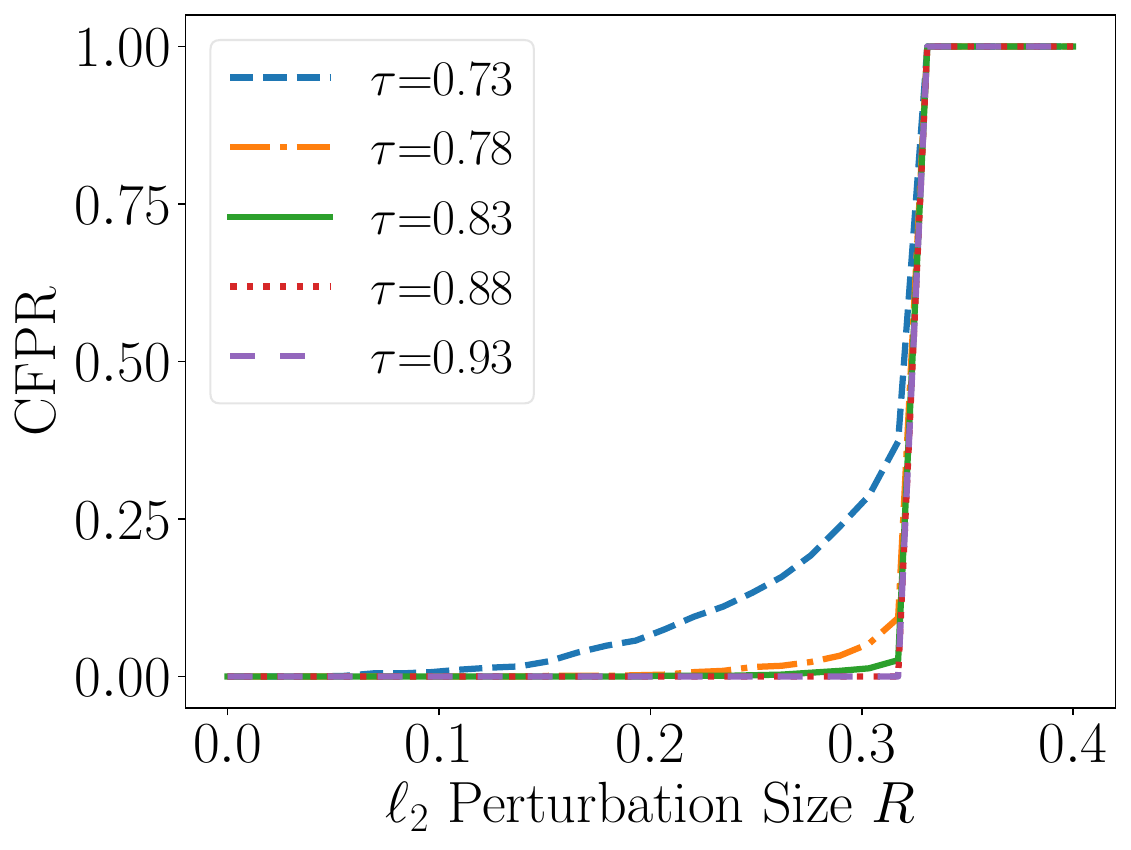}}
\caption{\label{figure:tau--other datasets} Impact of detection threshold $\tau$ for Midjourney and DALL-E datasets.}
\end{figure}

\begin{figure}[!t]
\centering
\subfloat[Midjourney]{\includegraphics[width=0.23\textwidth]{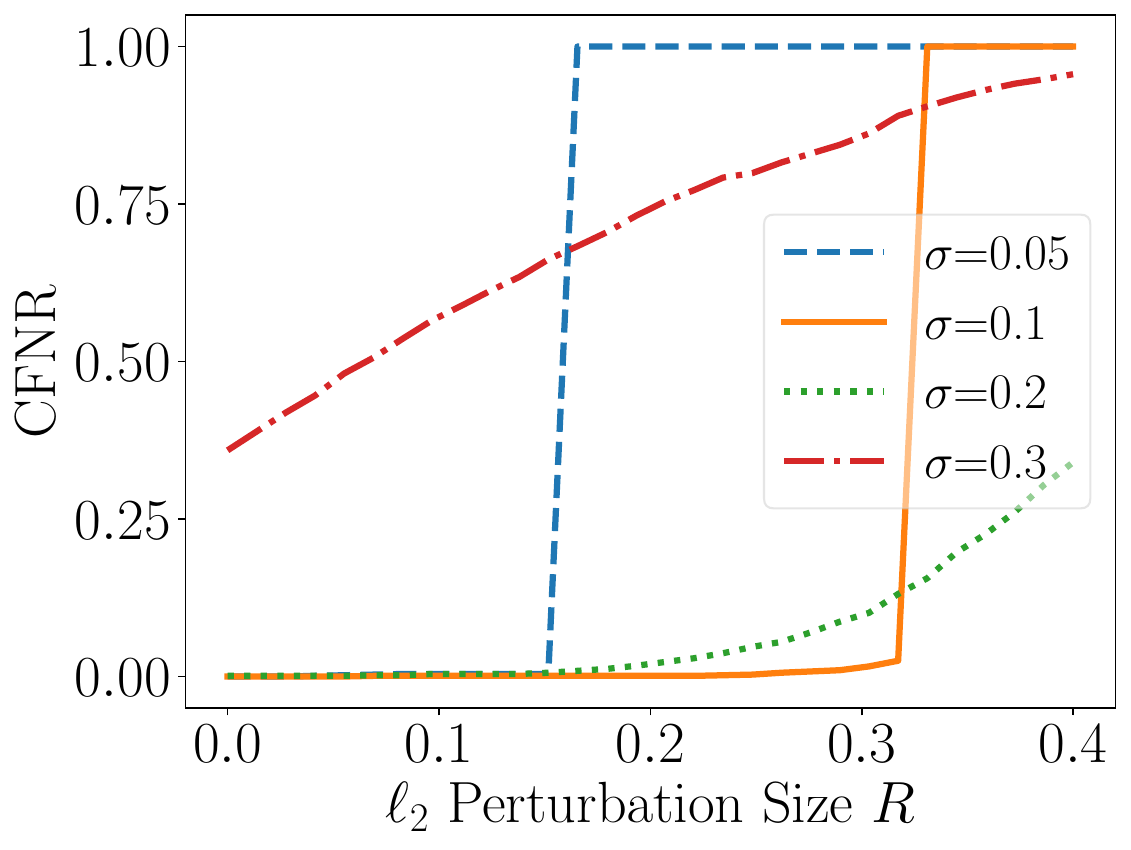}}
\subfloat[Midjourney]{\includegraphics[width=0.23\textwidth]{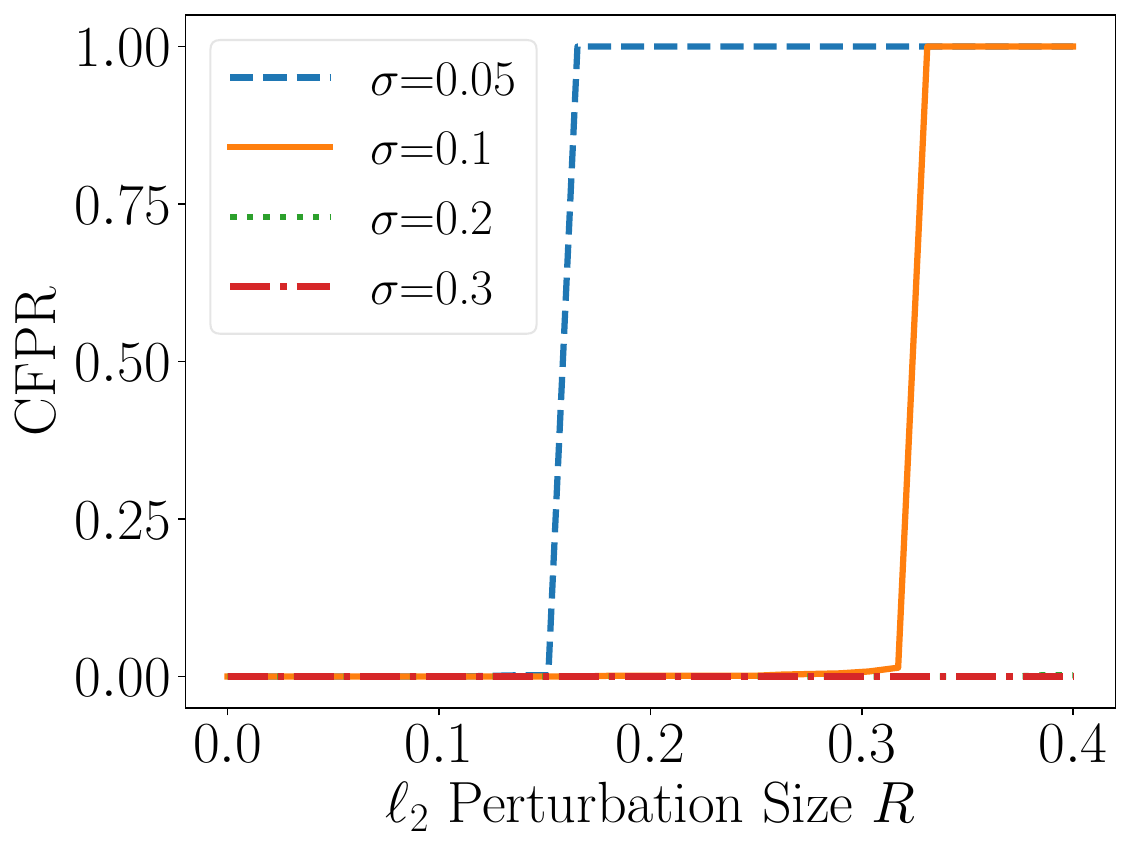}}
\subfloat[DALL-E]{\includegraphics[width=0.23\textwidth]{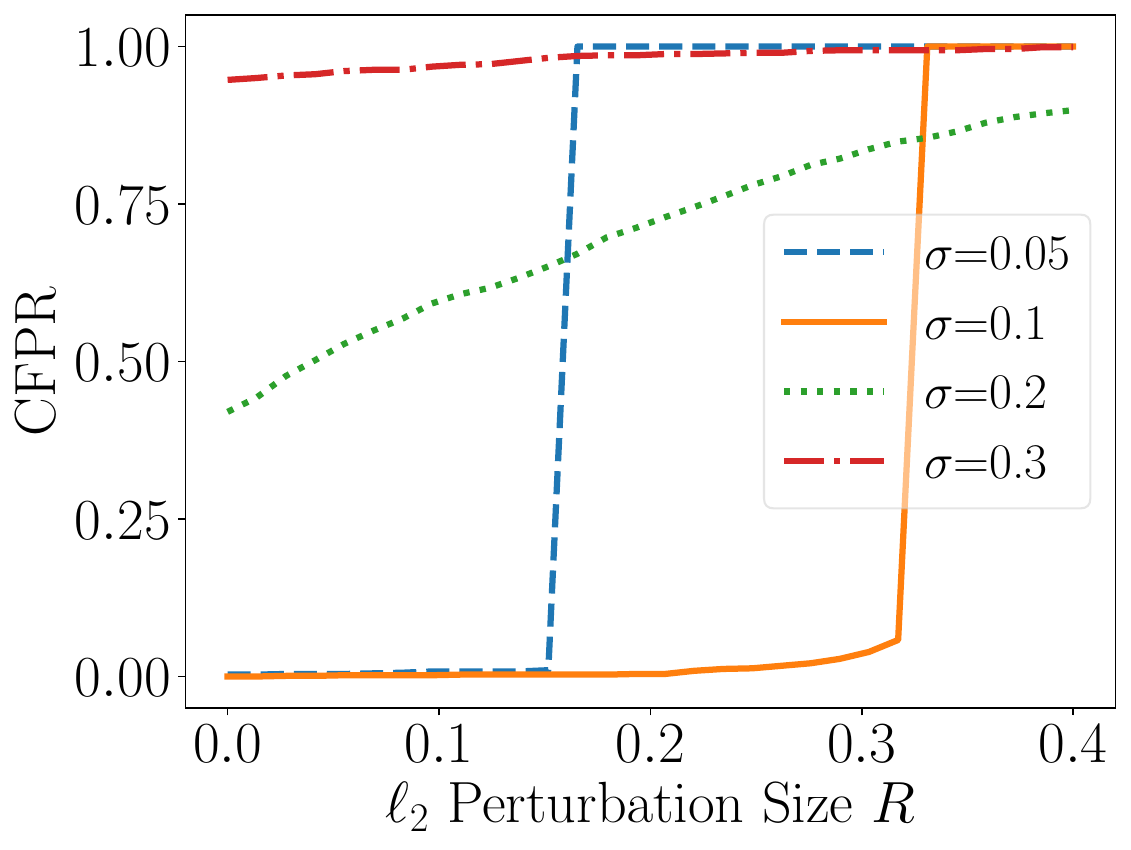}}
\subfloat[DALL-E]{\includegraphics[width=0.23\textwidth]{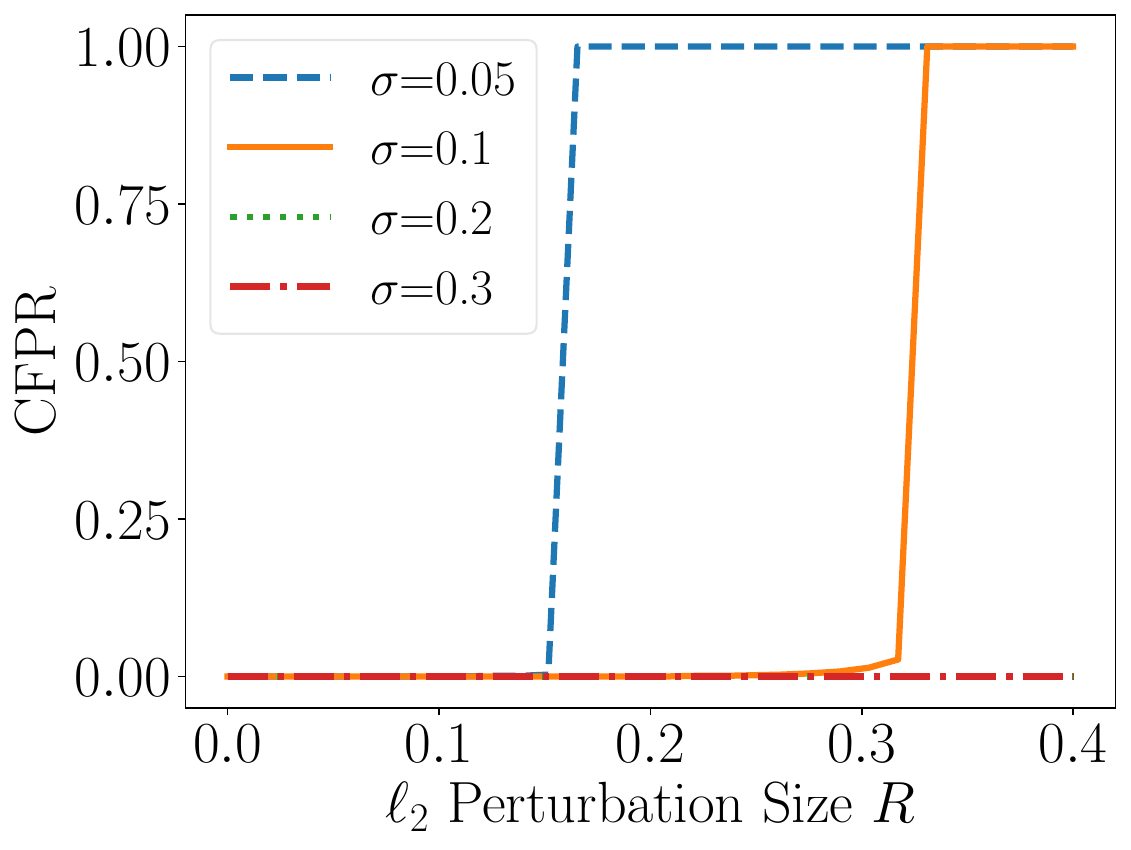}}
\caption{\label{figure:sigma other datasets} Impact of smoothing Gaussian noise standard derivation $\sigma$ for Midjourney and DALL-E datasets.}
\vspace{-3mm}
\end{figure}

\begin{figure}[!t]
\centering
\subfloat{\includegraphics[width=0.247\textwidth]{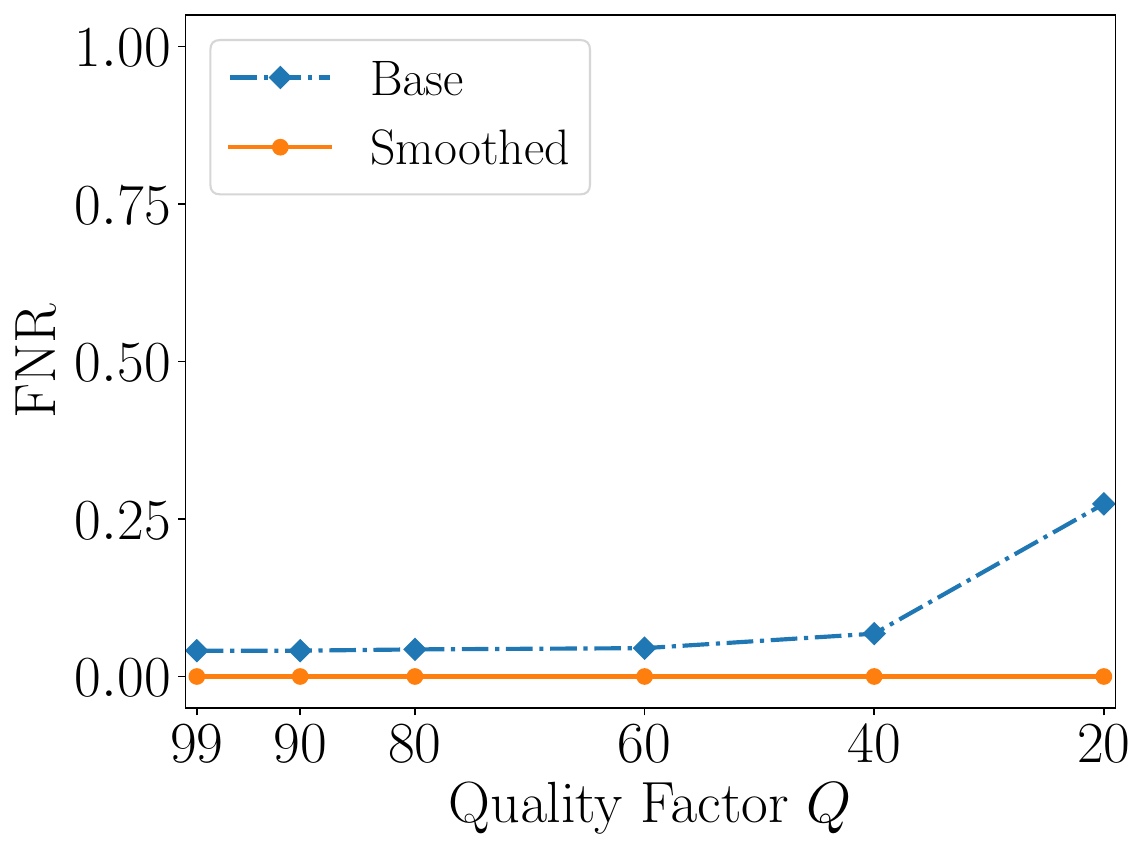}}
\subfloat{\includegraphics[width=0.235\textwidth]{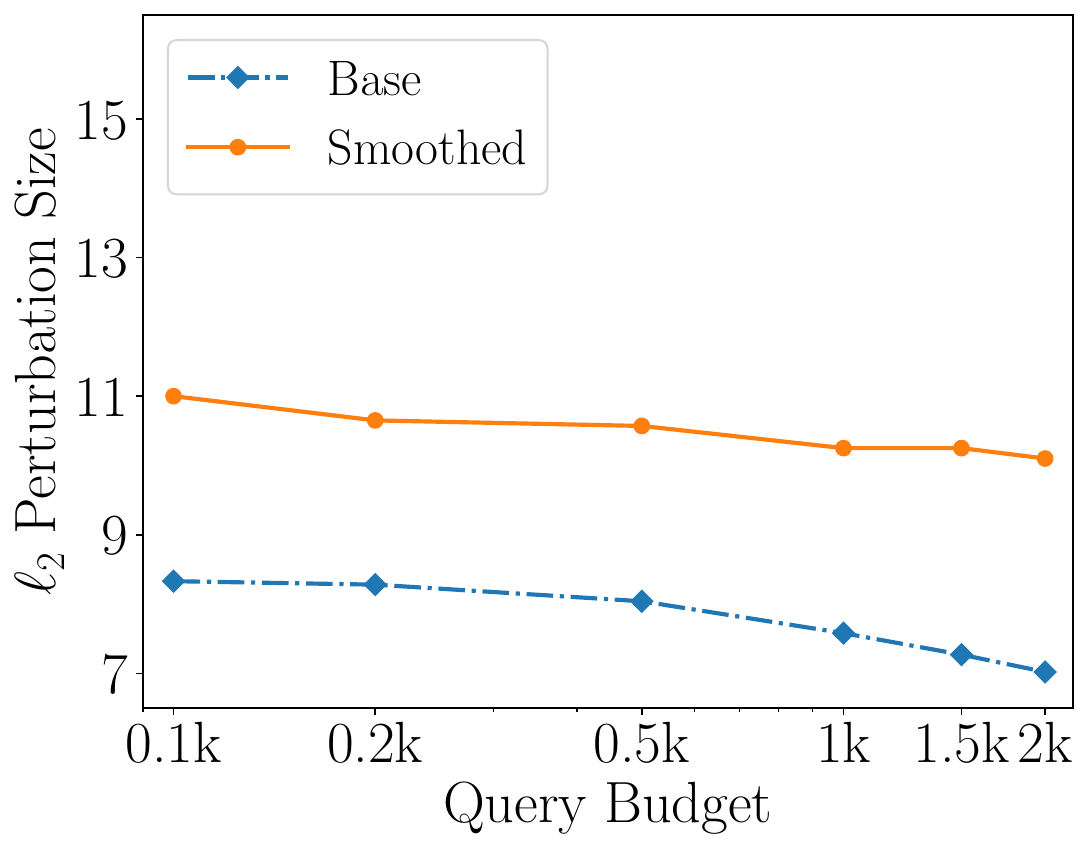}}
\subfloat{\includegraphics[width=0.24\textwidth]{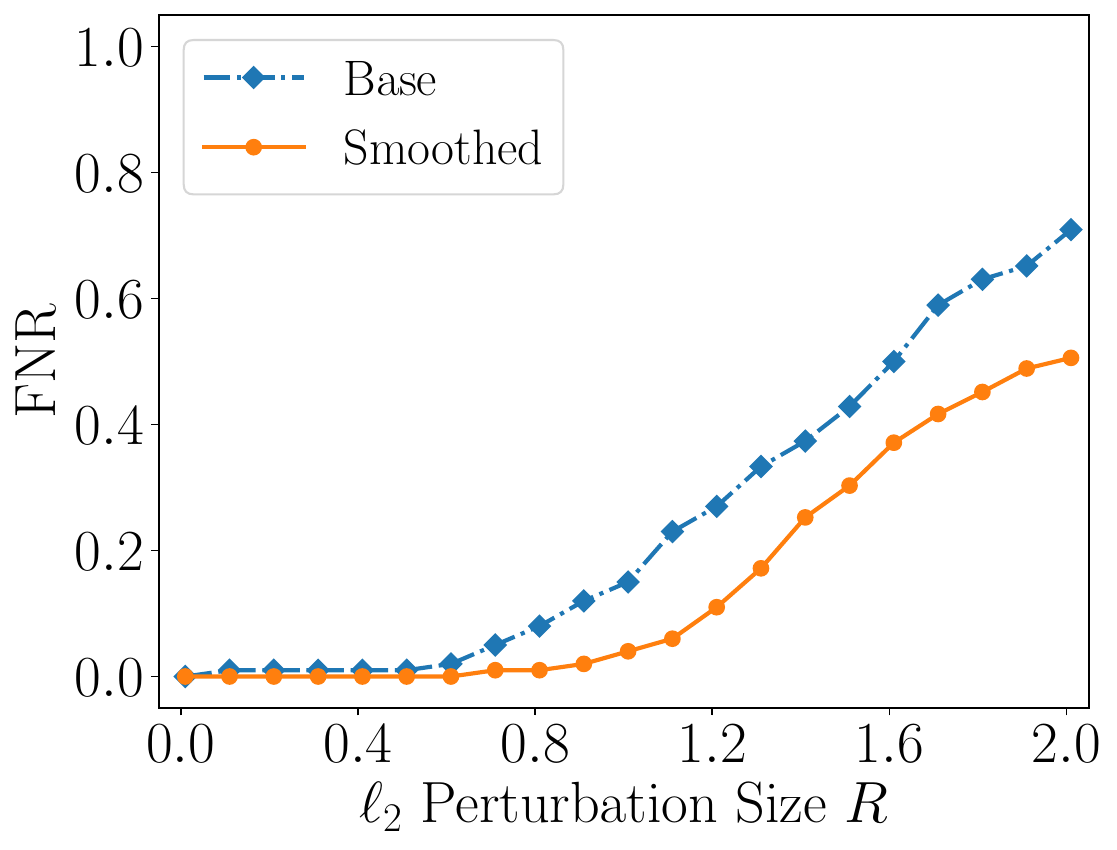}}
\subfloat{\includegraphics[width=0.24\textwidth]{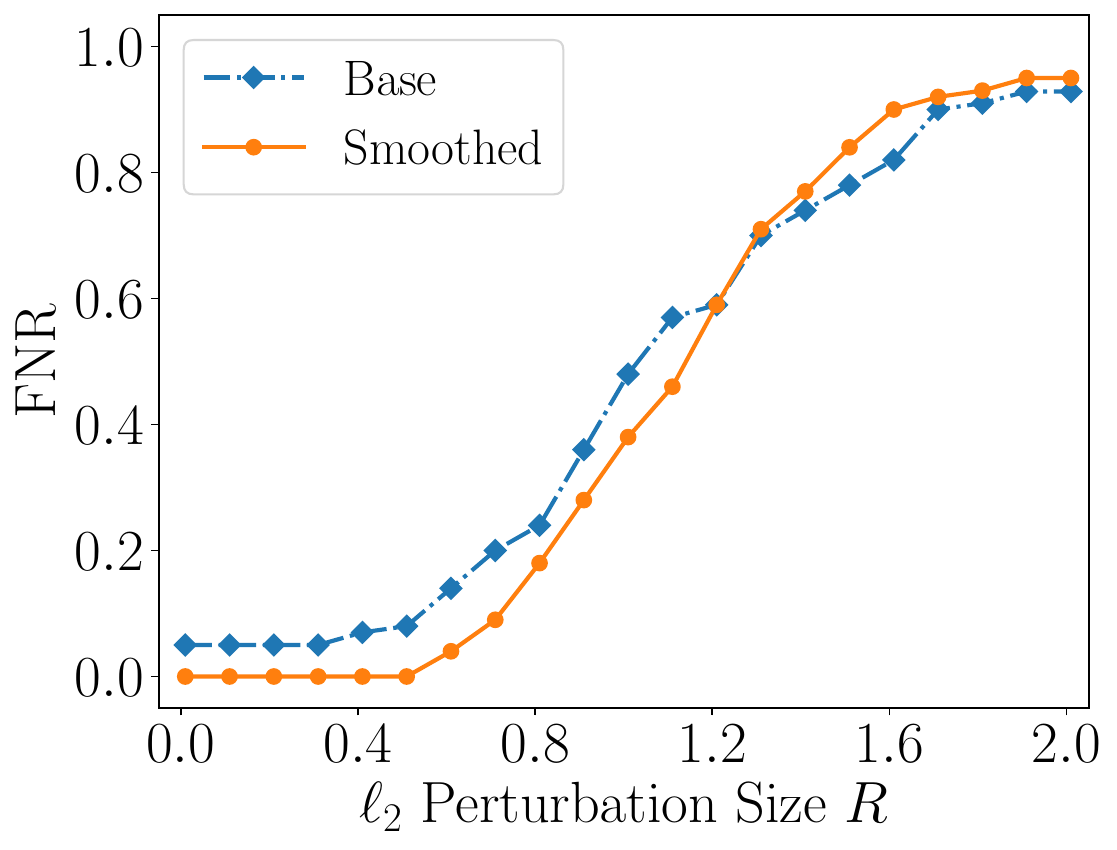}}

\subfloat[JPEG compression]{\includegraphics[width=0.247\textwidth]{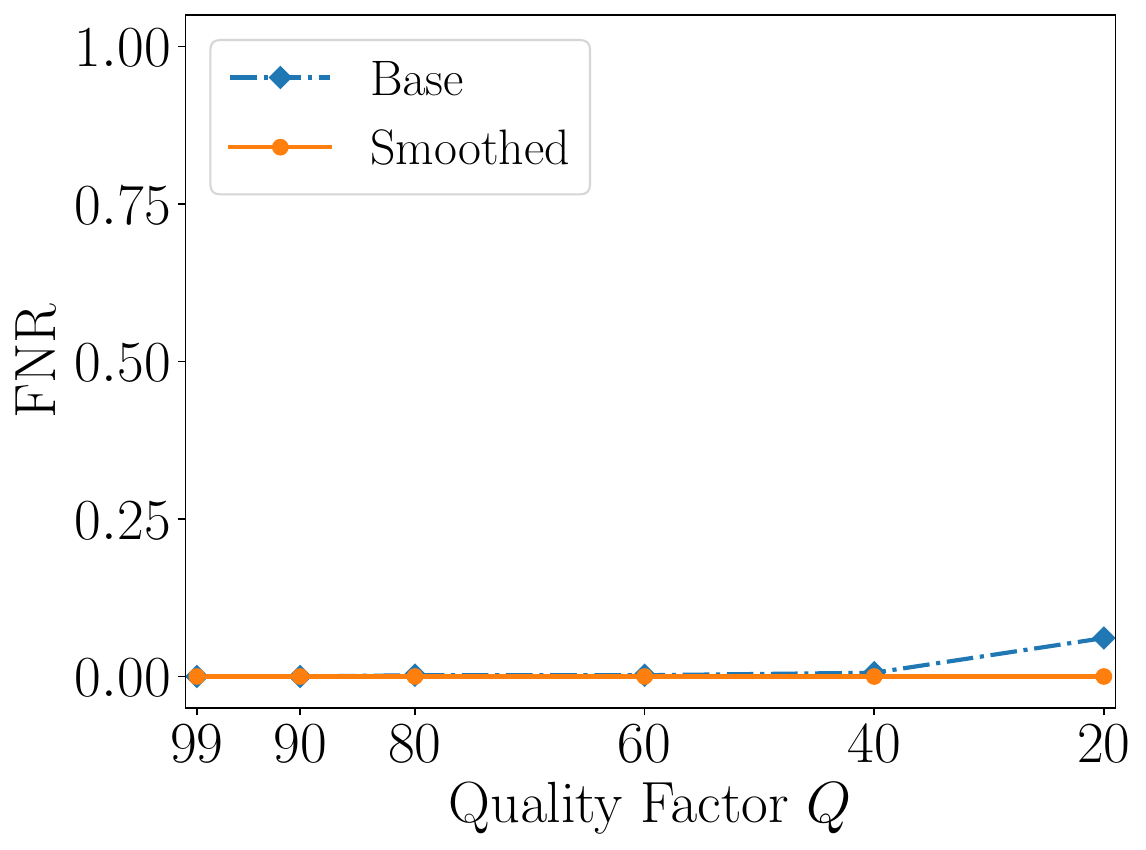}}
\subfloat[Black-box]{\includegraphics[width=0.235\textwidth]{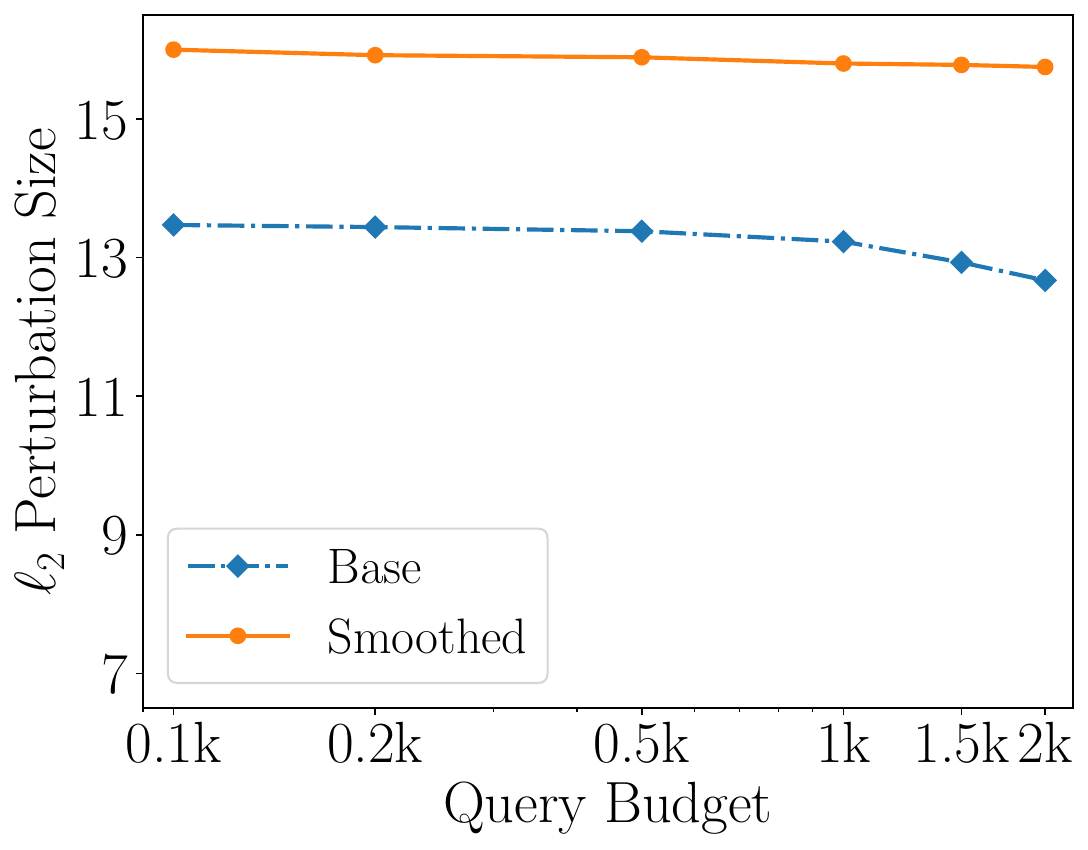}}
\subfloat[White-box]{\includegraphics[width=0.24\textwidth]{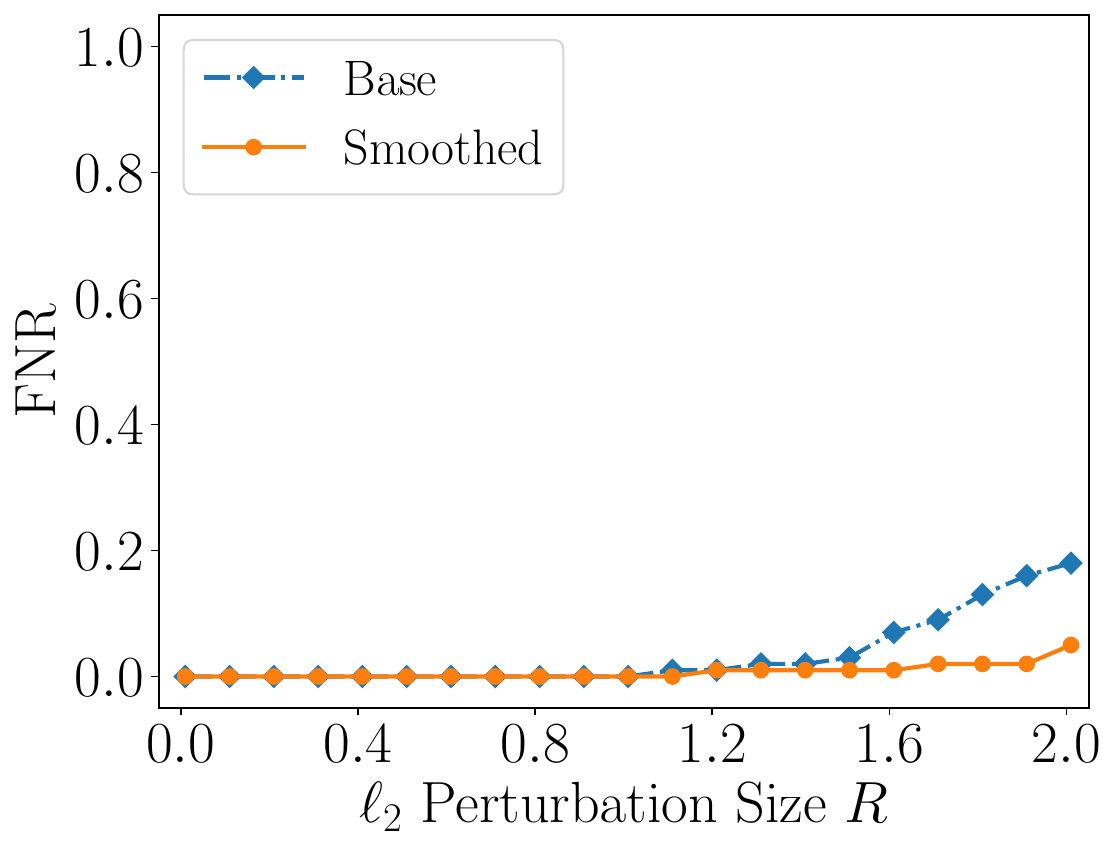}}
\subfloat[Adaptive white-box]{\includegraphics[width=0.24\textwidth]{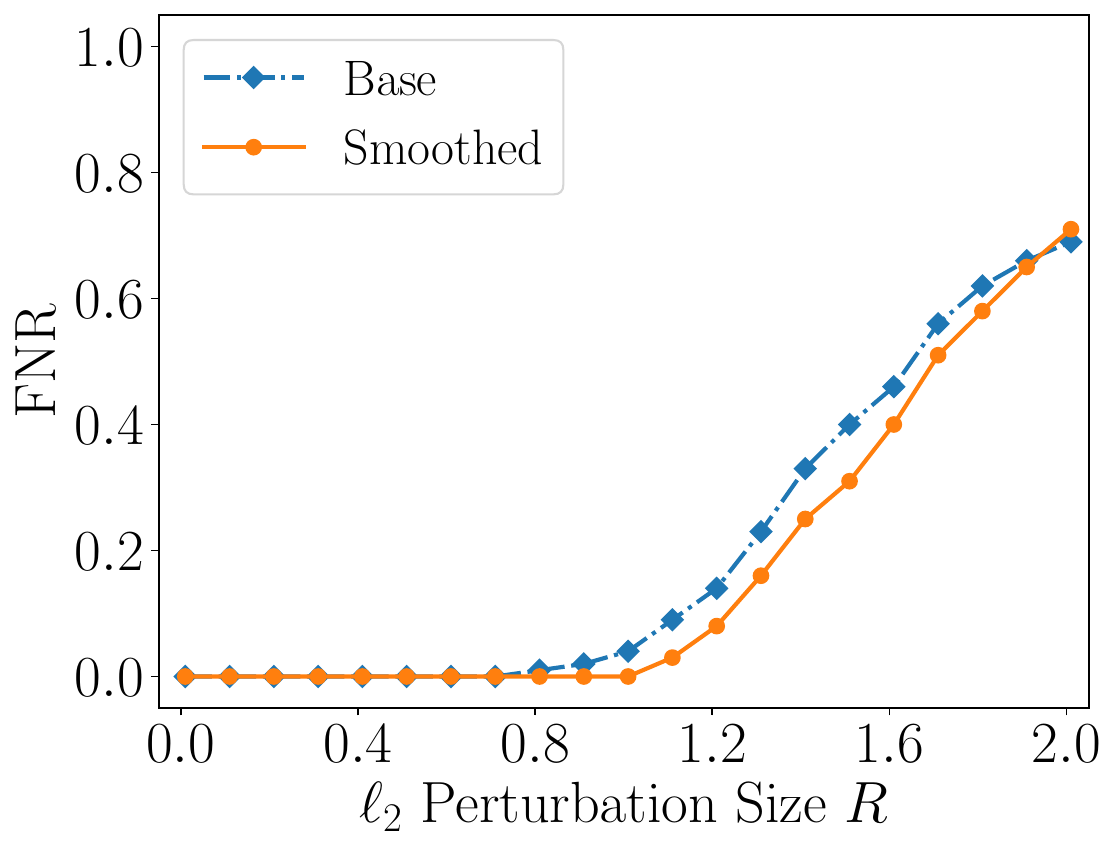}}
\caption{Results of base vs. smoothed watermarking under the 4 removal attacks. First row: Midjourney. Second row: DALL-E.}
\label{empiricalFNR-othertwo}
\vspace{-3mm}
\end{figure}

\begin{figure}[!t]
\centering
\subfloat{\includegraphics[width=0.247\textwidth]{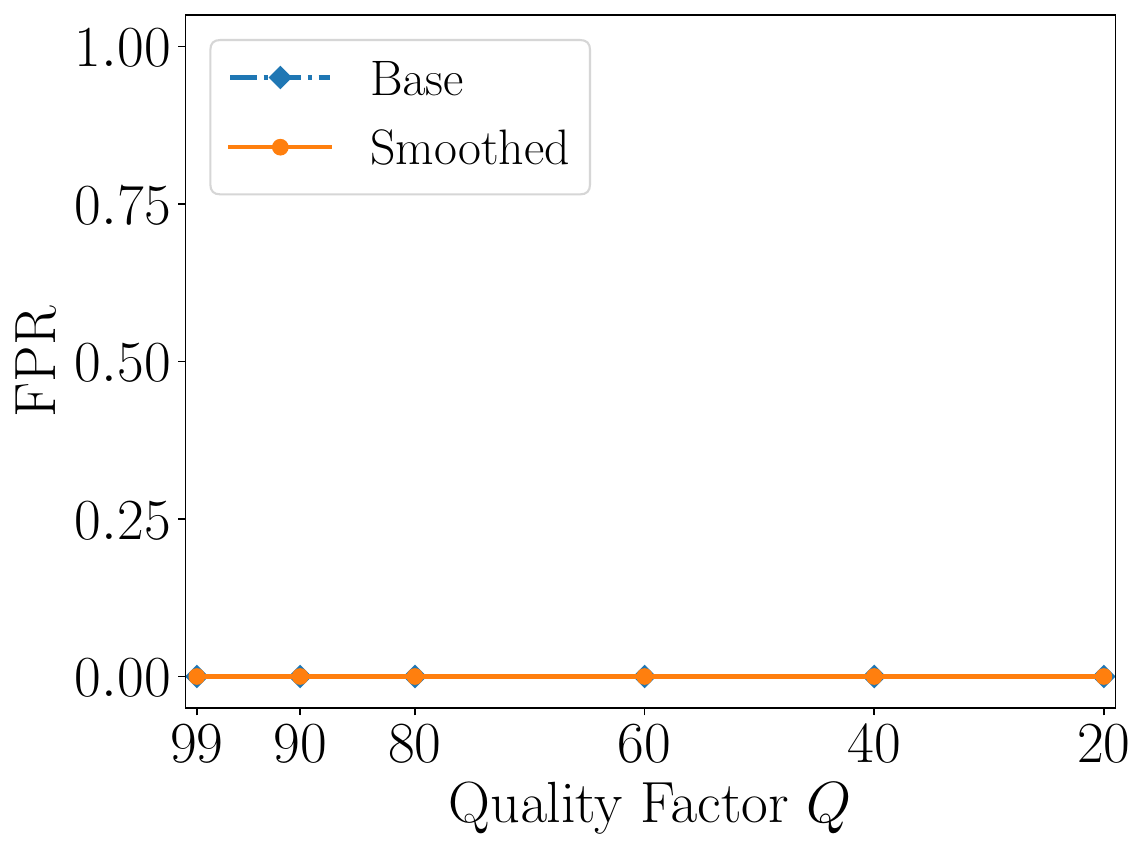}}
\subfloat{\includegraphics[width=0.235\textwidth]{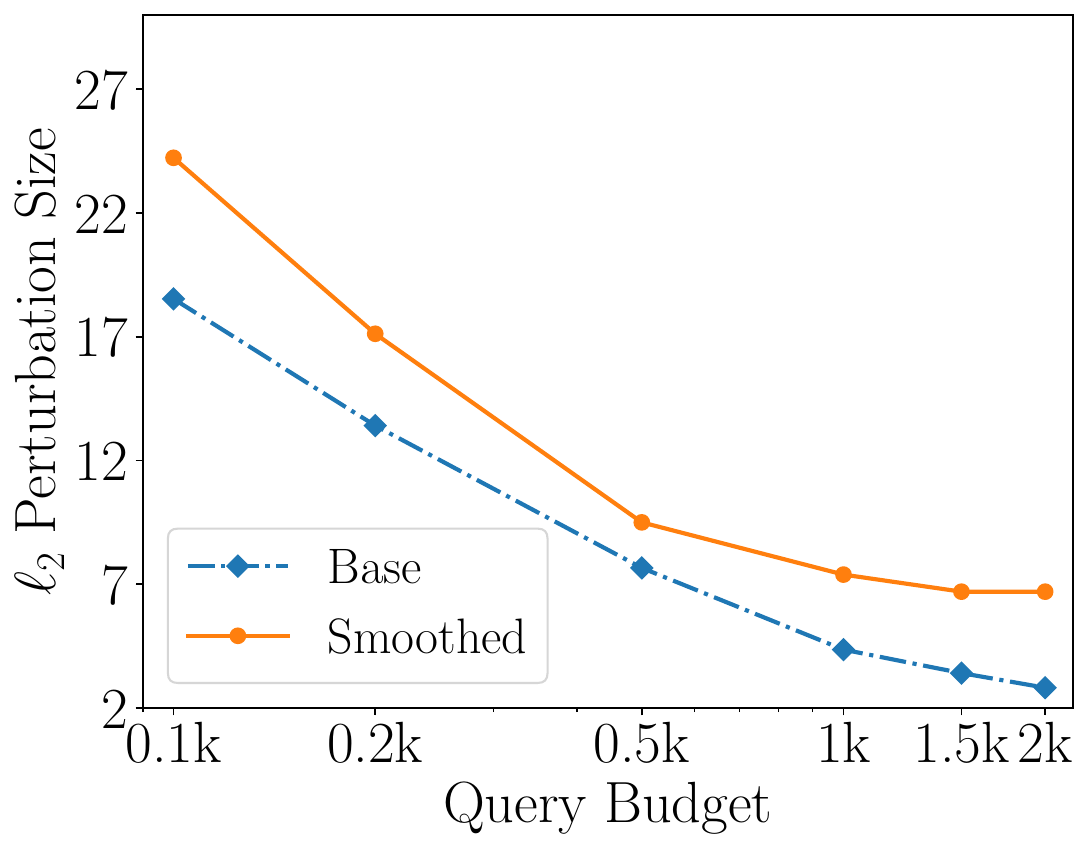}}
\subfloat{\includegraphics[width=0.24\textwidth]{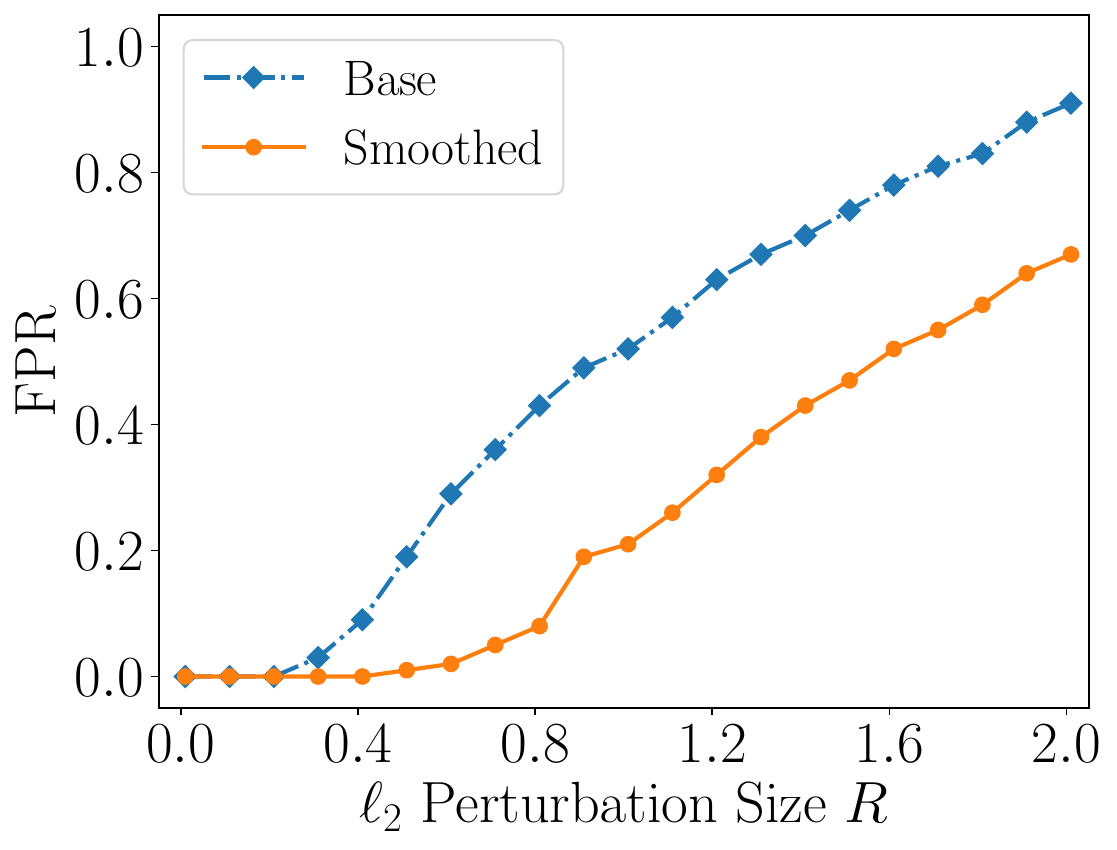}}
\subfloat{\includegraphics[width=0.24\textwidth]{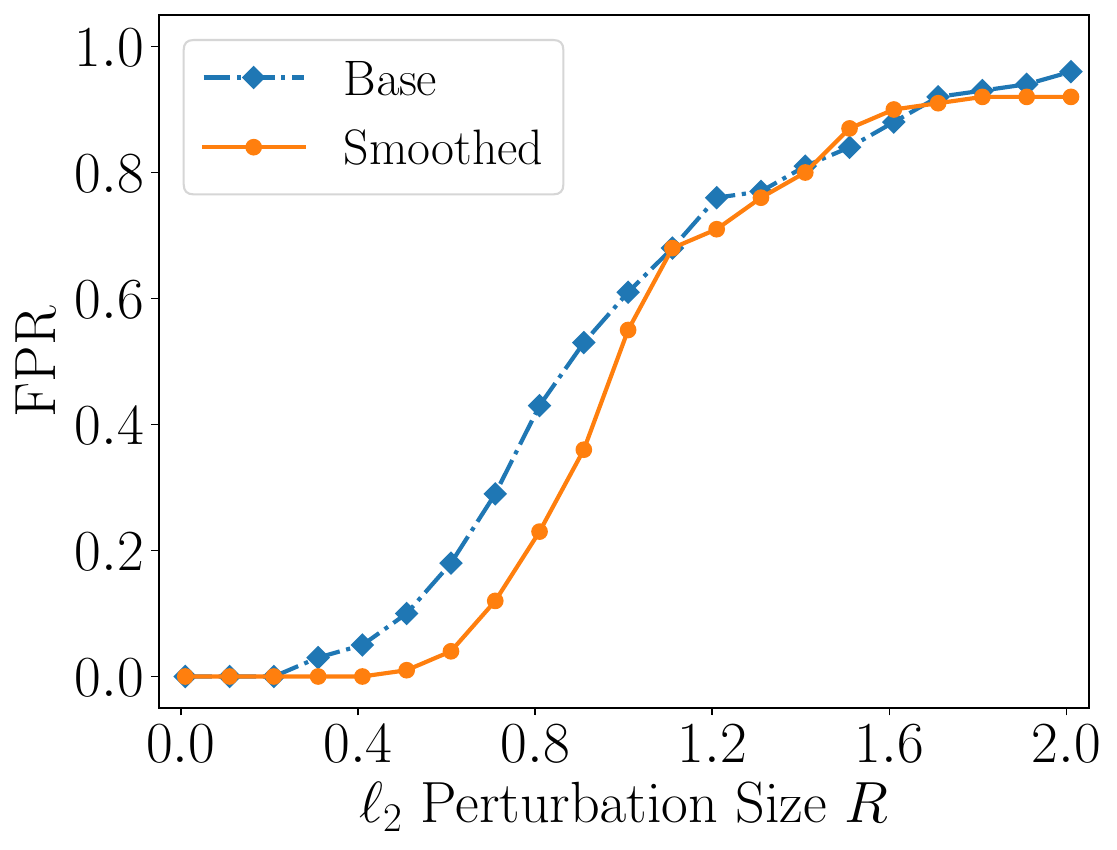}}

\subfloat{\includegraphics[width=0.247\textwidth]{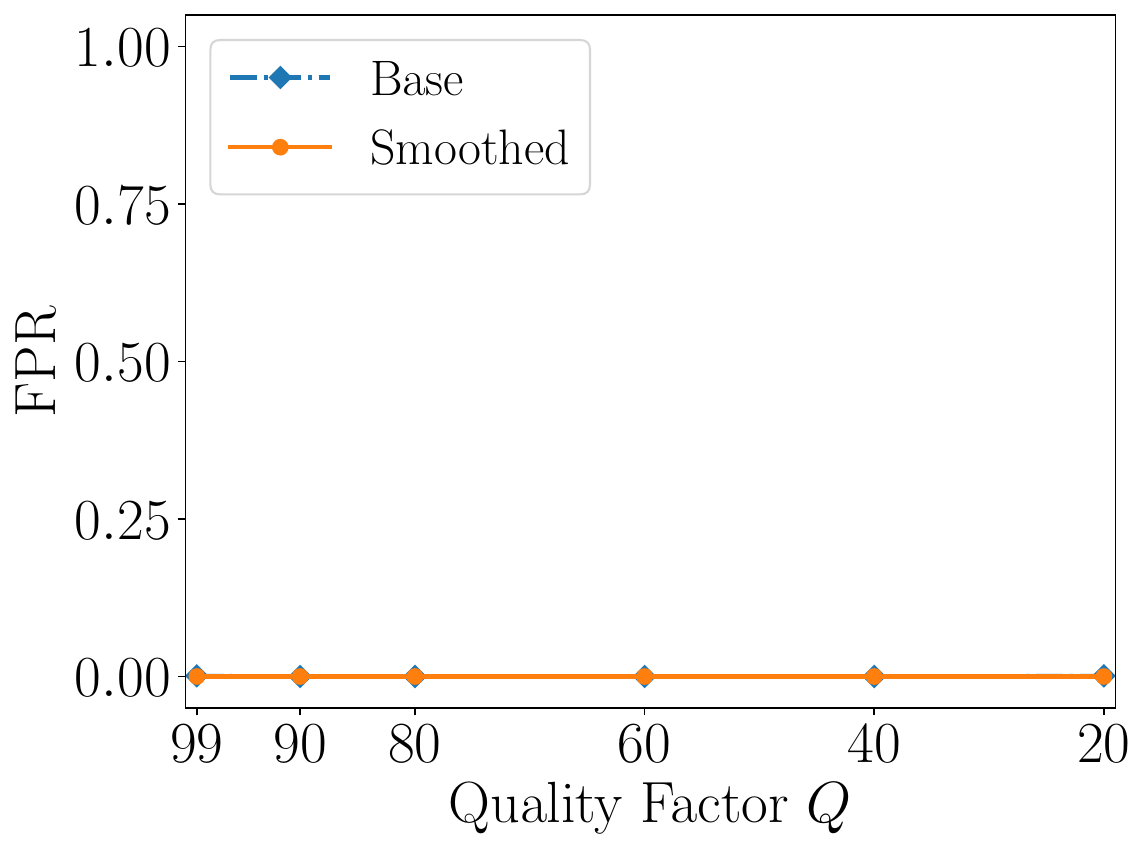}}
\subfloat{\includegraphics[width=0.235\textwidth]{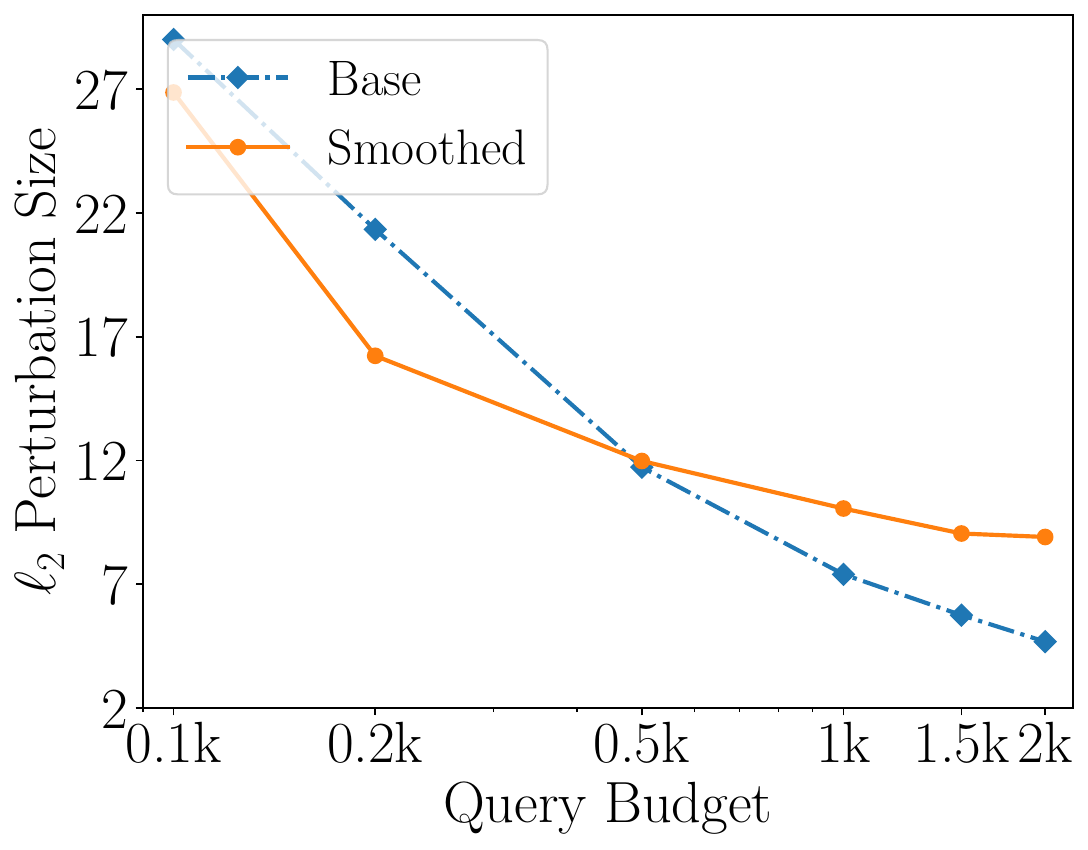}}
\subfloat{\includegraphics[width=0.24\textwidth]{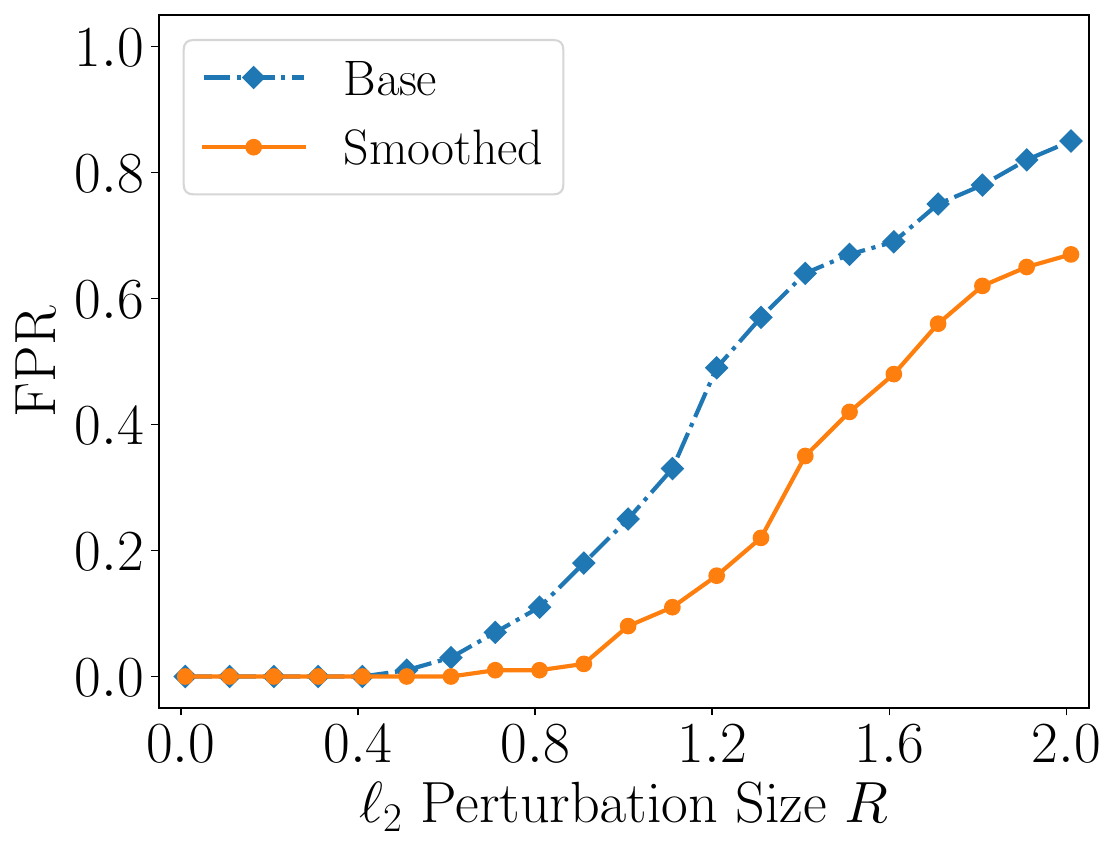}}
\subfloat{\includegraphics[width=0.24\textwidth]{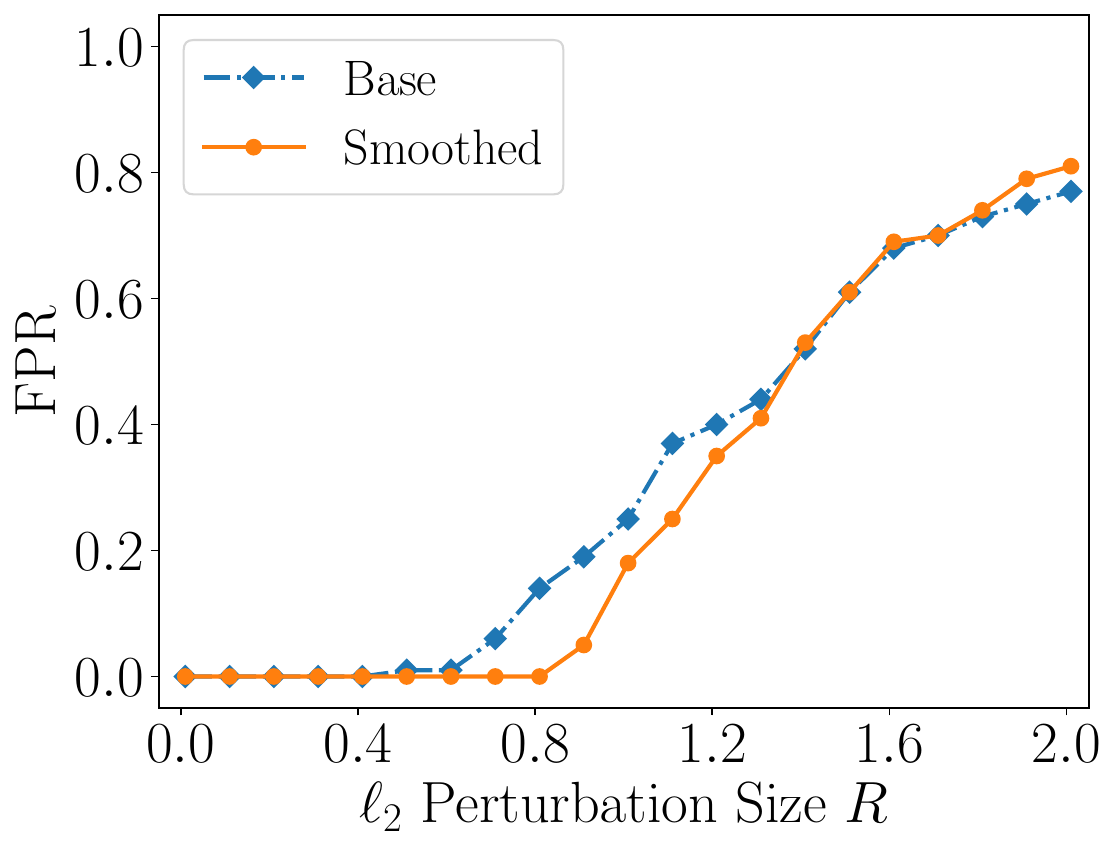}}

\subfloat[JPEG compression]{\includegraphics[width=0.247\textwidth]{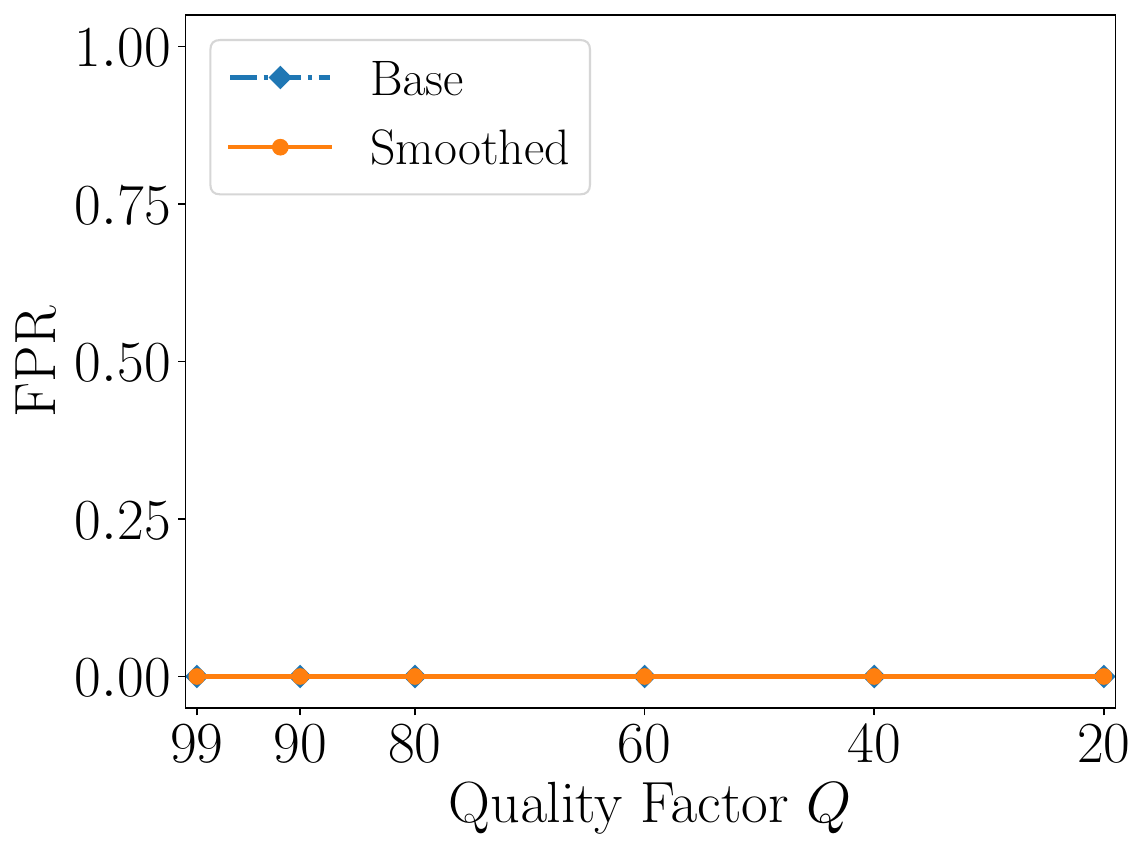}}
\subfloat[Black-box]{\includegraphics[width=0.235\textwidth]{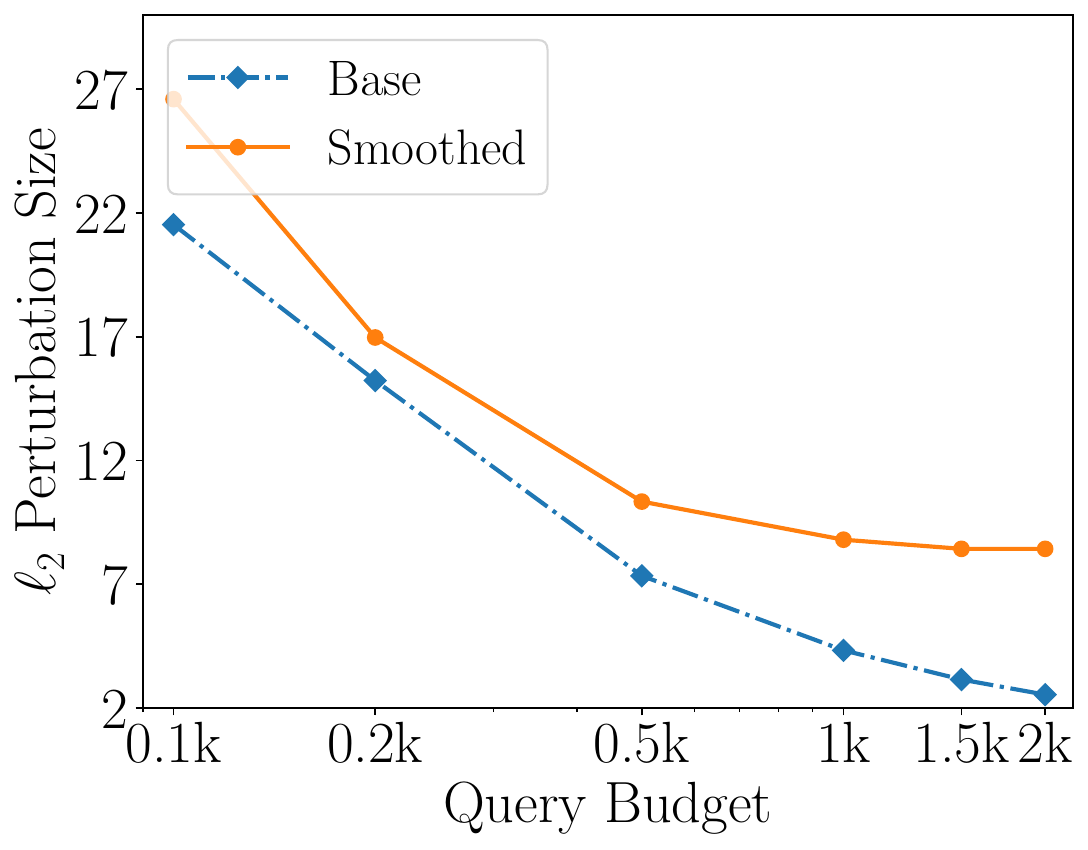}}
\subfloat[White-box]{\includegraphics[width=0.24\textwidth]{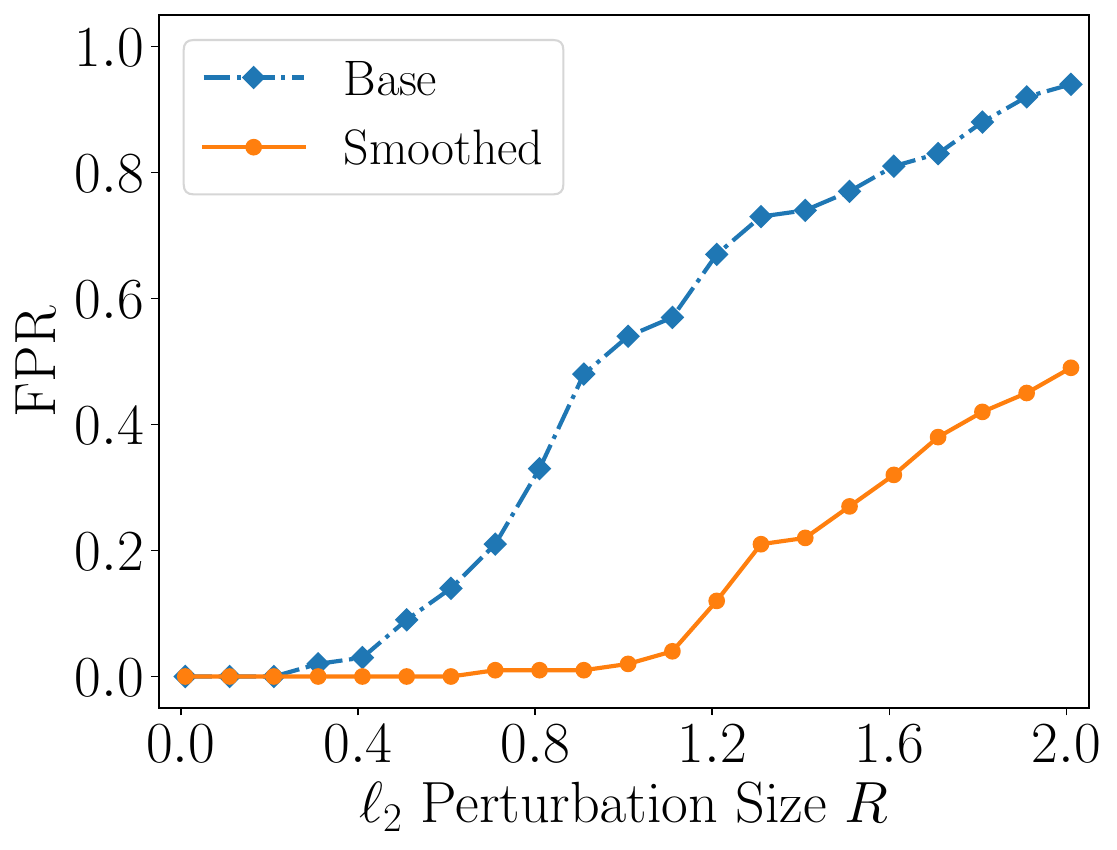}}
\subfloat[Adaptive white-box]{\includegraphics[width=0.24\textwidth]{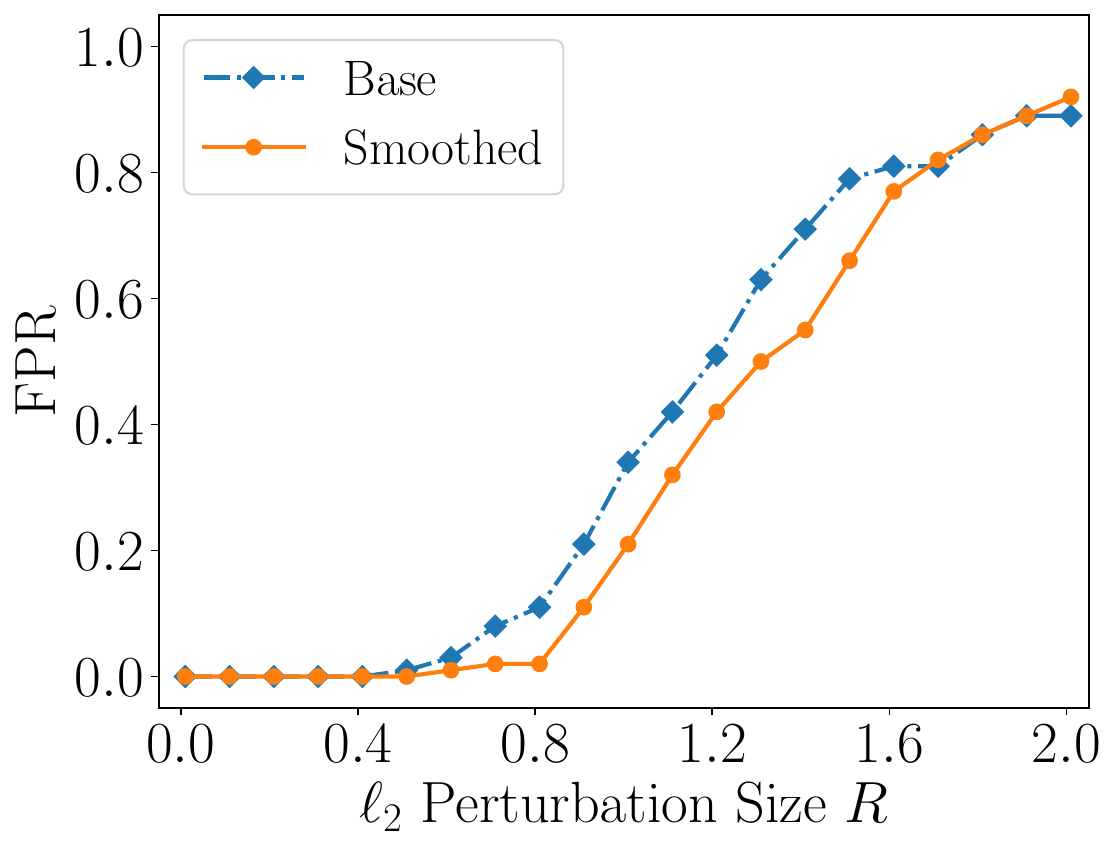}}
\caption{\label{fig:forgery attack}Results of base vs. smoothed watermarking under the 4 forgery attacks. First row: Stable Diffusion. Second row: Midjourney. Third row: DALL-E.}
\label{empiricalFPR}
\vspace{-3mm}
\end{figure}

\section{\label{proof:theorem1}Proof of Theorem~\ref{theorem:multi-class}}
Given an image $x$ with the perturbation $\delta$ added to it, our smoothed decoder $D_s$ decodes a watermark $D_s(x+\delta)$. Following Equation~\ref{equ:cohen}, we calculate $r_i(x)$ for $i$th bit in the watermark as follows:
\begin{align}
    r_i(x)=\sigma \Phi^{-1} (\underline{p_{l_i}}), 
\end{align}
where $\Phi^{-1}$ is the inverse cumulative distribution function of the standard Gaussian, and $\underline{p_{l_i}}$ is a lower bound of $\text{Pr}(D(x+\epsilon)[i]=D_s(x)[i])$, i.e., $\underline{p_{l_i}} \leq \text{Pr}(D(x+\epsilon)[i]=D_s(x)[i])$, $\epsilon \sim \mathcal{N}(0, \sigma^{2}I)$. If the perturbation size is smaller than $r_i(x)$, we have:
\begin{align}
    D_s(x+\delta)[i] = D_s(x)[i], \quad \forall \|\delta\|_2<r_i(x).
\end{align}
Given the ground-truth watermark $w_t$, when the added perturbation $\delta$ is $\ell_2$-norm bounded by $R$, we derive a lower bound $\underline{BA}(x)$ for $BA(D_s(x+\delta), w_t)$ as follows:
\begin{align}
    BA(D_s(x+\delta), w_t) &= \frac{1}{m} \sum_{i=1}^{m} \mathbb{I}(D_s(x+\delta)[i]=w_t[i]) \\
    &= \frac{1}{m} \sum_{i=1}^{m} \mathbb{I}(D_s(x+\delta)[i]=w_t[i]) \\
    &\geq \frac{1}{m} \sum_{i=1}^{m} \mathbb{I}(D_s(x+\delta)[i]=w_t[i]) \cdot \mathbb{I}(r_i(x) \geq R) \\
    &= \frac{1}{m} \sum_{i=1}^{m} \mathbb{I}(D_s(x)[i]=w_t[i]) \cdot \mathbb{I}(r_i(x) \geq R), 
\end{align}
where $\mathbb{I}$ is the indicator function. Also, we derive an upper bound $\overline{BA}(x)$ for $BA(D_s(x+\delta), w_t)$ as follows:
\begin{align}
    BA(D_s(x+\delta), w_t) &= \frac{1}{m} \sum_{i=1}^{m} \mathbb{I}(D_s(x+\delta)[i]=w_t[i]) \\
    &= 1 - \frac{1}{m} \sum_{i=1}^{m} \mathbb{I}(D_s(x+\delta)[i]=\neg w_t[i]) \\
    &\leq 1 - \frac{1}{m} \sum_{i=1}^{m} \mathbb{I}(D_s(x+\delta)[i]=\neg w_t[i]) \cdot \mathbb{I}(r_i(x) \geq R) \\
    &= 1 - \frac{1}{m} \sum_{i=1}^{m} \mathbb{I}(D_s(x)[i]=\neg w_t[i]) \cdot \mathbb{I}(r_i(x) \geq R), 
\end{align}
where $\neg w_t$ means flipping each bit of the watermark $w_t$.

\section{\label{proof:theorem2}Proof of Theorem~\ref{theorem:multi-label}}
Given the smoothed multi-label classifier $g$, we define $\underline{e}(x)$ and $\overline{e}(x)$ as follows:
\begin{align}
    \underline{e}(x) &= \sup \{e \in \mathbb{N} \mid \left|L \cap g(x+\delta)\right| \geq e, \forall \|\delta\|_2<R \}, \\
    \overline{e}(x) &= \sup \{e \in \mathbb{N} \mid \left|\mathcal{Y}/L \cap g(x+\delta)\right| \geq e, \forall \|\delta\|_2<R \}, 
\end{align}
where $L$ is the set of indices of ones in $w_t$, i.e., $L=\{i\in \mathcal{Y}|w_t[i]=1\}$, and $\mathcal{Y}=\{1,2,\cdots,m\}$. Specifically, the smoothed decoder $D_s$ returns $k$ labels (corresponding to $k$ bits that are predicted to be 1), and other $m-k$ labels correspond to $m-k$ bits that are predicted to be 0. As defined in Section~\ref{building smoothed decoder}, the watermark $D_s(x + \delta)$ decoded by $D_s$ for $x+\delta$ is as follows:
\begin{align}
    D_s(x+\delta)[i] = 
    \begin{cases}
    1, & \text{if } i \in g(x+\delta), \\
    0, & \text{if } i \notin g(x+\delta).
    \end{cases}
\end{align}
Therefore, for any perturbation $\delta$ whose $\ell_2$-norm is bounded by $R$, we derive a lower bound $\underline{BA}(x)$ for $BA(D_s(x+\delta), w_t)$ as follows:
\begin{align}
    &BA(D_s(x+\delta), w_t) = \frac{1}{m} \sum_{i=1}^{m} \mathbb{I}(D_s(x+\delta)[i]=w_t[i]) \\
    =& 1 - \frac{1}{m} \sum_{i=1}^{m} \mathbb{I}(i \notin g(x+\delta)) \cdot \mathbb{I}(i \in L) + \mathbb{I}(i \in g(x+\delta)) \cdot \mathbb{I}(i \in \mathcal{Y}/L) \\
    =& 1 - \frac{1}{m} \sum_{i=1}^{m} (\left|L\right| - \left|L \cap g(x+\delta)\right| + \left|g(x+\delta)\right| - \left|L \cap g(x+\delta)\right|) \\
    =& 1 - \frac{1}{m} (\|w_t\|_1 + k - 2\left|L \cap g(x+\delta)\right|) \\
    \geq& 1 - \frac{1}{m} (\|w_t\|_1 + k - 2\underline{e}(x)).
\end{align}
Also, for any perturbation $\delta$ whose $\ell_2$-norm is bounded by $R$, we derive an upper bound $\underline{BA}(x)$ for $BA(D_s(x+\delta), w_t)$ as follows:
\begin{align}
    &BA(D_s(x+\delta), w_t) = 1 - \frac{1}{m} \sum_{i=1}^{m} \mathbb{I}(D_s(x+\delta)[i]=\neg w_t[i]) \\
    =& \frac{1}{m} \sum_{i=1}^{m} \mathbb{I}(i \notin g(x+\delta)) \cdot \mathbb{I}(i \in \mathcal{Y}/L) + \mathbb{I}(i \in g(x+\delta)) \cdot \mathbb{I}(i \in L) \\
    =& \frac{1}{m} \sum_{i=1}^{m} (\left|\mathcal{Y}/L\right| - \left|\mathcal{Y}/L \cap g(x+\delta)\right| + \left|g(x+\delta)\right| - \left|\mathcal{Y}/L \cap g(x+\delta)\right|) \\
    =& \frac{1}{m} (m - \|w_t\|_1 + k - 2\left|\mathcal{Y}/L \cap g(x+\delta)\right|) \\
    \leq& \frac{1}{m} (m - \|w_t\|_1 + k - 2\overline{e}(x)).
\end{align}

\section{Proof of Theorem~\ref{theorem:regression}}
Following Equation~\ref{equ:median smoothing}, given the added perturbation $\delta$, the ground-truth watermark $w_t$, the decoder $D$, and the smoothed decoder $D_s$, we define $g(x+\delta)=BA(D_s(x+\delta),w_t)$ and $f(x)=BA(D(x),w_t)$. When $\delta$ is $\ell_2$-norm bounded by $R$, we have:
\begin{align}
    \sup \{y& \in \mathbb{R} \mid \text{Pr}\left(BA(D(x+\epsilon), w_t) \leq y\right) \leq \Phi(-\frac{R}{\sigma})\} \leq BA(D_s(x+\delta),w_t) \nonumber \\
    &\leq \inf \{y \in \mathbb{R} \mid \text{Pr}\left(BA(D(x+\epsilon), w_t) \leq y\right) \geq \Phi(\frac{R}{\sigma})\}, \quad \forall \|\delta\|_2<R,
\end{align}
where $\epsilon \sim \mathcal{N}(0, \sigma^2 I)$ and $\Phi$ is the cumulative distribution function of the standard Gaussian.

\section{\label{appendix:calculate e}Calculation of $\underline{e}(x)$ and $\overline{e}(x)$}
We have estimated the lower bounds $\underline{p_s}$ for $s \in \{i \in \mathcal{Y}|w_t[i]=1\}$ and upper bounds $\overline{p_t}$ for $t \in \{i \in \mathcal{Y}|w_t[i]=0\}$ in Section~\ref{estimating bound}. To calculate $\underline{e}(x)$ and $\overline{e}(x)$, we follow Jia et al.~\cite{jia2022multiguard}. We denote that $d=\|w_t\|_1$. Without loss of generality, we assume $\underline{p_{s_1}} \geq \underline{p_{s_2}} \geq \cdots \geq \underline{p_{s_d}}$ for lower bounds $\underline{p_s}$ and $\overline{p_{t_1}} \geq \overline{p_{t_2}} \geq \cdots \geq \underline{p_{t_d}}$ for upper bounds $\overline{p_t}$. Then, we have:
\begin{align}
&\underline{e}(x)=\underset{e^{\prime}=1,2, \cdots, \min \{d, k\}}{\operatorname{argmax}} e^{\prime} \\
\text { s.t. } &\max \left\{\Phi\left(\Phi^{-1}\left(\underline{p_{s_{e^{\prime}}}}\right)-\frac{R}{\sigma}\right), \max _{u=1}^\eta \frac{k^{\prime}}{u} \cdot \Phi\left(\Phi^{-1}\left(\frac{p_{S_u}}{k^{\prime}}\right)-\frac{R}{\sigma}\right)\right\} \nonumber \\
> &\min \left\{\Phi\left(\Phi^{-1}\left(\overline{p_{t_\mu}}\right)+\frac{R}{\sigma}\right), \min _{v=1}^\mu \frac{k^{\prime}}{v} \cdot \Phi\left(\Phi^{-1}\left(\frac{p_{T_v}}{k^{\prime}}\right)+\frac{R}{\sigma}\right)\right\},
\end{align}
where $\Phi$ and $\Phi^{-1}$ are the cumulative distribution function and its reverse of the standard Gaussian distribution, $\eta=d-e^{\prime}+1$, $\mu=k-e^{\prime}+1$, $p_{S_u}=\sum_{l=e^{\prime}}^{e^{\prime}+u-1} \underline{p_{s_l}}$, $p_{T_v}=\sum_{l=\mu-v+1}^{\mu} \overline{p_{t_l}}$, $R$ is perturbation's $\ell_2$-norm bound, $\sigma$ is the standard derivation of Gaussian noise, $k^{\prime}$ is the number of returned labels for the base multi-label classifier $f$ defined in Section~\ref{building smoothed decoder}, and $k$ is the number of returned labels for the smoothed multi-label classifier $g$ defined in Section~\ref{building smoothed decoder}. The calculation of $\overline{e}(x)$ follows a similar procedure.

\section{\label{appendix:post-processing} Removal and Forgery Attacks}
We consider four post-processing methods that can be applied to both removal and forgery attacks. For removal attack, the attacker takes a watermarked image $x_w$ as input and tries to perturb $x_w$ such that the detector misclassifies the perturbed image as non-watermarked. For forgery attack, the attacker takes a non-watermarked image $x_n$ as input and tries to perturb $x_n$ such that the detector misclassifies the perturbed image as watermarked.

\myparatight{JPEG compression} JPEG compression is widely used in image transmission and compression. This operation compresses an image with a quality factor $Q$. In particular, a smaller quality factor $Q$ may introduce larger perturbation. JPEG compression can be used for both removal and forgery attacks. In particular, given an image (watermarked or non-watermarked), the attacker perturbs the image via JPEG such that the detector may classify the perturbed image as the opposite. Note that JPEG compression is unsuccessful for forgery attack, which is validated by our experimental results in Figure~\ref{fig:forgery attack}.

\myparatight{Black-box attack} In this attack, the attacker only has access to the detection API, which returns whether an image is watermarked or not. The attacker's goal is to evade the detector by adding a perturbation to the watermarked or non-watermarked image such that the perturbed image is misclassified as the opposite. Specifically, the attacker keeps querying the API and reduces the perturbation based on the binary result~\cite{jiang2023evading}. Ideally, the attacker eventually finds an imperceptible perturbation to evade the detector. For removal attack, the attacker initializes watermarked images via JPEG compression to obtain perturbed images that are misclassified as non-watermarked by the detector. For forgery attack, we assume that the attacker collects at least one watermarked image and uses it as an initial image. In the black-box forgery attack, we set $\tau=0.63$. This is because, we note that WEvade-B-Q performs successful forgery attack on base watermarking while fails to forge watermark on our smoothed watermarking when $\tau=0.83$ by default, which further demonstrates that our smoothed watermarking is more robust against the black-box attack.

\myparatight{White-box attack} In this attack, the attacker has access to watermark decoder's parameters. To evade the detector, the attacker designs a small perturbation via solving an optimization problem using gradient descent. For removal attack, the attacker does not need to know the ground-truth watermark~\cite{jiang2023evading}. In particular, the attacker selects a target watermark uniformly at random as the target watermark.  For forgery attack, we assume the attacker knows the ground-truth watermark and uses it as the target watermark. For both attacks, the attacker optimizes the perturbation $\delta$ to maximize the similarity between the target watermark and decoded watermark in each iteration.

\myparatight{Adaptive white-box attack} In adaptive white-box attack, the attacker further takes smoothing process into consideration when finding the perturbation. The attacker knows that there are $N$ bitwise accuracy (corresponds to $N$ noisy images) during detection and the final bitwise accuracy is the median one. Thus, The attacker tries to mimic such smoothing process in this attack. Specifically, the attacker samples $N^{\prime}$ Gaussian noises $\epsilon_1,\epsilon_2,\cdots,\epsilon_{N^{\prime}}$ and adds them to the perturbed image $x+\delta$ to obtain $x+\delta+\epsilon_1,x+\delta+\epsilon_2,\cdots,x+\delta+\epsilon_{N^{\prime}}$. Then, the attacker calculates $N^{\prime}$ bitwise accuracy with respect to the target watermark, takes the median, and updates the perturbation $\delta$ based on the gradient of this noisy image. We describe the adaptive attack in Algorithm~\ref{algorithm:whitebox}. In our experiments, $N^{\prime}$ is set as 100. For removal attack, the attacker randomly selects a bitstring as the target watermark. For forgery attack, the attacker uses the ground-truth watermark as the target watermark. For both attacks, the attacker optimizes the perturbation $\delta$ to maximize the similarity between the target watermark and decoded watermark (in the smoothing setting). Note that we use double-tailed detector~\cite{jiang2023evading} in our experiments, which is more robust than single-tailed detector.

\begin{algorithm}
\renewcommand{\algorithmicrequire}{\textbf{Input:}} 
\renewcommand{\algorithmicensure}{\textbf{Output:}}
    \caption{Adaptive white-box attack}
    \label{algorithm:whitebox}
    \begin{algorithmic}[1]
        \REQUIRE Image $x$, decoder $D$, target watermark 
        $w_T$, number of iteration $n\_iter$, learning rate $\alpha$, objective function $l$, perturbation bound $R$
        \ENSURE Perturbation $\delta$
        \STATE $\delta \gets$ 0
        \FOR{$i$ = 1 to $n\_iter$}
        
            \IF{adaptive attack}
                \STATE $\epsilon^{\prime} \gets$ Smooth$\_$noise($x$, $D$, $w_T$)
                \STATE $w \gets D(x+\epsilon^{\prime}+\delta)$
            \ELSE
                \STATE $w \gets D(x+\delta)$
            \ENDIF
            
            \STATE $\delta \gets \delta - \alpha \cdot \nabla_{\delta} l(w,w_T)$
            
            \IF{${\lVert \delta \rVert}_2 > R$}
                \STATE $\delta \gets \delta \cdot \frac{R}{{\lVert \delta \rVert}_2}$
            \ENDIF

            \IF{smoothed decoder}
                \STATE $\epsilon^{*} \gets$ Smooth$\_$noise($x$, $D$, $w_T$)
                \STATE $w \gets D(x+\epsilon^{*}+\delta)$
            \ELSE
                \STATE $w \gets D(x+\delta)$
            \ENDIF
            \IF{$BA(w,w_T) \geq 1-\epsilon$}
                \STATE return $\delta$
            \ENDIF
        \ENDFOR
        \STATE return $\delta$
    \end{algorithmic}
\end{algorithm}

\begin{algorithm}   
\renewcommand{\algorithmicrequire}{\textbf{Input:}} 
\renewcommand{\algorithmicensure}{\textbf{Output:}}
    \caption{Smooth$\_$noise($x$, $D$, $w_T$)}
    \label{algorithm:smoothing}
    \begin{algorithmic}[1]
        \REQUIRE Image $x_w$, decoder $D$, watermark $w_T$, standard deviation $\sigma$, number of Gaussian noises $N^{\prime}$
        \ENSURE Gaussian noise $\epsilon$
        
        \STATE $\epsilon_1,\epsilon_2,\cdots,\epsilon_{N^{\prime}}$ $\gets$ $N^{\prime}$ Gaussian noises randomly sampled from $\mathcal{N}(0,\sigma^2I)$
        \STATE $BA_1,\cdots,BA_{N^{\prime}} \gets BA(D(x+\epsilon_{1}),w_T),\cdots,BA(D(x+\epsilon_{N^{\prime}}),w_T)$ 
        \STATE sort $BA_1,BA_2,\cdots,BA_{N^{\prime}}$
        \STATE pick $\epsilon$ among $\epsilon_1,\epsilon_2,\cdots,\epsilon_{N^{\prime}}$ such that $BA(D(x+\epsilon),w_T)$ is the median among $BA_1,BA_2\cdots,BA_{N^{\prime}}$
        \STATE return $\epsilon$
    \end{algorithmic}
\end{algorithm}

%% file: main.bbl
\begin{thebibliography}{10}
\providecommand{\url}[1]{\texttt{#1}}
\providecommand{\urlprefix}{URL }
\providecommand{\doi}[1]{https://doi.org/#1}

\bibitem{midjourney}
Midjourney. \url{https://www.midjourney.com} (2022)

\bibitem{stablediffusion}
Stable diffusion. \url{https://github.com/CompVis/stable-diffusion} (2022)

\bibitem{dalle2-database}
Dall-e 2 dataset. \url{https://dalle2.gallery} (2023)

\bibitem{dalle}
Dall-e 3. \url{https://openai.com/index/dall-e-3} (2023)

\bibitem{an2024benchmarking}
An, B., Ding, M., Rabbani, T., Agrawal, A., Xu, Y., Deng, C., Zhu, S., Mohamed, A., Wen, Y., Goldstein, T., et~al.: Benchmarking the robustness of image watermarks. arXiv preprint arXiv:2401.08573  (2024)

\bibitem{bansal2022certified}
Bansal, A., Chiang, P.y., Curry, M.J., Jain, R., Wigington, C., Manjunatha, V., Dickerson, J.P., Goldstein, T.: Certified neural network watermarks with randomized smoothing. In: International Conference on Machine Learning (2022)

\bibitem{bonferroni1936teoria}
Bonferroni, C.: Teoria statistica delle classi e calcolo delle probabilita. Pubblicazioni del R Istituto Superiore di Scienze Economiche e Commericiali di Firenze  (1936)

\bibitem{chiang2020detection}
Chiang, P.y., Curry, M., Abdelkader, A., Kumar, A., Dickerson, J., Goldstein, T.: Detection as regression: Certified object detection with median smoothing. In: Conference on Neural Information Processing Systems (2020)

\bibitem{clopper1934use}
Clopper, C.J., Pearson, E.S.: The use of confidence or fiducial limits illustrated in the case of the binomial. Biometrika  (1934)

\bibitem{cohen2019certified}
Cohen, J., Rosenfeld, E., Kolter, Z.: Certified adversarial robustness via randomized smoothing. In: International Conference on Machine Learning (2019)

\bibitem{deng2009imagenet}
Deng, J., Dong, W., Socher, R., Li, L.J., Li, K., Fei-Fei, L.: Imagenet: A large-scale hierarchical image database. In: IEEE/CVF Conference on Computer Vision and Pattern Recognition (2009)

\bibitem{stable-diffusion-watermark}
Esser, P.: Stable diffusion invisible watermark. \url{https://github.com/CompVis/stable-diffusion/blob/main/scripts/tests/test_watermark.py} (2022)

\bibitem{fang2023denol}
Fang, H., Chen, K., Qiu, Y., Liu, J., Xu, K., Fang, C., Zhang, W., Chang, E.C.: Denol: A few-shot-sample-based decoupling noise layer for cross-channel watermarking robustness. In: ACM Multimedia (2023)

\bibitem{fernandez2023stable}
Fernandez, P., Couairon, G., J{\'e}gou, H., Douze, M., Furon, T.: The stable signature: Rooting watermarks in latent diffusion models. In: International Conference on Computer Vision (2023)

\bibitem{google-synthid}
Gowal, S., Kohli, P.: Identifying ai-generated images with synthid. \url{https://deepmind.google/discover/blog/identifying-ai-generated-images-with-synthid} (2023)

\bibitem{hu2024stable}
Hu, Y., Jiang, Z., Guo, M., Gong, N.: Stable signature is unstable: Removing image watermark from diffusion models. arXiv preprint arXiv:2405.07145  (2024)

\bibitem{hu2024transfer}
Hu, Y., Jiang, Z., Guo, M., Gong, N.: A transfer attack to image watermarks. arXiv preprint arXiv:2403.15365  (2024)

\bibitem{jia2019certified}
Jia, J., Cao, X., Wang, B., Gong, N.Z.: Certified robustness for top-k predictions against adversarial perturbations via randomized smoothing. In: International Conference on Learning Representations (2019)

\bibitem{jia2022multiguard}
Jia, J., Qu, W., Gong, N.: Multiguard: Provably robust multi-label classification against adversarial examples. In: Conference on Neural Information Processing Systems (2022)

\bibitem{jiang2023ipcert}
Jiang, Z., Fang, M., Gong, N.Z.: Ipcert: Provably robust intellectual property protection for machine learning. In: International Conference on Computer Vision Workshop (2023)

\bibitem{jiang2024watermark}
Jiang, Z., Guo, M., Hu, Y., Gong, N.Z.: Watermark-based detection and attribution of ai-generated content. arXiv preprint arXiv:2404.04254  (2024)

\bibitem{jiang2023evading}
Jiang, Z., Zhang, J., Gong, N.Z.: Evading watermark based detection of ai-generated content. In: ACM Conference on Computer and Communications Security (CCS) (2023)

\bibitem{lin2014microsoft}
Lin, T.Y., Maire, M., Belongie, S., Hays, J., Perona, P., Ramanan, D., Doll{\'a}r, P., Zitnick, C.L.: Microsoft coco: Common objects in context. In: European Conference on Computer Vision (2014)

\bibitem{lukas2023leveraging}
Lukas, N., Diaa, A., Fenaux, L., Kerschbaum, F.: Leveraging optimization for adaptive attacks on image watermarks. In: International Conference on Learning Representations (2024)

\bibitem{luo2020distortion}
Luo, X., Zhan, R., Chang, H., Yang, F., Milanfar, P.: Distortion agnostic deep watermarking. In: IEEE/CVF Conference on Computer Vision and Pattern Recognition (2020)

\bibitem{nie2022diffusion}
Nie, W., Guo, B., Huang, Y., Xiao, C., Vahdat, A., Anandkumar, A.: Diffusion models for adversarial purification. In: International Conference on Machine Learning (2022)

\bibitem{openai-c2pa}
OpenAI: C2pa in dall-e 3. \url{https://help.openai.com/en/articles/8912793-c2pa-in-dall-e-3} (2024)

\bibitem{pereira2000robust}
Pereira, S., Pun, T.: Robust template matching for affine resistant image watermarks. IEEE transactions on image Processing  (2000)

\bibitem{saberi2023robustness}
Saberi, M., Sadasivan, V.S., Rezaei, K., Kumar, A., Chegini, A., Wang, W., Feizi, S.: Robustness of ai-image detectors: Fundamental limits and practical attacks. arXiv preprint arXiv:2310.00076  (2023)

\bibitem{sharma2018conceptual}
Sharma, P., Ding, N., Goodman, S., Soricut, R.: Conceptual captions: A cleaned, hypernymed, image alt-text dataset for automatic image captioning. In: Annual Meeting of the Association for Computational Linguistics (2018)

\bibitem{tancik2020stegastamp}
Tancik, M., Mildenhall, B., Ng, R.: Stegastamp: Invisible hyperlinks in physical photographs. In: IEEE/CVF Conference on Computer Vision and Pattern Recognition (2020)

\bibitem{Executive-Order}
{THE WHITE HOUSE}: {Executive Order on the Safe, Secure, and Trustworthy Development and Use of Artificial Intelligence.} \url{https://www.whitehouse.gov/briefing-room/presidential-actions/2023/10/30/executive-order-on-the-safe-secure-and-trustworthy-development-and-use-of-artificial-intelligence} (2023)

\bibitem{midjourney-database}
Turc, I., Nemade, G.: Midjourney user prompts \& generated images (250k). \url{https://www.kaggle.com/ds/2349267} (2022)

\bibitem{wang2021watermark}
Wang, R., Lin, C., Zhao, Q., Zhu, F.: Watermark faker: towards forgery of digital image watermarking. In: IEEE International Conference on Multimedia and Expo (2021)

\bibitem{wang2022diffusiondb}
Wang, Z.J., Montoya, E., Munechika, D., Yang, H., Hoover, B., Chau, D.H.: {{DiffusionDB}}: {{A}} large-scale prompt gallery dataset for text-to-image generative models. In: Annual Meeting of the Association for Computational Linguistics (2023)

\bibitem{wen2024tree}
Wen, Y., Kirchenbauer, J., Geiping, J., Goldstein, T.: Tree-rings watermarks: Invisible fingerprints for diffusion images. In: Conference on Neural Information Processing Systems (2023)

\bibitem{yoo2022deep}
Yoo, I., Chang, H., Luo, X., Stava, O., Liu, C., Milanfar, P., Yang, F.: Deep 3d-to-2d watermarking: embedding messages in 3d meshes and extracting them from 2d renderings. In: IEEE/CVF Conference on Computer Vision and Pattern Recognition (2022)

\bibitem{zhang2020udh}
Zhang, C., Benz, P., Karjauv, A., Sun, G., Kweon, I.S.: Udh: Universal deep hiding for steganography, watermarking, and light field messaging. In: Conference on Neural Information Processing Systems (2020)

\bibitem{zhao2023protecting}
Zhao, X., Wang, Y.X., Li, L.: Protecting language generation models via invisible watermarking. In: International Conference on Machine Learning (2023)

\bibitem{zhao2023invisible}
Zhao, X., Zhang, K., Su, Z., Vasan, S.V., Grishchenko, I., Kruegel, C., Vigna, G., Wang, Y.X., Li, L.: Invisible image watermarks are provably removable using generative ai. arXiv preprint arXiv:2306.01953  (2023)

\bibitem{zhu2018hidden}
Zhu, J., Kaplan, R., Johnson, J., Fei-Fei, L.: Hidden: Hiding data with deep networks. In: European Conference on Computer Vision (2018)

\end{thebibliography}
